\documentclass[11pt,reqno]{amsart}
\usepackage{amsmath,amssymb,amsthm,amsfonts}
\usepackage[letterpaper]{geometry}

\usepackage{booktabs} 
\usepackage{hyperref}
\usepackage[authoryear,square]{natbib}
\setcitestyle{nosort}

\usepackage{stmaryrd}
\usepackage{picinpar,moresize,xfrac,graphpap,dcolumn,wrapfig,graphicx}
\DeclareGraphicsExtensions{.png,.pdf,.jpg}
\usepackage{subcaption}
\usepackage{stackengine}

\usepackage{tikz,pgfplots}
\usepackage{tikz-cd}
\usepackage{pgf}
\usepgfplotslibrary{colormaps}
\usetikzlibrary{pgfplots.colormaps}
\usetikzlibrary{arrows,automata,positioning,backgrounds}
\usepackage{rotating}
\pgfplotsset{compat=newest}
\pgfplotsset{plot coordinates/math parser=false}
\newlength\figureheight
\newlength\figurewidth

\usepackage{soul,array,calc,url,ragged2e,graphpap}
\usepackage{booktabs} 
\usepackage{tabularx}
\usepackage{colortbl}
\usepackage{multirow}
\usepackage{hyphenat}
\usepackage{transparent}
\usepackage{enumerate}
\usepackage{mathrsfs}
\usepackage{mdframed}
\usepackage{acro}

\usepackage[figure]{hypcap}
\usepackage{mathtools}
\mathtoolsset{centercolon}
\usepackage{nicefrac}
\usepackage{units}
\usepackage[normalem]{ulem}
\usepackage{cancel}
\usepackage{pbox}
\usepackage{multicol}
\usepackage{cases}
\usepackage[all]{xy}
\usepackage{nicematrix}

\usepackage{adjustbox}
\usepackage{wrapfig}
\AfterEndEnvironment{wrapfigure}{\setlength{\intextsep}{0mm}}

\usepackage{cleveref}
\crefmultiformat{subequation}%
{\edef\crefstripprefixinfo{#1}(#2#1#3}%
{,#2\crefstripprefix{\crefstripprefixinfo}{#1}#3)}%
{,#2\crefstripprefix{\crefstripprefixinfo}{#1}#3}%
{,#2\crefstripprefix{\crefstripprefixinfo}{#1}#3)}

\makeatletter
\DeclareFontFamily{OMX}{MnSymbolE}{}
\DeclareSymbolFont{MnLargeSymbols}{OMX}{MnSymbolE}{m}{n}
\SetSymbolFont{MnLargeSymbols}{bold}{OMX}{MnSymbolE}{b}{n}
\DeclareFontShape{OMX}{MnSymbolE}{m}{n}{
    <-6>  MnSymbolE5
   <6-7>  MnSymbolE6
   <7-8>  MnSymbolE7
   <8-9>  MnSymbolE8
   <9-10> MnSymbolE9
  <10-12> MnSymbolE10
  <12->   MnSymbolE12
}{}
\DeclareFontShape{OMX}{MnSymbolE}{b}{n}{
    <-6>  MnSymbolE-Bold5
   <6-7>  MnSymbolE-Bold6
   <7-8>  MnSymbolE-Bold7
   <8-9>  MnSymbolE-Bold8
   <9-10> MnSymbolE-Bold9
  <10-12> MnSymbolE-Bold10
  <12->   MnSymbolE-Bold12
}{}
\let\llangle\@undefined
\let\rrangle\@undefined
\DeclareMathDelimiter{\llangle}{\mathopen}%
                     {MnLargeSymbols}{'164}{MnLargeSymbols}{'164}
\DeclareMathDelimiter{\rrangle}{\mathclose}%
                     {MnLargeSymbols}{'171}{MnLargeSymbols}{'171}
\makeatother

\usepackage{fancyvrb,listings}
\usepackage{algorithm}
\usepackage{algorithmicx}
\usepackage[noend]{algpseudocode}

\algrenewcommand\alglinenumber[1]{\footnotesize #1:}
\makeatletter  
 \renewcommand{\ALG@name}{\small Algorithm} 
\makeatother 
\algdef{SE}[DOWHILE]{Do}{doWhile}{\algorithmicdo}[1]{\algorithmicwhile\ #1}%

\newtheoremstyle{mine}{3pt}{3pt}{\itshape}{}{\bfseries}{.}{.5em}{}
\theoremstyle{mine}
\newtheorem{theorem}{Theorem}[section]
\newtheorem{lemma}{Lemma}[section]
\newtheorem{proposition}{Proposition}[section]
\newtheorem{corollary}{Corollary}[section]
\newtheorem{definition}{Definition}[section]

\newtheorem{remark}{Remark}[section]

\def\ie{\emph{i.e.}}

\def\etc{\emph{etc.}}

\def\RR{\mathbb{R}}
\def\SS{\mathbb{S}}

\def\bA{\mathbf{A}}

\def\bH{\mathbf{H}}

\def\bM{\mathbf{M}}

\def\bW{\mathbf{W}}

\def\cB{\mathcal{B}}
\def\cC{\mathcal{C}}
\def\cD{\mathcal{D}}

\def\cF{\mathcal{F}}
\def\cG{\mathcal{G}}
\def\cH{\mathcal{H}}

\def\cU{\mathcal{U}}

\def\cX{\mathcal{X}}

\def\sfE{\mathsf{E}}

\def\sfV{\mathsf{V}}

\def\sfe{\mathsf{e}}

\def\ba{\mathbf{a}}
\def\bb{\mathbf{b}}

\def\bff{\mathbf{f}}

\def\bk{\mathbf{k}}
\def\bl{\mathbf{l}}

\def\br{\mathbf{r}}

\def\bv{\mathbf{v}}

\def\bx{\mathbf{x}}

\def\bbeta{\boldsymbol{\beta}}

\def\brho{\boldsymbol{\rho}}

\def\bsigma{\boldsymbol{\sigma}}

\def\btau{\boldsymbol{\tau}}

\def\fX{\mathfrak{X}}

\usepackage{bbm}
\DeclareSymbolFont{bbold}{U}{bbold}{m}{n}
\DeclareSymbolFontAlphabet{\mathbbold}{bbold}

\def\tr {\operatorname{tr}}
\def\det{\operatorname{det}}
\def\End{\operatorname{End}}

\def\im {\operatorname{im}}
\def\ker{\operatorname{ker}}
\def\coker{\operatorname{coker}}

\def\id{\operatorname{id}}

\def\cof{\operatorname{cof}}

\DeclareMathOperator{\SE}{SE}

\renewcommand{\so}{\mathfrak{so}} 

\DeclareMathOperator{\se}{\mathfrak{se}}

\DeclareMathOperator{\Isom}{Isom}

\DeclareMathOperator{\isom}{\mathfrak{isom}}

\DeclareMathOperator{\Emb}{Emb}

\DeclarePairedDelimiterX\braket[2]{\langle}{\rangle}{#1\,\delimsize\vert\,\mathopen{}#2}

\DeclareMathOperator{\FFF}{I}
\DeclareMathOperator{\SFF}{I\!I}

\newcommand{\grad}{\mathop{\mathrm{grad}}\nolimits}
\newcommand{\sgrad}{\mathop{\mathrm{sgrad}}\nolimits}
\newcommand{\curl}{\mathop{\mathrm{curl}}\nolimits}
\newcommand{\rot}{\mathop{\mathrm{rot}}\nolimits}
\renewcommand{\div}{\mathop{\mathrm{div}}\nolimits}
\newcommand{\LD}{\mathop{\mathscr{L}}\nolimits}
\newcommand{\Hess}{\mathop{\mathrm{Hess}}\nolimits}

\newcommand{\GC}{\mathop{{\mathrm{GC}}}\nolimits}

\newcommand{\sym}
{\mathop{{\mathrm{sym}}}\nolimits}
\newcommand{\alt}
{\mathop{{\mathrm{alt}}}\nolimits}
\newcommand{\kos}
{\mathop{{\mathrm{kos}}}\nolimits}

\newcommand{\Sym}{{\rm Sym}^2}
\newcommand{\SymPos}{{\rm Sym}_{>0}^2}
\newcommand{\Symm}{{\rm Sym}^3}

\newcommand{\clamped}{{\rm clamped}}
\newcommand{\tractionfree}{{\substack{{\rm traction}\\{\rm free}}}}
\newcommand{\GCBC}{{\rm GCBC}}
\newcommand{\free}{{\rm free}}

\def\ip{i}

\def\GC{\mathrm{GC}}

\definecolor{b1}{rgb}{0.158099,0.313781,0.636957}
\definecolor{b2}{rgb}{0.525367,0.691857,0.998936}
\definecolor{g1}{rgb}{0.256000,0.640000,0.576000}
\definecolor{g2}{rgb}{0.559573,0.800781,0.760580}
\definecolor{p1}{rgb}{0.416000,0.192000,0.640000}
\definecolor{p2}{rgb}{0.680880,0.561880,0.799881}
\definecolor{r1}{rgb}{0.800000,0.200000,0.000000}
\definecolor{r2}{rgb}{1.000000,0.702595,0.603461}
\definecolor{k1}{gray}{0}
\definecolor{k2}{gray}{0.7}

\newcommand{\vcat}[2]{\begin{bmatrix}#1\\#2\end{bmatrix}}

\title{An Elasticity Complex for Kirchhoff--Love Shells}

\author{Hang Yin}
\author{Albert Chern}

\newcommand{\HY}[1]{{\footnotesize\color[rgb]{0.6,0.3,0.5}\textbf{HY} #1}}

\begin{document}

\begin{abstract}
We develop an elasticity complex for Kirchhoff--Love shell theory, based on the linearized Gauss--Codazzi equations as compatibility conditions for shell strains. The resulting shell complex forms a fine resolution of the sheaf of infinitesimal rigid motions, so its cohomology groups are determined by the corresponding sheaf cohomology. Remarkably, the complex also reveals an Airy potential representation of equilibrium stress fields. The framework further identifies natural boundary conditions for both kinematic and dynamic variables. Finally, we establish a Hodge--Morrey--Friedrichs decomposition for strain and stress fields on shells with boundary.
\end{abstract}


\maketitle

\section{Introduction}
\label{sec:Introduction}

The \emph{elasticity complex} has been a central mathematical tool in two- and three-dimensional Euclidean linear elasticity. It organizes the tensor fields associated with displacements, forces, strains, and stresses into sequences of vector spaces connected by differential operators that form a \emph{chain complexes}. These complexes share the same algebraic structure as the de Rham complex of differential forms and exterior derivatives, enabling topological analysis and structure preserving discretizations.

The standard elasticity complex on a three-dimensional domain takes the form
\begin{align}
\label{eq:ElasticityComplex3D}
\begin{aligned}
    \left\{\substack{\text{rigid}\\\text{motions}}\right\}
    \hookrightarrow\{\text{displacements}\}\xrightarrow{\nabla^{\rm sym}}\left\{\substack{\text{strain}\\\text{tensors}}\right\}
    \xrightarrow{\curl\curl}
    \left\{\substack{\text{stress}\\\text{tensors}}\right\}
    \xrightarrow{\div}\left\{\text{forces}\right\},
\end{aligned}
\end{align}
whereas in two dimensions, there are two mutually adjoint complexes:
\begin{align}
\label{eq:StrainComplex2D}
    \left\{\substack{\text{rigid}\\\text{motions}}\right\}
    \hookrightarrow
    &\{\text{displacements}\}\xrightarrow{\nabla^{\rm sym}}\left\{\substack{\text{strain}\\\text{tensors}}\right\}\xrightarrow{\rot\rot}\left\{\text{curvatures}\right\},\\
\label{eq:StressComplex2D}
    &\left\{
    \substack{\text{Airy}\\\text{potentials}}
    \right\}
    \xrightarrow{\Hess^\bot}
    \left\{\substack{\text{stress}\\\text{tensors}}\right\}
    \xrightarrow{\div}\left\{\text{forces}\right\}.
\end{align}
These diagrams satisfy the properties that the composition of any two consecutive maps vanishes, and conversely every tensor field in the kernel of a map locally lies in the image of the preceding map.  Together, these chain-complex and the stronger local-exactness properties encapsulate the \emph{Saint-Venant compatibility condition} for strain tensors (a vanishing \(\curl\curl\) implies that the strain tensor is locally the symmetrized gradient of a displacement field), as well as the \emph{Airy representation} of divergence-free stress fields as the rotated Hessian of a scalar potential (the stress function). 
A standard technique for deriving \eqref{eq:ElasticityComplex3D}--\eqref{eq:StressComplex2D} is the Bernstein--Gelfand--Gelfand (BGG) construction, which algebraically derives differential complexes for symmetric tensors from the de Rham complex.

These elasticity complexes are foundational framework for understanding the structure of tensor fields in two- and three-dimensional elasticity.  A natural question is whether such a framework can be extended to the theory of curved thin shells.

The answer is not immediately apparent from the standard formulations of the elasticity complexes. 
Unlike the de Rham complex, which is defined on general manifolds,  \eqref{eq:ElasticityComplex3D}--\eqref{eq:StressComplex2D} fundamentally concerns tensor fields in Euclidean space.  For example, geometrically, the \(\rot\rot\) operator in \eqref{eq:StrainComplex2D} measures the variation of Gaussian curvature induced by a strain tensor viewed as an infinitesimal variation of the Riemannian metric.  The corresponding Saint-Venant compatibility condition states that a metric variation induced by a displacement field within the Euclidean plane must have vanishing Gaussian curvature variation.  Without the Euclidean assumption, compatibility condition generally no longer exists: generic metric variations with nonzero curvature variations can be realized by some reparameterization displacement wherever the curvature is non-constant.  From the BGG perspective, the construction also relies on isomorphisms induced by the Euclidean structure.

In this article, we develop an elasticity complex for curved thin shells. We adopt the Kirchhoff--Love shell model, in which the shell strain depends only on the first and second fundamental forms of the surface, while allowing the shell to have arbitrary geometry.  Our derivation begins by reformulating the classical elasticity complexes to emphasize the duality between kinematic and kinetic variables.  One complex consists of kinematic fields, such as displacements and strains, while its dual consists of kinetic fields, including forces and stresses.  
More generally, the sequence of integrability conditions forms a \emph{fine resolution} of the \emph{sheaf of rigid motion on the given surface}.  
Passing to the dual complex then turns each integrability condition into an Airy-like potential representation for equilibrium fields.
This geometric viewpoint provides a systematic route from the integrability conditions of shell kinematics to potential representations of shell stresses and forces, extending the classical relationship between Saint-Venant compatibility and Airy stress functions to curved surfaces.

The resulting elasticity complex for thin shells takes the form
\begin{align}
\label{eq:ShellComplexOverview}
    \begin{aligned}
        \xymatrix@C=12pt{
        \left\{\substack{\text{global rigid}\\\text{motions}}\right\}
        \ar@{^{(}->}[r]
        \ar@{.}[d]
        &
        \left\{\substack{\text{displacement}\\\text{fields}}\right\}
        \ar[r]^-{D}
        \ar@{.}[d]
        &
        \left\{\substack{\text{variations of 1st \&}\\\text{2nd fundamental}\\\text{forms}}\right\}
        \ar[r]^-{\GC}
        \ar@{.}[d]
        &
        \left\{\substack{\text{Gauss--Codazzi}\\\text{residuals}}\right\}
        \ar@{->>}[r]
        \ar@{.}[d]
        &
        \left\{\substack{\text{global}\\\text{residual}\\\text{classes}}\right\}
        \ar@{.}[d]
        \\
        \left\{\substack{\text{total}\\\text{forces \&}\\\text{moments}}\right\}
        &
        \left\{\substack{\text{force}\\\text{fields}}\right\}
        \ar@{->>}[l]
        &
        \left\{\substack{\text{stretching \&}\\\text{bending stress}\\\text{fields}}\right\}
        \ar[l]_-{-D^*}
        &
        \left\{\substack{\text{vector}\\\text{Airy}\\\text{potentials}}\right\}
        \ar[l]_-{\GC^*}
        &
        \left\{\substack{\text{global}\\\text{Airy gauge}\\\text{modes}}\right\}
        \ar@{_{(}->}[l]
        }
    \end{aligned}
\end{align}
Here, \(D\) maps an infinitesimal displacement field (\(\RR^3\)-valued function on the surface) to the induced variations of the first and second fundamental forms (a pair of symmetric \(2\)-by-\(2\) tensor fields). The operator \(\GC\) is the linearization of the three Gauss--Codazzi equations, which are the compatibility conditions for the first and second fundamental forms.  In particular, \(\GC\) maps a pair of symmetric tensor fields to an \(\so(3)\)-valued 2-form (\(\RR^3\)-valued function after suitable identifications). 
The initial and terminal spaces of global rigid motions and residual classes are both isomorphic to the 6-dimensional space \(\se(3)\), provided that the domain consists of one connected component without clamped boundary.  On the bottom row, the corresponding spaces of global gauge fields and total momenta are \(\se(3)^*\).

The two rows of \eqref{eq:ShellComplexOverview} are formally adjoint to each other.  In particular, the space of vector Airy potentials is dual to the space of Gauss--Codazzi residuals. 
Applying the adjoint operator \(\GC^*\) to an arbitrary \(\RR^3\)-valued Airy potential produces a pair of stretching and bending stress fields in equilibrium. Conversely, every pair of equilibrial stretching and bending stress fields locally admits such an Airy representation.  This is a substantial reduction in the number of variables: the stretching and bending stresses together have six independent components at each point, whereas a vector Airy potential has only three components.

An important feature of \eqref{eq:ShellComplexOverview} is that its construction is purely geometric and does not require any constitutive models for the material.  In particular, the Airy-like representation of equilibrated stresses applies to any Kirchhoff--Love shells, linear or nonlinear, independent of the material.
Constitutive strain--stress relations can be introduced separately through Hodge-star-like operators connecting the kinematic and kinetic complexes. These operators encode the material response by mapping variations of the fundamental forms to the corresponding stretching and bending stresses. When the constitutive operator is independent of the deformation, the resulting theory describes a linear elastic shell. More generally, the operator may depend on the current embedding and arise as the linearization of a nonlinear constitutive relation. In this setting, the kinematic complex can be viewed as the linearization of the nonlinear finite deformation sequence consisting of surface embeddings, their first and second fundamental forms, and the Gauss--Codazzi compatibility equations. Thus, the same geometric framework also underlies fully nonlinear Kirchhoff--Love shell theories, with the elasticity complex describing their infinitesimal structure.

\section{Related Work}

\subsection{Foundations of Shell Theory}

The mathematical foundations of shell elasticity have been extensively developed by Ciarlet and collaborators \citep{ciarlet2000mathematical}. Their work established appropriate Sobolev spaces and Korn-type inequalities for the well-posedness of shell elasticity problems.  In particular, \cite{ciarlet2008saint} showed that the linearized Gauss--Codazzi equations are necessary and sufficient conditions for the shell strains on simply-connected surfaces.  Our approach builds on these analytical foundations while organizing the operators into a geometric elasticity complex whose global cohomology, including on multiply connected surfaces, is characterized through sheaf-theoretic techniques.  This systematic framework further reveals an Airy potential representation of equilibrium stresses and provides natural boundary conditions for Hodge decompositions on shells.

A related line of work concerns the \emph{static--geometric analogy} in shell theory.  Originating in the work of Gol'denweiser and Lur'e in the 1940s, this analogy reveals a formal correspondence between the equations of static equilibrium and geometric compatibility and motivates various stress-function representations of equilibrium solutions \cite{Goldenveizer1961Book}.
The analogy was further developed by
\cite{calladine1977static}  shallow shells, demonstrating the relationship between the classical Airy representation and Gaussian curvature.  
The duality between the classical Airy representation and the linearized Riemann curvature operator was revisited by \cite{pommaret2015airy} from the perspective of differential complexes.  While the Airy--curvature duality is well understood in the elasticity complexes (see below), it was also emphasized by \cite{Lloria:2026:LEC}, who proposed an organizing elasticity complex similar to ours. 
However, these characterizations of the Airy stress operator have largely been developed in Euclidean elasticity or shallow-shell settings, rather than for general curved shells.

A parallel line of work developed stress-function representations directly for curved shells.  In particular, six-component stress-function formulations for shell force and moment resultants were investigated in \cite{pietraszkiewicz1968multivalued,simmonds1972nonlinear,pietraszkiewicz2016resultant}. More recently, \cite{adiels2025airy} revisited the construction of stress potentials through integrated force and moment resultants with the Beltrami--G\"unther stress-function framework.  
Unlike these resultant-based formulations, our approach derives the Airy representation as the formal adjoint of the linearized Gauss--Codazzi operator.  

\subsection{Elasticity complex and finite elements}

In the theory of elasticity complexes, differential complexes provide a natural framework for describing geometric compatibility.   \cite{angoshtari2015differential} studied differential complexes for continuum mechanics in two and three dimensions, as well as on manifolds with constant sectional curvature, and suggested using the linearized Gauss--Codazzi conditions as compatibility shell complex. However, similar to \cite{ciarlet2008saint}, the dual complex is not explicitly given.

Modern research on elasticity complexes has been driven by the development of structure-preserving finite element spaces \cite{ArnoldFEEC,ArnoldHu2020,Hu2026}. 
In shell elasticity, mixed finite element methods have been developed for both linear and nonlinear models, including approaches based on Regge interpolation to alleviate membrane locking~\cite{NeunteufelSchoberl2019HHJshells,NeunteufelSchoberl2021membranelocking,NeunteufelSchoberl2024HHJTDNNS}. Regge finite elements have also been used to approximate Gaussian and Riemannian curvature from piecewise smooth metrics, with accompanying convergence analyses~\cite{GNSW2023curvature,GNSW2024liftedGauss,GNSW2023Riemann}. Related Sobolev spaces of matrix fields involving weak second-order differential operators have been investigated within the framework of 2-complexes~\cite{GHS2025twocomplex}.

\subsection{Computer Graphics}

In computer graphics, considerable effort has been devoted to developing geometry-aware discrete models of thin shells. One foundational approach is the Discrete Elastic Shells model of \cite{Grinspun:2003:DS}. 
More recently, sophisticated bending models have been introduced to improve convergence properties \cite{Chen2026BB}. 
Airy stress functions have also been employed in form finding for the design of self-supporting structures \cite{dG2013Masonry,Deuss:2014,Vouga:2019}. However, these approaches typically rely on the classical two-dimensional Airy representation of horizontal stress components, rather than a representation of the full stress field on a general curved shell.
In particular, \cite{dG2013Masonry} introduced simplicial masonry structures for the design of self-supporting surfaces of arbitrary topology. They identified harmonic components of the underlying truss networks and recognized their potential as additional degrees of freedom for structural design. \cite{dG2014Tensor} developed one of the first discrete decompositions of symmetric tensor fields on triangle meshes, although the decomposition is orthogonal only in the absence of curvature. More recently, \cite{Miki2025} employed the Schaefer--Gurtin stress-function representation to account for topological obstructions in the form finding of self-supporting structures with holes. In contrast, our framework addresses general Kirchhoff--Love shells, incorporating both membrane and bending stresses on surfaces of arbitrary genus and with boundary. The resulting elasticity complex provides a unified treatment of geometric compatibility, equilibrium stress representations, and topological harmonic components.

\subsection{Topological shells}

More recently, topological and homological methods have emerged in the study of structural mechanics.  \cite{Copperband2025origami} developed a cosheaf-homological framework for rigid origami kinematics.  In a related direction, \cite{nassar2026topological} established a topological counting rule for shells, showing that the number of isometric deformation modes change in the presence of holes or handles.  While sharing the perspective that topology govers global mechanical constraints, our work develops a continuous elasticity complex for Kirchhoff--Love shells based on the linearized Gauss--Codazzi equations.
The resulting sheaf-theoretic framework explicitly characterizes the cohomology of the shell complex and yields Hodge--Morrey--Friedrichs decompositions for strain and stress fields, including boundary conditions.

\section{Sheaf Perspective of Elasticity Complexes}
In either 2D, 3D, or thin shells, elasticity complexes are chain complexes for deformation variables (displacements, strains, \etc) and their local compatibility conditions.
Sheaf theory provides a concise language for discussing such local exact sequences of compatibility conditions and offers tools for studying global topological obstructions.
Here, we review the sheaf theoretic language and use it to characterize general elasticity complexes.

\subsection{Preliminary}
A \emph{sheaf} is a domain-dependent function space.  For a familiar example, on a compact manifold \(M\), the notation \(L^2(\cdot)\) is the sheaf of \(L^2\) functions, so that for each open set \(U\subset M\), \(L^2(U)\) is the space of all \(L^2\) functions on \(U\).  For another example, \(\Omega^k\) is the sheaf of smooth \(k\)-forms, with \(\Omega^k(U)\) be the space of \(k\)-forms on the subdomain \(U\subset M\).  In general, a sheaf \(\cF\) on \(M\) assigns a vector space \(\cF(U)\) to each open set \(U\subset M\), with each element called a \emph{section}, together with a list of axioms that allow restrictions with consistency on overlaps and that a section \(\psi\in\cF(U)\) is uniquely determined by its restrictions to \(U_i\) for any open cover \(\bigcup_{i}U_i\supset U\).

A sheaf is called \emph{fine} if there are partitions of unity to glue local sections together.  
In our case, every sheaf will be a sheaf of sections of vector bundles over a manifold.  These sheaves are always fine sheaves.

A sequence of sheaves
\begin{align}
    \cdots \to \cF^{k-1} \xrightarrow{D_{k-1}} \cF^k \xrightarrow{D_k} \cF^{k+1} \to\cdots
\end{align}
is a sequence of linear maps \(D_k\) between the vector spaces \(\cF^k(U)\) for each open set \(U\).  The sequence is an \emph{exact sequence of sheaves} if \(D_k\circ D_{k-1} = 0\), and for each kernel section \(\psi\in\ker(D_k)\subset \cF^k(U)\), there exists an open cover \(\bigcup_{i}U_i\supset U\) so that \(\psi|_{U_i}\)  is in the image of \(D_{k-1}\) on every \(U_i\).  In other words, if a section vanishes by the operator \(D_k\), then it is locally in the image of preceding map \(D_{k-1}\).
That is, an exact sequence of sheaves is a sequence of function spaces that is locally exact.

A local exact sequence does not need to be globally exact.  Inserting the entire domain \(M\) into the sheaves of an exact sequence yields a chain complex of global sections:
\begin{align}
    \cdots \to \cF^{k-1}(M) \xrightarrow{D_{k-1}} \cF^k(M) \xrightarrow{D_k} \cF^{k+1}(M) \to\cdots
\end{align}
The non-exactness of this chain complex is described by the cohomology group
\begin{align}
    H^k(\cF^\bullet(M))\coloneqq {\ker(D_k\colon \cF^k(M)\to \cF^{k+1}(M))\over \im(D_{k-1}\colon \cF^{k-1}(M) \to \cF^k(M))}.
\end{align}
The remarkable fact about this cohomology group is that if \(\cF^\bullet\) are fine (or more generally acyclic) sheaves, which will be our case for sheaves of sections of vector bundles, the cohomology is determined by the initial sheaf of the exact sequence of sheaves.  This is called the \emph{abstract de Rham--Weil theorem}.

Take the de Rham complex for example. On a manifold \(M\), one begins by considering the sheaf \(\underline{\RR}\) of constant functions, defined by \(\underline{\RR}(U) = \{c\colon U\to \RR\,|\, \text{$c$ is locally constant}\}\); in particular \(\underline{\RR}(U)\cong \RR\) if \(U\) is connected.  Then this constant sheaf, the sheaves of differential forms, and the exterior derivatives \(d_\bullet\) together form an exact sequence of fine sheaves:
\begin{align}
    0\to \underline{\RR} \to \Omega^0\xrightarrow{d_0}\Omega^1\xrightarrow{d_1}\Omega^2\xrightarrow{d_2}\cdots
\end{align}
This is called a \emph{fine resolution} of the constant sheaf \(\underline{\RR}\).  The abstract de Rham--Weil theorem then states that the sheaf cohomology \(H^k(\Omega^\bullet(M))\) depends only on \(\underline\RR\) and is independent of the choice of the fine resolution of \(\underline\RR\): 
\begin{align}
    H^k(\Omega^\bullet(M))\cong H^k(M;\underline{\RR})
\end{align} 
and in the case when \(\underline\RR(U)\cong\RR\) for connected \(U\), we have \(H^k(M;\underline\RR)\cong H^k(M;\RR)\), which is the \(k\)-th cohomology group of \(M\) with coefficients in \(\RR\).

In general, given a sheaf \(\cF\), and suppose \(\cF^\bullet\) is any fine or soft resolution of \(\cF\)
\begin{align}
    0\to\cF\to\cF^0\to\cF^1\to\cF^2\to\cdots
\end{align}
then the cohomology of the chain complex \(\cF^0(M)\to\cF^1(M)\to\cF^2(M)\to\cdots\) is independent of the resolution
\begin{align}
    H^k(\cF^\bullet(M))\cong H^k(M;\cF)
\end{align}
with \(H^k(M;\cF)\) called \emph{\(k\)-th the sheaf cohomology group of \(M\) with coefficients in \(\cF\)}.

\subsection{Elasticity Complexes as Fine Resolutions of Rigid Body Sheaves}
\label{sec:ElasticityComplexesAsFineResolutions}
We use the sheaf theoretic language to describe the general structure of linear elasticity complexes.

In continuum mechanics, the state of a continuum body is described by an embedding \(x\colon M\to W\), where \(M\) is a manifold describing the material domain, while \(W\) is a Riemannian manifold representing the world space, typically \(W\cong\RR^n\).  Given each state \(x\in\Emb(M;W)\), the tangent space \(T_x\Emb(M; W)\) collects all infinitesimal deformations, each of which is a vector field taking values in \(TW\):
\begin{align}
\label{eq:TangentOfEmbedding}
    T_x\Emb(M;W) \cong \Gamma(x^*TW).
\end{align}
Within this space of displacement fields, one takes the natural modeling assumption that rigid body motions, induced by infinitesimal isometries on \(W\), corresponds to those displacements that have no deformation.
At each \(x\in\Emb(M;W)\), define the sheaf \(\underline{\isom}\) of rigid body motions as, for each connected open set \(U\subset M\),
\begin{align}
    \underline{\isom}(U)\coloneqq \{\bv\in \Gamma_U(x^*TW)\,|\,\bv = \bk\circ x\text{ for some Killing vector field \(\bk\) on \(W\)}\}.
\end{align}
This rigid body sheaf plays a similar role as the constant sheaf in the de Rham complex.  They host displacements of no deformation.  Let \(\fX^0\) denote the sheaf of displacements:
\begin{align}
    \fX^0(U)\coloneqq \Gamma_U( x^*TW)
\end{align}
and so we have a beginning of an exact sequence \(0\to \underline{\isom}\to \fX^0\).  

Now, extend it to a fine resolution of sheaves:
\begin{align}
\label{eq:GeneralResolutionOfIsom}
    0\to \underline{\isom}\to\fX^0\xrightarrow{D_0}\fX^1\xrightarrow{D_1}\fX^2\to\cdots.
\end{align}
The first operator \(D_0\colon \fX^0\to\fX^1\) measures deformations or \emph{strains} of a given displacement fields.  Exactness states that a displacement field is in the kernel of \(D_0\) if and only if it is locally a rigid body motion.  Hence, \(\fX^1\) is interpreted as the sheaf of \emph{strain fields}.  The next operator \(D_1\) measures the compatibility of the strain fields;  that is, \(D_1S\) for some strain field \(S\in\fX^1(M)\) is the obstruction that the strain field arises from a displacement.  The type of \(D_1S\in \fX^2(M)\) is usually regarded as a curvature measure.
This compatibility condition and subsequent sheaves and maps are best demonstrated in the concrete example of the classical Euclidean elasticity complex in \Cref{sec:EuclideanElasticityComplex}.

By inserting the domain \(M\) into the sequence of sheaves \eqref{eq:GeneralResolutionOfIsom}, we obtain a chain complex
\begin{align}
\label{eq:GeneralKinematicComplex}
    0\to \underline{\isom}(M)\hookrightarrow \fX^0(M) \xrightarrow{D_0}\fX^1(M)\xrightarrow{D_1}\fX^2(M)\to\cdots
\end{align}
This is the \emph{kinematic} part of an elasticity complex where each space is associated to variables about deformations.  
The \emph{dynamic} (\emph{kinetic}) variables (such as stresses and forces) and operators (such as the divergence operator that maps stress fields to force fields) are derived concepts from the kinematic complex.  Precisely, the dynamic complex arises as the adjoint of the kinetmatic complex:
\begin{align}
\label{eq:GeneralDynamicComplex}
    0\gets \underline{\isom}(M)^*\leftarrow\fX^0(M)^* \xleftarrow{D_0^*}\fX^1(M)^*\xleftarrow{D_1^*}\fX^2(M)^*\gets\cdots
\end{align}
The dual space \(\fX^0(M)^*\) of the space \(\fX^0(M)\) displacement fields are the space of force fields, using the integral dual pairing.  Similarly, the dual \(\fX^1(M)^*\) of the space of strain fields is the space of stress fields.  The adjoint \(D_0^*\) of \(D_0\) is (minus) the divergence operator aggregating stress fields to the pointwise net force.  The dual \(\fX^{2}(M)^*\) is the space of stress potentials, representing stress fields in \(\ker(D_0^*)\) by the \(D_1^*\) of a stress potential.
Finally, the projection \(\fX^0(M)^*\to\underline{\isom}(M)^*\) adjoint to the inclusion \(\underline{\isom}(M)\hookrightarrow\fX^0(M)\) maps a force field to the total moments (Noether charges) associated to the isometry on \(W\) for each connected component of \(M\).

\subsection{Euclidean Elasticity Complex and Calabi Complex}
\label{sec:EuclideanElasticityComplex}
    For classical elasticity complexes in Euclidean space, \(M\) is \(n\)-dimensional and \(W\cong\RR^n\).  The rigid body sheaf is given by \(\underline{\isom} = \underline{\se(n)}\) defined by
\begin{align}
    \underline{\se(n)}(U) = \{\bv\colon M\to\RR^n\,|\,\bv = \bA x + \bb, \bA\in\so(n),\bb\in\RR^n\}
\end{align}
for each connected subdomain \(U\subset M\).  The sheaf \(\fX^0\) of displacements is \(\fX^0(U) = C^\infty(U;\RR^n)\), the sheaf of \(\RR^n\)-valued functions.
The sheaf \(\fX^1\) of strain fields is given by the sheaf of symmetric bilinear forms 
\begin{align}
    \fX^1(U)=\Gamma_U(\Sym T^*M),
\end{align}
and the operator \(D_0\) is the symmetrized gradient that measures the rate of change of the induced metric \(D_0(\bv)\coloneqq {1\over 2}\LD_\bv \langle\cdot,\cdot\rangle_{\RR^n}\), \((D_0\bv)(\ba,\bb) = {1\over 2}(\langle \nabla_\ba\bv,\bb\rangle+\langle\nabla_\bb\bv,\ba\rangle)\) for each \(\ba,\bb\in TM\).  

An infinitesimal change of metric \(\mathring g\in \fX^1(M)\) locally arises from a displacement within the Euclidean space if and only if the induced curvature vanishes.  Hence, the operator \(D_1\) that encodes the compatibility condition measures the induced change of Riemann curvature, with \(\fX^2\) being the sheaf of algebraic curvature tensors
\begin{align}
\label{eq:CurvatureSheafKNProduct}
    \fX^2(U) &= \Gamma_U(\owedge^2\Sym T^*M) \\
    &=\left\{ R\in\Gamma_U(\Sym\wedge^2T^*M)\,\Big|\, R_{ijk\ell}+R_{jki\ell}+R_{kij\ell}=0 \right\}.
    \label{eq:CurvatureSheafFirstBianchi}
\end{align}
where \(\owedge\) in \eqref{eq:CurvatureSheafKNProduct} is the Kulkarni--Nomizu product, which gives rise to symmetric bilinear form-valued 2-forms satisfying the first Bianchi identity as characterized in \eqref{eq:CurvatureSheafFirstBianchi}.  Explicitly, for \(\mathring R = D_1\mathring g\), we have the linearized curvature-from-metric expression
\begin{align}
    \mathring R_{ijk\ell} = {1\over 2}(\partial_i\partial_k\mathring g_{j\ell} + \partial_j\partial_\ell \mathring g_{ik} - \partial_i\partial_\ell \mathring g_{jk} - \partial_j\partial_k\mathring g_{i\ell}) = {1\over 2}(\Hess\owedge \mathring g)_{ijk\ell}
\end{align}
which is known as the classical Saint-Venant compatibility operator.  

In both two dimension and three dimenions (\(n=2,3\)), the first Bianchi identity holds automatically.  Thus, in these cases \(\owedge^2\Sym T^*M = \Sym\wedge^2 T^*M\).

In 2D (\(n=2\)), the fibers of \(\wedge^2 T^*M\) is one-dimensional, and thus sections of \(\Sym\wedge^2 T^*M\) can be represented by a scalar field, in the same way that a Riemann curvature reduces to a Gaussican curvature 2-form.

In 3D (\(n=3\)), the bundle \(\Sym\wedge^2 T^*M\) of symmetric bilinear form-valued 2-forms are often identified with the bundle \(\Sym TM\) of symmetric bivectors via \(\wedge^2 T^*M\cong TM\), \(\ip_{\bv}\mu \leftrightarrow\bv\),  using a volume form \(\mu\in\Gamma(\wedge^3 T^*M)\).

In dimensions \(n\geq 3\), the curvature variations have a differential compatibility conditions: the second Bianchi identity.  This leads to the next map in the resolution \(D_2\colon \fX^2\to\fX^3\).  Explicitly, for \(\mathring B = D_2\mathring R\), we have
\begin{align}
    \mathring B_{ijk\ell m} = \partial_{i}\mathring R_{jk\ell m} + \partial_j\mathring R_{ki\ell m} + \partial_{k}\mathring R_{ij\ell m},
\end{align}
with \(\mathring B\) be of the type \(\mathring B\in \Gamma(\wedge^3 T^*M \otimes \wedge^2 T^*M)\).
In 3D (\(n=3\)), under the volume-form-induced identifications \(\owedge^2\Sym T^*M\cong \Sym TM\) and \(\wedge^3T^*M\otimes\wedge^2 T^*M\cong TM\), the operator \(D_2\) becomes the divergence operator \(\mathring B^i=\partial_m\mathring R^{mi}\).

For this reason, in 3D, the target space \(\fX^2(M)\) of the Saint-Venant operator \(D_1\) is often identified with the space of stress fields \(\Gamma_M(\Sym TM)\), and \(\fX^3(M)\) is identified with the force fields, with the divergence operator that sends stress fields to force fields taking the role of \(D_2\), the second Bianchi operator.
However, for developing a more general theory, we refrain ourselves from making this identification.  The space of stress fields will be kept as the dual \(\fX^1(M)^*\) of the space of strain field \(\fX^1(M)\); up to a volume form for the integral dual pairing, the stress space is indeed \(\Gamma_M(\Sym T^*M)^*\cong\Gamma_M(\Sym TM)\).

This classical elasticity complex structure generalizes to cases where the ambient space \(W\supseteq M\) is a Riemannian manifold that admits a nontrivial sheaf \(\underline{\isom}\) of Killing vector fields.  The most prominent example is manifolds with constant sectional curvature, in which case the sequence 
\begin{align}
\label{eq:CalabiComplex}
    0\to\underbrace{\underline{\isom}}_{\substack{\text{Killing}\\\text{fields}}}\to
    \underbrace{\Gamma(TM)}_{\substack{\text{displace-}\\\text{ments}}}
    \xrightarrow{D_0}
    \underbrace{\Gamma(\Sym T^*M)}_{\substack{\text{metric}\\\text{variations}}}
    \xrightarrow{D_1}
    \underbrace{\Gamma(\owedge^2\Sym T^*M)}_{\substack{\text{curvature}\\\text{variations}}}
    \xrightarrow{D_2}
    \underbrace{\fX^3}_{\substack{\text{2nd Bianchi}\\\text{residuals}}}
    \to\cdots
\end{align}
is called the \emph{Calabi complex}, whose cohomology groups are the \emph{Killing sheaf cohomology} \cite{Khavkine2017Calabi}.

\subsection{Nonlinear Elasticity}
Even though the theory of \Cref{sec:ElasticityComplexesAsFineResolutions} describes linear elasticity complex, its applicability extends to general \emph{nonlinear} elasticity systems, where deformations are not limited to small displacements.  Here, we discuss the general picture of nonlinear analogues of the elasticity complex and general constitute models for nonlinear hyperelasticity.  We will find that the equation of motion still boils down to the interplay between the linear complexes of \Cref{sec:ElasticityComplexesAsFineResolutions}. 

The linear kinematic elasticity complex \eqref{eq:GeneralKinematicComplex} and its Euclidean (or constant curvature) example \eqref{eq:CalabiComplex} are linearization of more general nonlinear elasticity systems.  In the example of \eqref{eq:CalabiComplex}, the Killing field (infinitesimal isometries) can be replaced isometry group actions, infinitesimal displacements by all embeddings \(\Emb(M;W)\), metric variations by space of all Riemannian metrics \(\{g\in\Gamma(\Sym T^*M)\,|\, g\text{ is positive definite}\}\), and the space of curvature variations is replaced by the space of all curvature tensors.  The map from each of these nonlinear spaces are replaced by the standard operators that take the induced metric of an embedding and the induced curvature from a metric; these operators are nonlinear in general. 
Let us now formalize this correspondence between nonlinear complexes and linear complexes.

Recall \eqref{eq:TangentOfEmbedding}
that \(\fX^0(U)\) is the tangent space \(T_ x\Emb(M;W)\).  Treating the entire state space \(\Emb(M;W)\) as the nonlinear space counterpart, consider the sheaf \(\cX^0\) of sets (instead of a sheaf of linear function space)
\begin{align}
    \cX^0(U) \coloneqq \Emb(U;W).
\end{align}
The sheaf \(\underline{\Isom}\) of groups, defined by that \(\underline{\Isom}(U)\cong\Isom_W\) (the group of isometries on \(W\)) for each connected \(U\), acts on \(\cX^0\) by left composition.  Then there is a map \(\cD_0\) that maps each \(\cX^0(U)\) to the space \(\cX^1(U)\) of finite strain fields over \(U\)
\begin{align}
    \underline{\Isom}\to \cX^0\xrightarrow{\cD_0}\cX^1
\end{align}
so that \(\cD_0 x = \cD_0 y\) if and only if \( x, y\) are related by (connected-componentwise) isometry actions.  That is, the finite strain field encodes true deformations distinguishing configurations not related by isometries.   Taking linearization amounts to setting \(\fX^1(U) = T_{g}\cX^1(U)\) for each finite strain data \(g\in \cX^1(U)\) and \(D_0 = d\cD_0|_{ x}\) for each \( x\in \cX^0(U)\), which yields the local exactness sequence \(0\to\underline{\isom}\to\fX^0\xrightarrow{D_0}\fX^1\).
Next, the geometric data \(g\in\cX^1(U)\) often come with integrability conditions, captured by a map (often nonlinear) \(\cD_1\colon\cX^1(U)\to\cX^2(U)\) where \(\cX^2(U)\) collects the space of residual fields for incompatibility.  The resolution of sheaves of linear spaces \eqref{eq:GeneralResolutionOfIsom} is then obtained by the tangent maps of the sequence of geometric integrability conditions
\begin{align}
\label{eq:GeneralNonlinearResolutionOfIsom}
    \underbrace{\underline{\Isom}}_{\substack{\text{comonentwise}\\\text{isometry}\\\text{actions}}}\to
    \underbrace{\cX^0}_{\substack{\text{embeddings}\\\text{of material}\\\text{in world}}}
    \xrightarrow{\cD_0}
    \underbrace{\cX^1}_{\substack{\text{induced}\\\text{finite}\\\text{strains}}}
    \xrightarrow{\cD_1}
    \underbrace{\cX^2}_{\substack{\text{curvature}}}
    \xrightarrow{\cD_2}
    \underbrace{\cX^3}_{\substack{\text{incompatibility}\\\text{of curvature}}}
    \to\cdots
\end{align}
with \(\fX^\bullet = T\cX^\bullet\) and \(D_\bullet = d\cD_\bullet\).

As a technical note, we require the sheaves \(\cX^1, \cX^2,\ldots\) to be fiber bundles (\ie, each \(\cX^i(U)\) is the set of sections of a fiber bundle over \(U\)) with fibers being possibly nonlinear manifolds, so that \(\fX^1,\fX^2,\ldots\) are sheaves of vector bundles through fiberwise linearization.  

Any nonlinear resolution \eqref{eq:GeneralNonlinearResolutionOfIsom} sets up the stage for a general nonlinear elasticity constitute model.
A hyperelastic model assigns an energy density \(\Psi\) per fiber of \(\cX^1\).  This constitutive model defines a total elastic potential energy \(\cU\) for each state \( x\in\cX^0(M) = \Emb(M;W)\) by \(\cU( x)\coloneqq \int_M \Psi(\cD_0 x)\). Now, even for this general nonlinear model, the force \(\bff\coloneqq - {\delta\cU\over\delta x}|_{ x}\) at each state \( x\in\cX^0(M)\) is derived by taking a linear perturbative analysis of \(\cU\).
Explicitly, 
\begin{align}
    \left\llangle{\delta\cU\over\delta x}\Big|_{ x},\mathring x\right\rrangle = \int_M (\Psi'|_{\cD_0 x})(\underbrace{d\cD_0|_{ x}}_{D_0}\mathring x),\quad\text{for \(\mathring x\in T_ x\cX^0(M) = \fX^0(M)\).}
\end{align}
Here \(\Psi'\) is the fiberwise derivative of the fiberwise function \(\Psi\).
This shows that \(\bff = - {\delta\cU\over\delta x}|_{ x} =  -D_0^*(\Psi'|_{\cD_0 x})\), which is minus the divergence (\(D_0^*\)) of the stress tensor \(d\Psi|_{\cD_0 x}\in \fX^1(M)^*\), where \(\fX^1(M) = T_{\cD_0 x}\cX^1(M)\).  The map \(\Psi'\colon \cX^1(M)\to T\cX^1(M)\), \(\cD_0 x\mapsto \Psi'|_{\cD_0 x}\), is the stress--strain relation.

\section{Smooth Complex for Thin Shells}
Thin shells are the limit of elastic bodies taking the form of a surface whose thickness is much smaller than other length scale, such as its curvature radius.  In this limit, the state of the shell is represented by its central surface.  A Kirchhoff--Love shell model is an elastic shell whose elastic energy depends only on the first and second fundamental forms of the surface.  In particular, one assumes that the transversal shear is negligible.  Then, Kirchhoff--Love shell models fit in the framework where the material space \(M\) is a 2D surface while the world space is the Euclidean 3-space \(W = \RR^3\).

\subsection{Nonlinear Shell Complex}

Starting with the sheaf \(\underline{\SE(3)}\) of isometry actions, the sequence \eqref{eq:GeneralNonlinearResolutionOfIsom} for differential geometry of surfaces is
\begin{align}
\label{eq:NonlinearShellComplex}
    \underbrace{\underline{\SE(3)}}_{\substack{\text{comonentwise}\\\text{rigid body}\\\text{transforms}}}\to
    \underbrace{\cX^0}_{\substack{\text{surface}\\\text{embeddings}}}
    \xrightarrow{\cD_0=\cD}
    \underbrace{\cX^1}_{\substack{\text{first \& second}\\\text{fundamental}\\\text{forms}}}
    \xrightarrow{\cD_1=\cG\cC}
    \underbrace{\cX^2}_{\substack{\text{Gauss--Codazzi}\\\text{residuals}}}.
\end{align}
Here, \(\cD_0\), denoted also by \(\cD\), maps a given surface embedding \( x\) to the induced first and second fundamental forms \((g_ x,\SFF_ x)\).  The next map \(\cD_1\), denoted by \(\cG\cC\), encodes the Gauss--Codazzi equations in the form of \(\cG\cC(g,\SFF) = 0\), which are the compatibility conditions for a given fundamental form data \((g,\SFF)\).  \emph{Bonnet's fundamental theorem of surfaces} state that if the Gauss--Codazzi equations \(\cG\cC(g,\SFF)\) is satisfied by a given set of fundamental forms \((g,\SFF)\), then locally there exists an immersion \( x\) realizing the fundamental forms \((g_ x,\SFF_ x) = (g,\SFF)\) and this realizing surface is unique up to rigid body congruence.  
In the following, we express \(\cD, \cG\cC\) explicitly.

The space of surface embeddings and fundamental forms are
\begin{align}
    \cX^0(M) = \Emb(M;\RR^3),\quad \cX^1(M) = \Gamma\left(\SymPos T^*M \oplus \Sym T^*M\right)
\end{align}
where \(\SymPos T^*M\subset \Sym T^*M\) denote the cone of positive definite symmetric bilinear forms.  

Given a surface \( x\in\cX^0(M)\), define the deformation gradient
\begin{align}
    F\coloneqq d x \in \Omega^1(M;\RR^3)
\end{align}
and its unit normal
\begin{align}
    \vec n\in C^\infty(M;\SS^2)\quad\text{so that}\quad \langle\vec n,F(\cdot)\rangle_{\RR^3} = 0,
\end{align}
for which one either assumes \(M\) is orientable, or regards \(\vec n\) as a twisted form.
Then the first and second fundamental forms from the map \(\cD_0\colon x\mapsto (g_ x\in\Gamma(\SymPos T^*M),\SFF_ x\in \Gamma(\Sym T^*M))\) is given by, for each \(X,Y\in TM\),
\begin{align}
    &g_ x(X,Y) \coloneqq \langle FX,FY\rangle_{\RR^3},\\
    &\SFF_ x(X,Y)\coloneqq \langle d\vec n (X),FY\rangle_{\RR^3}.
\end{align}
By treating both symmetric bilinear forms as maps from vectors to covectors, these induced fundamental forms can be written as
\begin{align}
    g_ x = F^\intercal\flat_{\RR^3}F,\quad\SFF_{ x} = F^\intercal\flat_{\RR^3} d\vec n = (d\vec n)^\intercal \flat_{\RR^3}F.
\end{align}

The Gauss--Codazzi equations for a given first and second fundamental forms \((g,\SFF)\in\Gamma(\SymPos T^*M \oplus \Sym T^*M)\) are expressed as follows.  
First, the metric \(g\) unqiuely defines a Levi-Civita connection \(\nabla\) for the tangent bundle.  The connection induces a curvature \(R^\nabla\in\Omega^2(M;\End(TM))\) by
\begin{align}
    R^\nabla Z = d^\nabla(\nabla Z),\quad Z\in\Gamma(TM).
\end{align}
Define the \emph{shape operator} using both operators fundamental forms
\begin{align}
    S \coloneqq g^{-1}\SFF \in \Gamma(\End(TM))\cong\Omega^1(M;TM),
\end{align}
which is an endomorphism field over \(M\), identified as a tangent-vector valued 1-form.
The compatibility conditions for \((g,\SFF)\) are
\begin{align}
\label{eq:GaussEquationOriginal}
    &R^\nabla = -\det(S)J\mu&&\text{(Gauss equation)}\\
\label{eq:CodazziEquationOriginal}
    &d^\nabla S = 0 &&\text{(Codazzi equation)}
\end{align}
where \(\mu\in\Omega^2(M)\) is the area form and \(J\in\Gamma(\End(TM))\) is the \(90^\circ\) rotation operator, both induced by \(g\), satisfying the relation
\begin{align}
\label{eq:muJConvention}
    \mu(X,Y) = g(JX,Y),\quad X,Y\in TM.
\end{align}
\begin{lemma}
    The Gauss equation \eqref{eq:GaussEquationOriginal} is equivalent to
    \begin{align}
    \label{eq:GaussEquation}
        gR^\nabla + {1\over 2}\SFF\owedge\SFF=0.
    \end{align}
\end{lemma}
Here, \((gR^\nabla)_{ijk\ell} = g_{\ell m}{(R^\nabla)_{ij}}^m_k\) is the fully covariant curvature tensor, and the Kulkarni--Nomizu product \(\owedge\) produces an algebraic curvature from two symmetric bilinear forms by \((A\owedge B)_{ijk\ell} = A_{ik}B_{j\ell}+A_{j\ell}B_{ik} - A_{i\ell}B_{jk}-A_{jk}B_{i\ell}\).
\begin{proof}
    The relation \eqref{eq:muJConvention} gives us \(g_{\ell m}{J^m}_k = \mu_{k\ell}\).  Therefore, multiplying the flat operator \(g\) from the left on both sides of \eqref{eq:GaussEquationOriginal} yields 
    \begin{align}
        (gR^\nabla)_{ijk\ell} = -\det(S)\mu_{ij}\mu_{k\ell}.
    \end{align}
    Since \(\det(S) = \det(\SFF)/\det(g)\) and \(\mu_{ij} = \sqrt{\det(g)}\epsilon_{ij}\), we get
    \begin{align}
        \det(S)\mu_{ij}\mu_{k\ell} = \det(\SFF)\epsilon_{ij}\epsilon_{k\ell} = \SFF_{ik}\SFF_{j\ell} - \SFF_{i\ell}\SFF_{jk} = {1\over 2}(\SFF\owedge\SFF)_{ijk\ell}.
    \end{align}
\end{proof}

\begin{lemma}
    The Codazzi equation \eqref{eq:CodazziEquationOriginal} is equivalent to
    \begin{align}
    \label{eq:CodazziEquation}
        d^\nabla\SFF = 0,
    \end{align}
    where \(\SFF\) is regarded as a covector-valued 1-form, and \(d^\nabla\) is the covariant exterior derivative using the Levi-Civita connection for covectors.
\end{lemma}
In coordinates, \eqref{eq:CodazziEquation} reads as \(\nabla_i\SFF_{jk} - \nabla_j\SFF_{ik} = 0\).
\begin{proof}
    The metric compatibility of \(\nabla\) ensures \(d^\nabla \SFF = d^\nabla (\flat_{g}S) = \flat_g d^\nabla S\).
\end{proof}

\begin{remark}[Levi-Civita connections on \(TM\) and \(T^*M\)]
    Since \eqref{eq:CodazziEquation} involves the Levi-Civita connection on covectors, we clarify its relationship with the connection on vectors.
    The Levi-Civita connection \(\nabla^{TM}\) on the tangent bundle induces a connection \(\nabla^{T^*M}\) on the cotangent bundle through
    \begin{align}
        d\langle \alpha|Z\rangle
        = \langle\nabla^{T^*M}\alpha | Z\rangle
        + \langle \alpha|\nabla^{TM}Z\rangle,
        \quad \alpha\in\Gamma(T^*M),\ Z\in\Gamma(TM).
    \end{align}
    In coordinates, if
    \((\nabla^{TM}Z)_i^{k} = \partial_i Z^k + \Gamma_{ij}^k Z^j\),
    then
    \((\nabla^{T^*M}\alpha)_{ij} = \partial_i\alpha_j - \Gamma_{ij}^k \alpha_k\).
    Note that the two connections involve the same Christoffel symbols \(\Gamma_{ij}^k\), but the connection matrices for covectors are the negative transposes of those for vectors.

    We omit the superscripts \(TM, T^*M\) on \(\nabla\) as the appropriate connection is clear from the type of tensor on which it acts.
\end{remark}

\begin{definition}[Gauss--Codazzi operator]
    Define the Gauss--Codazzi operator
    \begin{align}
        &\cG\cC\colon\underbrace{\Gamma\left(\SymPos T^*M \oplus \Sym T^*M\right)}_{\cX^1(M)}
        \to
        \underbrace{
        \Gamma\Big(\owedge^2\Sym T^*M \oplus (\wedge^2T^*M)\otimes T^*M\Big)}_{\cX^2(M)}
    \end{align}
    by
    \begin{align}
        &\cG\cC(g,\SFF)\coloneqq \bigg(\underbrace{gR^\nabla + {1\over 2}\SFF\owedge\SFF}_G, \underbrace{d^\nabla \SFF}_Q\bigg).
    \end{align}
    The quantities \(G\) and \(Q\) are called the Gauss--Codazzi curvatures or the (nonlinear) Gauss--Codazzi residuals.
\end{definition}
\begin{theorem}[Bonnet's Fundamental Theorem]
\label{thm:BonnetTheorem}
    The sequence \eqref{eq:NonlinearShellComplex} of sheaves is exact.  That is, \(\cG\cC\circ\cD = 0\), and conversely,
    if \((g,\SFF)\in\cX^1(M)\) satisfies \(\cG\cC(g,\SFF) = 0\), then there exists an open cover \(\bigcup_i U_i=M\) and \( x_i\in\cX^0(U_i)\) so that \((g,\SFF)|_{U_i} = \cD  x_i\), with each \( x_i\) unique up to an  \(\SE(3)\) action.
\end{theorem}

\subsection{Linearized Shell Complex}
The linearized shell complex is obtained by taking the tangent map of \eqref{eq:NonlinearShellComplex} at a surface \(x\in\cX^0(M)\) with fundamental forms \((g,\SFF) = \cD x\in\cX^1(M)\), which satisfies \(\cG\cC(g,\SFF) = 0\in \cX^2(M)\).
\begin{definition}
    Let \(\fX^0 \coloneqq T_x\cX^0\) (namely, \(\fX^0(U) = T_x(\cX^0(U))\)) be the sheaf of variations \(\mathring x\) at \(x\).  Similarly, let \(\fX^1 \coloneqq T_{(g,\SFF)}\cX^1\) be the sheaf of variations of the fundamental forms \((\mathring g,\mathring\SFF)\), and let \(\fX^2 \coloneqq T_0\cX^2\) be the sheaf of linearized Gauss--Codazzi residuals.  
\end{definition}

Each of these sheaves correspond to vector bundles:
\begin{align}
    &\fX^0(U) = \Gamma_U\left(x^* T\RR^3\right) = C^\infty(U;\RR^3),\\
    &\fX^1(U) = \Gamma_U\left(\Sym T^*M \oplus \Sym T^*M\right),\\
    &\fX^2(U) = \Gamma_U\Big(\owedge^2\Sym T^*M \oplus (\wedge^2T^*M)\otimes T^*M\Big).
\end{align}

\begin{definition}
    Define the linearized deformation operator \(D\coloneqq (d\cD)_x\colon \fX^0\to\fX^1\) and the linearized Gauss--Codazzi operaator  \(\GC \coloneqq (d\cG\cC)_{(g,\SFF)}\colon\fX^1\to\fX^2\).
\end{definition}

The following theorem states an important property about the linearized Gauss--Codazzi operator that there is no further local compatibility following \(\fX^2\).

\begin{theorem}\label{thm:GCLocalSurjectivity}
    The \(\GC\) operator is locally surjective.  That is, on a contractable \(U\subset M\), for any \((\mathring G,\mathring Q)\in \fX^2(U)\), there exists \((\mathring g,\mathring\SFF)\in\fX^1(U)\) so that \(\GC(\mathring g,\mathring\SFF) =  (\mathring G,\mathring Q)\).
\end{theorem}

Lastly, we linearize the sheaf \(\underline{\SE(3)}\) of componentwise rigid body transformations at the surface \(x\).

\begin{definition}
    Define the rigid motion sheaf \(\underline{\se(3)}\) by that for each connected open set \(U\subset M\),
    \begin{align}
        \underline{\se(3)}(U)\coloneqq \left\{\mathring x\colon U\to\RR^3\middle|\mathring x = \vec\omega\times x + \vec b\text{ for some \(\vec\omega,\vec b\in\RR^3\)}\right\}.
    \end{align}
\end{definition}
Then, directly following from the exactness of \eqref{eq:NonlinearShellComplex} (\Cref{thm:BonnetTheorem}), and the local surjectivity of \(\GC\) (\Cref{thm:GCLocalSurjectivity}), we have the following exact sequence of sheaves.
\begin{theorem}
The sequence of sheaves 
\begin{align}
    0\to\underline{\se(3)} \hookrightarrow \fX^0 \xrightarrow{D_0 = D}\fX^1\xrightarrow{D_1 = \GC}\fX^2\to 0
\end{align}
is a fine resolution of \(\underline{\se(3)}\).
\end{theorem}
\begin{proof}
    The exactness of \(0\to\underline{\se(3)} \hookrightarrow \fX^0 \xrightarrow{D}\fX^1\xrightarrow{\GC}\fX^2\) follows from taking variations of the Bonnet theorem.  The exactness of the end of the sequence \(\fX^1\xrightarrow{\GC}\fX^2\to 0\) follows from the local surjectivity of the linearized Gauss--Codazzi operator.
\end{proof}
\begin{corollary}
\label{cor:ShellCohomology}
The chain complex
\begin{align}
\label{eq:ShellCohomology}
    0\to\underline{\se(3)}(M)\hookrightarrow\fX^0(M)\xrightarrow{D_0 = D}\fX^1(M)\xrightarrow{D_1 = \GC}\fX^2(M)
\end{align}
has cohomology 
\begin{align}
\label{eq:ShellCohomology}
    H^i_{\rm shell}(M) = {\ker(D_i)\over \im(D_{i-1})}\cong H^i(M;\se(3)).
\end{align}
\end{corollary}
These cohomology groups arise from the abstract de Rham--Weil theorem.
Note that if \(M\) has boundary, where certain boundary conditions may apply to fields in \(\fX^i\), there can be different cohomology groups.  When there is boundary, we still formally obtain \eqref{eq:ShellCohomology} by de Rham--Weil, and refer to it as the \emph{absolute cohomology}, in contrast to any other relative cohomology with boundary conditions.

Now, \Cref{cor:ShellCohomology} implies that we have a characteristic class in
\begin{align}
    H^2(M;\se(3))\cong \bigoplus^{\substack{\text{$\#$ connected}\\
    \text{components}}}\se(3)
\end{align}
that can be assigned for elements in \(\fX^2(M)\).  
This class describes a global conservation law for the Gauss--Codazzi operator.
While the linearized Gauss--Codazzi operator is locally surjective, it is not globally surjective as captured by the cohomology \(H^2(M;\se(3))\).  
\begin{corollary}
    A field Gauss--Codazzi curvature \((\mathring G,\mathring Q)\in\fX^2(M)\) is in the image of \(\GC\) if and only if the projection of \((\mathring G,\mathring Q)\) to \(H^2(M;\se(3))\) vanishes.
\end{corollary}
By capping off \eqref{eq:ShellCohomology} by the projection from \(\fX^2(M)\) to \(H^2(M;\se(3))\), we obtain the kinematic shell complex:
\begin{align}
\label{eq:KinematicShellComplex}
    0\to\underbrace{\underline{\se(3)}(M)}_{
    \substack{\text{global rigid}\\\text{motions}}
    }
    \hookrightarrow
    \underbrace{\fX^0(M)}_{
    \substack{\text{displace-}\\\text{ments $\mathring x$}}
    }
    \xrightarrow{D}
    \underbrace{\fX^1(M)}_{\substack{\text{strains}\\
    (\mathring g,\mathring \SFF)}
    }
    \xrightarrow{\GC}
    \underbrace{\fX^2(M)}_{
    \substack{\text{Gauss--Codazzi}\\\text{residuals $(\mathring G,\mathring Q)$}}
    }
    \twoheadrightarrow 
    \underbrace{H^2(M;\se(3))}_{
    \substack{\text{global}\\\text{residual}\\\text{classes}}
    }\to 0
\end{align}

Next, we construct the adjoint for the dual, dynamic part of the shell complex.  The formal dual complex consists of sections of dual vector bundles
\begin{subequations}
\begin{align}
    &\fX^0(U)^* = C^\infty(U;\RR^3)\\
    &\fX^1(U)^* = \Gamma_U(\Sym TM\oplus \Sym TM)\\
    &\fX^2(U)^* = C^\infty(U;\RR)\oplus\Gamma_U(TM)
\end{align}
\end{subequations}
through the following integral pairings
\begin{subequations}
\label{eq:DualPairingMain}
\begin{align}
    &\llangle f,\mathring x\rrangle_U\coloneqq\int_U \langle f,\mathring x\rangle_{\RR^3}\, \mu_g,\\
    &\llangle(\sigma,\tau),(\mathring g,\mathring\SFF)\rrangle_U\coloneqq \int_U (\sigma^{ij}\mathring g_{ij}+\tau^{ij}\mathring\SFF_{ij})\, \mu_{g},\\
    &\llangle(\varphi,\psi),(\mathring G,\mathring Q)\rrangle_U\coloneqq \int_U \varphi\mathring G_{ijk\ell}{(\mu_g^{-1})^{ij}(\mu_g^{-1})^{k\ell}\over 4}\mu_g + \int_U\psi^k\mathring Q_{ijk}{(\mu_g^{-1})^{ji}\over 2}\, \mu_g
\end{align}
\end{subequations}
for \(\mathring x\in\fX^0(U)\), \((\mathring g,\mathring\SFF)\in\fX^1(U)\), \((\mathring G,\mathring Q)\in\fX^2(U)\), \(f\in \fX^0(U)^*\), \((\sigma,\tau)\in \fX^1(U)^*\), and \((\varphi,\eta)\in \fX^2(U)^*\).  Here, \(\mu_g\) is the area 2-form induced by the metric \(g\).

We will adopt the notation that \(\mu^{ij} = (\mu_g)^{ij}\), \ie\@ the subscript \(g\) is omitted, and the symplectic form with upper index denotes its inverse: \(\mu^{ij}\mu_{jk} = \delta_k^i\).  We will also use the notation \(\int(\cdots)\mu_g = \int(\cdots)\, dA_g\) for integration measure interchangeably to avoid confusion whenever the integrand also involves expression involving \(\mu\).

\begin{remark}[Orientability]
    Here the area element is identified as a 2-form by assuming \(M\) is orientable.  If \(M\) is non-orientable, \(\mu\) is regarded as a suitable twisted form.
\end{remark}

\begin{remark}[Explicit index relation among \(g,\mu,J\)]
    From the relationship \eqref{eq:muJConvention} between \(g,\mu\) and \(J\), we have
    \begin{align}
    \label{eq:muJConventionIndex}
        \mu_{ij} = g_{jm}J_i^m,\quad \mu^{ij} = -J_m^ig^{mj}.
    \end{align}
\end{remark}

The dual variables in \eqref{eq:DualPairingMain} have physical meanings.
The variable \(f\) is force field.  The symmetric bivectors \(\sigma,\tau\) are the stretching stress and bending stress.  The variables \((\varphi,\psi)\) are the Airy potentials.
\begin{remark}[Factor of \(2\) in \(\sigma\)]
    In classical convention, the strain rate \(E = {1\over 2}\mathring g\) uses half of the rate of change of the induced metric, which is to be paired with stress \(\sigma_{\rm conv}\) as \(\llangle \sigma_{\rm conv},E\rrangle\).  Our stretching stress \(\sigma={1\over 2}\sigma_{\rm conv}\) is half of the conventional stress, nevertheless yielding the same strain--stress pairing \(\llangle \sigma_{\rm conv},E\rrangle = \llangle \sigma, \mathring g\rrangle\).
\end{remark}

\begin{definition}
    The formal adjoints \(D^*\colon \fX^1(U)^*\to\fX^0(U)^*\) and \(\GC^*\colon\fX^2(U)^*\to\fX^1(U)^*\) of \(D\) and \(\GC\) are defined so that
    \begin{align}
        &\llangle D^*(\sigma,\tau),\mathring x\rrangle_U = \llangle (\sigma,\tau),D\mathring x\rrangle_U,\\
        &\llangle \GC^*(\varphi,\psi),(\mathring g,\mathring\SFF)\rrangle_U = \llangle (\varphi,\psi),\GC(\mathring g,\mathring\SFF)\rrangle_U
    \end{align}
    for all \(\mathring x\in\fX^0(U)\) and \((\mathring g,\mathring \SFF)\in\fX^1(U)\) with compact support in \(U\).
\end{definition}
Using the formal adjoints for the differential complex \eqref{eq:KinematicShellComplex}, we obtain the dynamic shell complex
\begin{align}
    \label{eq:DynamicShellComplex}
    0\leftarrow\underbrace{\underline{\se(3)}(M)^*}_{
    \substack{\text{net force}\\\text{\& torque}}
    }
    \twoheadleftarrow
    \underbrace{\fX^0(M)^*}_{
    \substack{\text{minus force}\\\text{fields \(-f\)}}
    }
    \xleftarrow{D^*}
    \underbrace{\fX^1(M)^*}_{\substack{\text{stretching \&}\\\text{bending stress}\\ (\sigma,\tau)}
    }
    \xleftarrow{\GC^*}
    \underbrace{\fX^2(M)^*}_{
    \substack{\text{Airy potentials}\\ (\varphi,\psi)}
    }
    \hookleftarrow 
    \underbrace{H^0(M;\se(3)^*)}_{
    \substack{\text{gauge}\\\text{modes}}
    }\leftarrow 0
\end{align}

This dual shell complex gives us remarkable information.  The space \(\underline{\se(3)}(M)^*\) is assigning an \(\se(3)^*\) element per connected component of \(M\).  The adjoint of the inclusion \(\underline{\se(3)}\hookrightarrow \fX^0(M)\) is the projection that maps a force field \(f\in\fX^0(M)\) to the total \(\se(3)^*\) moment (net force and torque) per component.

The force arises from a pair of stretching and bending stresses \(-f = D^*(\sigma,\tau)\) automatically satisfies that it has no net force and torque.

The stresses produced by Airy potentials \((\sigma,\tau) = \GC^*(\varphi,\psi)\) are automatically at equilibrium \(D^*(\sigma,\tau) = 0\).  The Airy potential representation for equilibrial stresses have a gauge degree of freedom, exactly given by an \(\se(3)^*\) mode per connected component, mirroring the presence of characteristic class for Gauss--Codazzi residuals.

\subsection{Explicit Formulae and Green's Identities for the Shell Complex}
\label{sec:ExplicitFormulaeForShellComplex}
This section lists the explict formulae for \(D\), \(\GC\), \(D^*\), \(\GC^*\), and the associated Green's identities including the boundary terms.

We adopt a few symmetrization notations.  For any vector space \(V\), let
\begin{alignat}{2}
    &\sym\colon V\otimes V\to \Sym V, &\qquad&(\sym A)^{ij}\coloneqq \textstyle{1\over 2}(A^{ij}+A^{ji}),\\
    &\alt\colon V\otimes V\to \wedge^2 V, &\qquad& (\alt A)^{ij}\coloneqq \textstyle{1\over 2}(A^{ij} - A^{ji}).
\end{alignat}
We shall add Kulkarni--Nomizu product to the list
\begin{align}
    &\owedge\colon \Sym V\times \Sym V\to \owedge^2\Sym V\subseteq \Sym(\wedge^2 V)\\
    & (A\owedge B)^{ijk\ell} = A^{ik}B^{j\ell} + A^{j\ell}B^{ik} - A^{i\ell}B^{jk} - A^{jk}B^{i\ell}.
\end{align}
In addition to these standard symmetrization operators, we introduce the following \emph{Koszul symmetrization operator}.%
\footnote{It is named after the Koszul formula for the Levi-Civita connection.}
\begin{definition}[Koszul symmetrization operator]
    Define the automorphism
    \begin{align}
        &\kos\colon V\otimes(\Sym V)\to V\otimes (\Sym V),\\
        &(\kos A)^{m|ij}\coloneqq \textstyle{1\over 2}(A^{i|mj}+A^{j|mi} - A^{m|ij}).
    \end{align}
\end{definition}
One can check that \(\kos\) is invertible, with
\begin{align}
    (\kos^{-1}B)^{m|ij} = B^{i|mj}+B^{j|mi},
\end{align}
and is formally self-adjoint
\begin{align}
\langle \kos A|C\rangle = \langle A|\kos C\rangle,\quad 
    (\kos A)^{m|ij}C_{m|ij} = A^{m|ij}(\kos C)_{m|ij}.
\end{align}
The operator \(\kos\) has two eigenvalues: \({1\over 2}\) and \(-1\).  The eigenspaces are characterized by
\begin{align}
    &\kos A = \tfrac12 A\quad\iff\quad A\in\Symm V,\\
    &\kos A = -A\quad\iff\quad A^{k|ij}+A^{i|jk}+A^{j|ki} = 0.
\end{align}
 In fact, the space \(V\otimes(\Sym V)\) is split by the short exact sequence
\begin{align}
    0\to\Symm V \hookrightarrow V\otimes (\Sym V)\xrightarrow{\alt_{12}} (\wedge^2 V)\otimes V \to 0
\end{align}
where \((\alt_{12}A)^{ij|k} = {1\over 2}(A^{i|jk} - A^{j|ik})\), and \(\kos\) is diagnalized by this splitting: its \(1/2\)-eigenspace is \(\Symm V = \ker(\alt_{12})\), whereas the \(-1\) eigenspace is isomorphic to \((\wedge^2 V)\otimes V\).  In particular, \(\kos(\alt_{12}A) = \alt_{12}(\kos A) = -\alt_{12}A\).

One final operator convention is the contracted wedge product between \(A\in V^*\otimes(\Sym V^*)\) and an endomorphism \(L\in V\otimes V^*\),
\begin{align}
    &\wedge\colon (V^*\otimes(\Sym V^*))\times (V\otimes V^*)\to (\wedge^2 V^*)\otimes V^*\\
    &(A\wedge L)_{ij|m} = A_{k|mi}L^k_j - A_{k|mj}L^k_i,
\end{align}
which generalizes \(\alt_{12}\) in the sense that \(\alt_{12}\) is the special case \(\alt_{12}A = -{1\over 2}A\wedge\id\).

\begin{theorem}[Deformation map]
\label{thm:DeformationMap}
    An explicit formula for \((\mathring g,\mathring\SFF) = D\mathring x\) is given by
    \begin{align}
        \begin{cases}
            \mathring g = 2\sym(F^\intercal\flat_{\RR^3} d\mathring x)\\
            \mathring\SFF = \langle n,\Hess\mathring x\rangle,
        \end{cases}
        \quad\text{or in indices:}\quad
    \begin{cases}
        \mathring g_{ij} = \delta_{ab}(\partial_i\mathring x^a F_j^b + F_i^a\partial_j\mathring x^b)\\
        \mathring\SFF_{ij} = n_a\nabla_j\partial_i\mathring x^a.
    \end{cases}
    \end{align}
\end{theorem}


\begin{theorem}[Linearized Gauss--Codazzi map]
\label{thm:LinearizedGaussCodazzi}
    An explicit formula for \((\mathring G,\mathring Q) = \GC(\mathring g,\mathring\SFF)\) is given by
    \begin{align}
        \begin{cases}
            \mathring G = {1\over 2}(\nabla\nabla\owedge\mathring g) - {1\over 4}K(g\owedge g)\tr_g(\mathring g) +  \SFF\owedge\mathring\SFF\\
            \mathring Q = d^\nabla\mathring\SFF - \kos(\nabla\mathring g)\wedge S.
        \end{cases}
    \end{align}
    In indices:
    \begin{align}
        \begin{cases}
            \mathring G_{ijk\ell}=
            \begin{aligned}[t]
                &
                \textstyle
                {1\over 2}(\nabla_i\nabla_k\mathring g_{j\ell} + \nabla_j\nabla_\ell\mathring g_{ik} - \nabla_i\nabla_\ell\mathring g_{jk} - \nabla_j\nabla_k\mathring g_{i\ell})
                \\
                &
                \textstyle
                -{K\over 2}(g_{ik}g_{j\ell} - g_{i\ell}g_{jk})g^{mn}\mathring g_{mn}\\
                &\textstyle
                +\SFF_{ik}\mathring\SFF_{j\ell} + \SFF_{j\ell}\mathring\SFF_{ik} - \SFF_{i\ell}\mathring\SFF_{jk} - \SFF_{jk}\mathring\SFF_{i\ell},
            \end{aligned}
            \\
            Q_{ij|k} = 
    \nabla_i\mathring\SFF_{jk}-\nabla_j\mathring\SFF_{ik} - (\kos\nabla\mathring g)_{m|ki}S_j^m + (\kos\nabla\mathring g)_{m|kj}S_i^m,\\
    \quad\text{where}\quad
    (\kos\nabla\mathring g)_{m|ij} = {1\over 2}(\nabla_i\mathring g_{mj} + \nabla_j\mathring g_{mi} - \nabla_m\mathring g_{ij}).
        \end{cases}
    \end{align}
\end{theorem}

\begin{theorem}[Stress-to-force map]
\label{thm:StressToForce}
    An explicit formula for \(-f = D^*(\sigma,\tau)\) is given by
    \begin{align}
        f = \nabla\cdot\Big(F(2\sigma + S\tau)\Big) + (\nabla\nabla:\tau)n + FS(\nabla\cdot\tau),
    \end{align}
    which has tangential and normal components
    \begin{align}
    \begin{cases}
        f^\parallel = \nabla\cdot(2\sigma + S\tau) + S(\nabla\cdot\tau)\\
        f^\bot = \nabla\nabla:\tau - \tr(2\sigma S + \tau S^2).
    \end{cases}
    \end{align}
    In indices:
    \begin{align}
        f^a = \nabla_i\Big(F_k^a(2\sigma^{ki}+S_j^k\tau^{ij}) + n^a\nabla_j\tau^{ij}\Big).
    \end{align}
\end{theorem}

For the next formula for \(\GC^*\), it is useful to decompose the second fundamental form \(\SFF\) into its trace and trace-free parts.
Let \(H = {1\over 2}\tr(S) = {1\over 2}\tr_g(\SFF)\) be the mean curvature, and \(q = \SFF - Hg\) be the trace-free part of \(\SFF\), known as the Hopf differential.
Define the bilinear form%
\footnote{Geodesic torsions \(\tau_{\rm g}\) should not be confused with bending stress tensors denoted by \(\tau\).}
\begin{align}
\label{eq:GeodesicTorsionForm}
    \tau_{\rm g}(\cdot,\cdot)\coloneqq q(J\cdot,\cdot),
\end{align}
which is symmetric and trace free.
It is the symmetric bilinear form polarization of the geodesic torsion quadratic form \(\tau_{\rm g}(X) = \SFF(JX,X)\); that is, \(\tau_{\rm g}(X,X) = \tau_{\rm g}(X)\).  Let
\begin{align}
\label{eq:GeodesicTorsionBivector}
    T_{\rm g} \coloneqq \tau_{\rm g}^{\sharp\sharp},\quad T_{\rm g}^{ij} = (\tau_{\rm g})_{i'j'}g^{i'i}g^{j'j}
\end{align}
be the associated symmetric bivector.

\begin{theorem}[Stress from Airy potential]
\label{thm:Airy}
    An explicit formula for \((\sigma,\tau) = \GC^*(\varphi,\psi)\), \(\varphi\in C^\infty(M)\), \(\psi\in\Gamma(TM)\), is given by
    \begin{align}
        \begin{cases}
            \sigma = {1\over 2\det(g)}{\cof(\Hess\varphi)} + \sym\left(\sgrad^{\nabla}(H\psi)\right)
    -{(\nabla\cdot\psi)T_{\rm g} + \nabla_\psi T_{\rm g}\over 2}-{1\over 2}K\varphi g^{-1},
            \\
            \tau = {1\over\det(g)}{\cof(\SFF)}\varphi - \sym(\sgrad^\nabla\psi).
        \end{cases}
    \end{align}
    In indices:
    \begin{align}
        \begin{cases}
            \sigma^{ij} = 
            {1\over 2}\mu^{ik}\mu^{j\ell}\partial_k\partial_\ell\varphi + {1\over 2}(\mu^{im}\nabla_m(H\psi^j)+\mu^{jm}\nabla_m(H\psi^i)) - {1\over 2}\nabla_m(\psi^mT_{\rm g}^{ij}) - {1\over 2}K\varphi g^{ij}\\
            \tau^{ij} = \mu^{ik}\mu^{j\ell}\SFF_{k\ell}\varphi - {1\over 2}(\mu^{im}\nabla_m\psi^j + \mu^{jm}\nabla_m\psi^i).
        \end{cases}
    \end{align}
\end{theorem}

The following Green's identities for \(D\) and \(\GC\) is the integration-by-parts formula including boundary terms.  We let \(t\) be the tangent vector of the boundary curve \(\partial M\).  While \(n\) is the normal of the surface, we let \(b = t\times n\) be the vector that is tangent to the surface but normal to the boundary curve.  Let \(t\) be oriented counterclockwise with respect to \(n\), so that \(b\) is the outward normal.

Let \(\kappa_{\rm n}=\SFF(t,t)\) be the normal curvature and \(\tau_{\rm g} = -\SFF(t,b) = \tau_{\rm g}(t,t)\) be the geodesic torsion along the boundary.  Let \(\kappa_{\rm g} = -\langle \nabla_tt,b\rangle\) be the geodesic curvature of the boundary.  They are the curvature and twist of the boundary as a framed curve:
\begin{align}
    {d\over ds}
    \begin{bmatrix}
        t\\
        n\\
        b
    \end{bmatrix}
    =
    \begin{bmatrix}
        & -\kappa_{\rm n} & -\kappa_{\rm g}\\
        \kappa_{\rm n} & & -\tau_{\rm g}\\
        \kappa_{\rm g}&\tau_{\rm g}
    \end{bmatrix}
    \begin{bmatrix}
        t\\ n\\ b
    \end{bmatrix}.
\end{align}

\begin{theorem}[Green's Identity for \(D\)]
\label{thm:GreenIdentityD}
For each \(\mathring x\in\fX^0(M)\) and \((\sigma,\tau)\in\fX^1(M)^*\), with \(-f\coloneqq D^*(\sigma,\tau)\) and \((\mathring g,\mathring\SFF)\coloneqq D\mathring x\), we have
\begin{align}
\nonumber
    \llangle-f,\mathring x\rrangle
    =
    &\llangle \sigma,\mathring g\rrangle + \llangle \tau,\mathring\SFF\rrangle
    - \oint_{\partial M}2\mathring x^\parallel\cdot(\sigma+S\tau)b\, ds\\
    &- \oint_{\partial M}\mathring x^
    \bot[2\partial_t(\tau_{tb}) + (\nabla_b\tau)_{bb} + \kappa_{\rm g}(\tau_{bb}-\tau_{tt})]\, ds + \oint_{\partial M} (\partial_b\mathring x^\bot)\tau_{bb}\, ds.
\end{align}
where \(\mathring x^\parallel\in\Gamma(TM)\) and \(\mathring x^\bot\in C^\infty(M)\) are the tangential and normal component of the \(\RR^3\)-valued field \(\mathring x = dx(\mathring x^\parallel) + \mathring x^\bot n\).
\end{theorem}

\begin{theorem}[Green's Identity for \(\GC\)]
\label{thm:GreenIdentityGC}
    For each \((\mathring g,\mathring\SFF)\in\fX^1(M)\) and \((\varphi,\psi)\in\fX^2(M)^*\), with \((\sigma,\tau)\coloneqq\GC^*(\varphi,\psi)\) and \((\mathring G,\mathring Q)\coloneqq \GC(\mathring g,\mathring\SFF)\), we have
\begin{align}
    \nonumber
    \llangle\varphi,\mathring G\rrangle + \llangle\psi,\mathring Q\rrangle &=
    \llangle\sigma,\mathring g\rrangle + \llangle\tau,\mathring\SFF\rrangle \\
    \nonumber
    &\quad
    -{1\over 2}\oint_{\partial M}(\nabla_b\mathring g)(t,t)\varphi\, ds\\
    \nonumber
    &\quad+
    \oint_{\partial M}
    \left(
    \begin{bmatrix}
        -{\partial_b\varphi\over 2}-{\kappa_{\rm g}\over 2}\varphi\\
        &{\kappa_{\rm g}\over 2}\varphi
    \end{bmatrix}
    +
    \begin{bmatrix}
        H\psi_t +{\tau_{\rm g}\over 2}\psi_b & {\kappa_{\rm n}\over 2}\psi_b\\
        {\kappa_{\rm n}\over 2}\psi_b &
        -{\tau_{\rm g}\over 2}\psi_b
    \end{bmatrix}
    \right)
    :
    \begin{bmatrix}
        \mathring g_{tt} & \mathring g_{tb}\\
        \mathring g_{bt}&
        \mathring g_{bb}
    \end{bmatrix}\\
    &\quad
    +\oint_{\partial M}
    \begin{bmatrix}
        \psi_t & {1\over 2}\psi_b\\
        {1\over 2}\psi_b
    \end{bmatrix}
    :
    \begin{bmatrix}
        \mathring \SFF_{tt} & \mathring \SFF_{tb}\\
        \mathring \SFF_{bt}&
        \mathring \SFF_{bb}
    \end{bmatrix},
\end{align}
where \(\psi_t = \langle t,\psi\rangle\), \(\psi_b = \langle b,\psi\rangle\).  Equivalently, with the boundary terms reorganized, we have
\begin{align}
\nonumber
    \llangle\varphi,\mathring G\rrangle + \llangle\psi,\mathring Q\rrangle =
    \llangle\sigma,\mathring g\rrangle + \llangle\tau,\mathring\SFF\rrangle + {1\over 2}\oint_{\partial M}\Big(
    &\varphi [-(\nabla_b\mathring g)_{tt}+\kappa_{\rm g}(\mathring g_{bb}-\mathring g_{tt})] - \partial_b\varphi\mathring g_{tt}\\
    &{}+\psi_t(2H\mathring g_{tt} + 2\mathring\SFF_{tt})\\
    \nonumber
    &{}+\psi_b(2\kappa_{\rm n}\mathring g_{tb} + \tau_{\rm g}(\mathring g_{tt} - \mathring g_{bb}) + 2\mathring\SFF_{tb})
    \Big)\, ds
\end{align}
\end{theorem}

Formulae of Theorems~\ref{thm:DeformationMap}--\ref{thm:GreenIdentityGC} are derived in \Cref{sec:Derivation}.

\subsection{Derivation for the Explicit Formulae and Green's Identities}
\label{sec:Derivation}

In this section, we prove Theorems~\ref{thm:DeformationMap}--\ref{thm:GreenIdentityGC}.  That is, we derive the explicit formulae for \(D,\GC\), their adjoints, and the boundary terms in integration by parts.

\subsubsection{Road maps}
We begin with the computational graph of the nonlinear \(\cD\) and \(\cG\cC\) operators:
\begin{align}
\label{eq:NonlinearNetwork}
\cD = 
\begin{aligned}
    \xymatrix{
    x
    \ar[r]
    &
    F
    \ar[r]
    \ar[d]
    \ar[rd]
    &
    g\\
    &
    n
    \ar[r]
    &
    \SFF
    }
\end{aligned}
,\quad
\cG\cC = 
\begin{aligned}
\xymatrix{
g
\ar@/^1pc/[rr]
\ar[r]
&
\nabla
\ar[r]
\ar[dr]|(0.35)\hole
&
G
\\
\SFF
\ar[urr]
\ar[rr]
&
&
Q
}
\end{aligned}
.
\end{align}
In the \(\cD\) diagram, we have \(F = dx\), \(g = F^\intercal\flat_{\RR^3}F\), \(\SFF = F^\intercal\flat_{\RR^3}dn\), and that the surface normal \(n\) is determined from \(F\) by \(\langle n,F\rangle_{\RR^3} = 0\) and \(|n|^2 = 1\).
In the \(\cG\cC\) diagram, \(\nabla\) is the Levi-Civita connection of metric \(g\), \(S = g^{-1}\SFF\), \(G = gR^{\nabla}+{1\over 2}\SFF\owedge\SFF\), and \(Q = d^\nabla \SFF\).
Each arrow indicates an explicit functional dependency.  The entire maps \(\cD\) and \(\cG\cC\) are the compositions of these arrows.

Finding \(D = d\cD\) and \(\GC = d\cG\cC\) amounts to taking the derivative of each arrow (as a linear tangent map) while replacing each nodal variables by their perturbations:
\begin{align}
\label{eq:LinearizedNetwork}
    D = 
\begin{aligned}
    \xymatrix{
    \mathring x
    \ar@{~>}[r]
    &
    \mathring F
    \ar[r]
    \ar[d]
    \ar[rd]
    &
    \mathring g\\
    &
    w
    \ar@{~>}[r]
    &
    \mathring \SFF
    }
\end{aligned}
,\quad
\GC = 
\begin{aligned}
\xymatrix{
\mathring g
\ar@/^1pc/[rr]
\ar@{~>}[r]
&
\mathring\Gamma
\ar@{~>}[r]
\ar[dr]|(0.35)\hole
&
\mathring G
\\
\mathring \SFF
\ar[urr]
\ar@{~>}[rr]
&
&
\mathring Q
}
\end{aligned}
.
\end{align}
Here, \(\rightsquigarrow\) indicates that the map is a first-order differential operator, whereas \(\rightarrow\) is a pointwise algebraic linear map.  Whether an arrow involves differentiation can be reasoned from the original expressions of each arrow in \eqref{eq:NonlinearNetwork}.  The \(\rightsquigarrow\) notation is helpful when we take the adjoint, as it indicates exactly which arrow involves an integration by parts producing boundary terms.

Note that \(\mathring n\) is represented by a tangent vector field \(w\in\Gamma(TM)\).  Since \(|n|^2=1\) implies that \(\mathring n\) must be tangential, we express \(\mathring n\) in terms of \(w\in\Gamma(TM)\) via
\begin{align}
    \mathring n = Fw.
\end{align}
One also notices that the perturbation of the connection derivative operator \(\nabla\) is \(\mathring\nabla = \mathring\Gamma\), the variation of the Christoffel symbol.  In fact, \(\mathring\Gamma\) is a \emph{tensor}, unlike the Christoffel symbols themselves.   To see this, note that \(\mathring\nabla\) is tensorial, deduced from the more general fact that the difference of any pair of affine connections is tensorial.  In coordinates, the Levi-Civita connection is expressed as \((\nabla_iX)^k = \partial_iX^k+\Gamma_{ij}^kX^j\) where \(\Gamma_{ij}^k\) is the coordinate-dependent Christoffel symbol.  Under variation (while fixing coordinates and \(X\)) one finds \((\mathring\nabla_i X)^k = \mathring\Gamma_{ij}^k X^j\) and thus the variational Christoffel symbol \(\mathring\Gamma\) is the tensorial \(\mathring\nabla\), which is now a coordinate-invariant object.  Specifically, it has the type
\begin{align}
    \mathring\Gamma\in\Omega^1(M;\End(TM))\quad\text{with symmetry}\quad \mathring\Gamma_{ij}^k = \mathring\Gamma_{ji}^k.
\end{align}

After we derive the expression of each linear map in \eqref{eq:LinearizedNetwork}, we take the adjoints of these maps to obtain the computational graph for the adjoint operators:
\begin{align}
\label{eq:AdjointNetwork}
    D^* = 
\begin{aligned}
    \xymatrix{
    -f
    &
    P
    \ar@{~>}[l]
    &
    \sigma
    \ar[l]\\
    &
    \nu
    \ar[u]
    &
    \tau
    \ar[lu]
    \ar@{~>}[l]
    }
\end{aligned}
,\quad
\GC^* = 
\begin{aligned}
\xymatrix{
\sigma
&
L
\ar@{~>}[l]
&
\varphi
\ar@{~>}[l]
\ar@/_1pc/[ll]
\ar[dll]
\\
\tau
&&
\psi
\ar[ul]|(0.65)\hole
\ar@{~>}[ll]
}
\end{aligned}
.
\end{align}
Here, each node is replaced by the corresponding dual variable.  In addition to the pairings introduced in \eqref{eq:DualPairingMain}, the pairings between the mutually dual variables are \(\llangle \cdot,\cdot\rrangle = \int_M\langle\cdot|\cdot\rangle\, dA_g\), with pointwise pairing
\begin{align}
    \langle P|\mathring F\rangle = P_a^i \mathring F_i^a,\quad \langle\nu|w\rangle = \nu_i w^i,\quad \langle L|\mathring\Gamma\rangle = L_{k}^{ij}\mathring\Gamma_{ij}^k.
\end{align}
We call \(P\in\Gamma(TM\otimes\RR^3)\) the Piola stress, \(\nu\in\Omega^1(M)\) the normal stress, and \(L\in\Gamma(TM\otimes\End(T^*M))\) the connection stress.

\subsubsection{Green's Identity for \(D\)}
Each of the following lemmas computes the derivative of each map in \eqref{eq:NonlinearNetwork} and its adjoint.
Some of these formulae follow from straightforward derivation, for which the proofs are omitted. 
\begin{lemma}[\({\delta F\over\delta x}\)]
    The variation \({\delta F\over\delta x}\) is given by \(\mathring F = d\mathring x\); that is, \(\mathring F^a_i = \partial_i\mathring x^a\).
\end{lemma}
\begin{lemma}[\(({\delta F\over\delta x})^*\)]
    The adjoint of \({\delta F\over\delta x}\) that maps \(P\) to \(-f=({\delta F\over\delta x})^*P\) is given by
    \begin{align}
        -f_a = -\nabla_iP^i_a,
    \end{align}
    with
    \begin{align}
    \llangle P,\mathring F\rrangle
         = \llangle -f,\mathring x\rrangle + \oint_{\partial M}P_a^i\mathring x^a b_i\, ds.
    \end{align}
\end{lemma}
\begin{lemma}[\({\delta g\over\delta F}\)]
    The variation \(\mathring g = ({\delta g\over \delta F})\mathring F\) is given by \(\mathring g = 2\sym(F^\intercal\flat_{\RR^3}\mathring F)\).  That is, \(\mathring g_{ij} = \delta_{ab}\mathring F_i^a F_j^b + \delta_{ab}F_i^a\mathring F_j^b\).
\end{lemma}
\begin{lemma}[\(({\delta g\over\delta F})^*\)]
    The adjoint of \(({\delta g\over\delta F})\) that maps \(\sigma\) to one contribution term \(P_\sigma = ({\delta g\over\delta F})^*\sigma\) (with \(\langle \sigma|\mathring g\rangle = \langle P_\sigma|\mathring F\rangle\)) is given by
    \begin{align}
        (P_\sigma)_a^i = 2\delta_{ab}F_j^b\sigma^{ij}.
    \end{align}
\end{lemma}
\begin{lemma}[\({\delta w\over\delta F}\)]
    The vector \(w\in TM\) representing the variation of normal \(\mathring n = Fw\) arising from a variation \(\mathring F\) of \(F\) is given by \(w^i = -g^{ij}n_a\mathring F_j^a\).
\end{lemma}
\begin{proof}
    Taking variation of \(\langle n,F\rangle_{\RR^3} = 0\) yields \(0=\langle \mathring n,F\rangle_{\RR^3} + \langle n,\mathring F\rangle_{\RR^3} = \langle Fw,F\rangle_{\RR^3} + \langle n,\mathring F\rangle_{\RR^3}\).  That is, \(\flat_g w = -\langle n,\mathring F\rangle_{\RR^3}\).
\end{proof}
\begin{lemma}[\(({\delta w\over\delta F})^*\)]
    The adjoint of \(({\delta w\over\delta F})\) that maps \(\nu\) to one contributing term \(P_\nu = ({\delta w\over\delta F})^*\nu\) in \(P\) (that is, \(\langle \nu|w\rangle = \langle P_\nu|\mathring F\rangle\)) is given by
    \begin{align}
        (P_\nu)_a^i = -\nu_j g^{ij}n_a.
    \end{align}
\end{lemma}
Next, note that \(\mathring\SFF\) has two terms \(\mathring\SFF = ({\delta\SFF\over\delta F})\mathring F + ({\delta\SFF\over\delta w})w\).  Let \(\mathring\SFF_F\coloneqq ({\delta\SFF\over\delta F})\mathring F\) and \(\mathring\SFF_n\coloneqq ({\delta\SFF\over\delta w}) w\)
\begin{lemma}[\({\delta\SFF\over\delta F}\)]
    \((\mathring\SFF_F)_{ij} = \delta_{ab}\mathring F_i^aF^b_kS^k_j\).
\end{lemma}
\begin{lemma}[\(({\delta\SFF\over\delta F})^*\)]
    The adjoint of \(({\delta\SFF\over\delta F})\) that maps \(\tau\) to one term \(P_{\tau} = ({\delta\SFF\over\delta F})^*\tau\) with 
    \(\langle \tau|\mathring\SFF_F\rangle = \langle P_{\tau}|\mathring F\rangle\) is given by
    \begin{align}
        (P_\tau)_a^i = \delta_{ab}F_k^bS_j^k\tau^{ij}
    \end{align}
\end{lemma}
\begin{lemma}[\({\delta\SFF\over\delta w}\)]
    \((\mathring\SFF_n)_{ij} = \delta_{ab}F_i^a\partial_j(F_k^bw^k)\).
\end{lemma}
\begin{lemma}[\(({\delta\SFF\over\delta w})^*\)]
    The adjoint of \(({\delta\SFF\over\delta w})\) that maps \(\tau\) to \(\nu = ({\delta\SFF\over\delta w})^*\tau\) is given by
    \begin{align}
        \nu_i = -g_{ij}(\nabla_m\tau^{jm})
    \end{align}
    with
    \begin{align}
        \llangle \mathring\SFF_n,\tau\rrangle = \llangle\nu,w\rrangle + \oint_{\partial M} g_{ik}w^k\tau^{ij}b_j\, ds
    \end{align}
\end{lemma}
\begin{lemma}[\(\nabla F\) is normal]
\label{lemma:NablaFIsNormal}
    \(\delta_{ab}(\nabla_iF_j^a)F_k^b = 0\).
\end{lemma}
\begin{proof}
    We have \(d_X(FY) = F(\nabla_XY)-\SFF(X,Y)n\) and \(F(\nabla_XY) = d_X(FY) - (\nabla_XF)Y\).  Therefore \((\nabla_XF)Y = \SFF(X,Y)n\).  In particular \(\nabla_iF_j^a\) is normal-valued, resulting in zero after taking \(\RR^3\)-inner product with \(F\).
\end{proof}

Now, we can assemble the above lemmas to obtain \(\mathring g,\mathring\SFF\) in terms of \(\mathring x\).  The composition according to the computation graph of \(D\) in \eqref{eq:LinearizedNetwork} yields
\begin{align}
    \mathring g_{ij} = \delta_{ab}(\mathring F_i^aF_j^b + F_i^a\mathring F_j^b) = \delta_{ab}(\partial_i\mathring x^a F_j^b + F_i^a\partial _j\mathring x^b),
\end{align}
and 
\begin{align}
    \mathring\SFF_{ij} &= (\mathring\SFF_{F})_{ij} + (\mathring\SFF_n)_{ij}
    =\delta_{ab}\partial_i\mathring x^a F_k^b S_j^k + \delta_{ab}F_i^a\partial_j(F_k^b \underbrace{w^k}_{-g^{km}n_c\partial_m\mathring x^c})\\
    &=
    \delta_{ab}\partial_i\mathring x^a F_k^b S_j^k - \delta_{ab} F_i^a g^{km}(\underbrace{(\nabla_jF_k^b)n_c\partial_m\mathring x^c}_{=0\text{ (\Cref{lemma:NablaFIsNormal})}} + F_k^b\nabla_j(n_c\partial_ m\mathring x^c))\\
    &=\delta_{ab}\partial_i\mathring x^a F_k^b S_j^k - \underbrace{(\partial_j n_c)}_{\delta_{cb}F_k^bS^k_j}\partial_i\mathring x^c - n_c\nabla_j\partial_i\mathring x^c = n_c\nabla_j\partial_i\mathring x^c.
\end{align}
For the adjoint \(D^*\), the composition according to \eqref{eq:AdjointNetwork} yields
\begin{align}
    -f_a &= -\nabla_iP_a^i = -\nabla_i\Big((P_{\sigma})_a^i + (P_{\tau})_a^i + (P_{\nu})_a^i\Big)\\
    &=-\delta_{ab}\nabla_i\Big(2F_j^b\sigma^{ij} + F_k^b S_j^k\tau^{ij} - n^b g^{ij}\nu_j\Big)\\
    &=-\delta_{ab}\nabla_i\Big(F_k^b(2\sigma^{ki}+S_j^k\tau^{ij}) + n^b\nabla_j\tau^{ij}\Big).
\end{align}
In index-free notation this can be expressed as
\begin{align}
    f = \nabla\cdot\Big(F(2\sigma + S\tau)\Big) + (\nabla\nabla:\tau)n + FS(\nabla\cdot\tau).
\end{align}
The first term splits into tangential and normal part by
\begin{align}
    \nabla\cdot(F(2\sigma + S\tau))=F[\nabla\cdot(2\sigma + S\tau)] - \tr(S(2\sigma + S\tau))n.
\end{align}
Therefore,
\begin{align}
    \begin{cases}
        f^\parallel = \nabla\cdot(2\sigma + S\tau) + S(\nabla\cdot\tau)\\
        f^\bot = \nabla\nabla:\tau - \tr(2\sigma S + \tau S^2).
    \end{cases}
\end{align}
To collect the boundary term, compute
\begin{align}
    \llangle -f,\mathring x\rrangle &= \llangle P,\mathring F\rrangle - \oint_{\partial M} b_iP_a^i\mathring x^a \, ds
    =\llangle P_\sigma + P_\tau + P_\nu, \mathring F\rrangle - \oint_{\partial M}b_iP_a^i\mathring x^a \, ds\\
    &=\llangle\sigma,\mathring g\rrangle + \llangle \tau,\mathring\SFF_F\rrangle + \llangle \nu, w\rrangle - \oint_{\partial M}b_iP_a^i\mathring x^a \, ds\\
    &=
    \llangle\sigma,\mathring g\rrangle + \underbrace{\llangle \tau,\mathring\SFF_F\rrangle + \llangle \tau, \mathring\SFF_n\rrangle}_{\llangle\tau,\mathring\SFF\rrangle} - \oint_{\partial M}b_iP_a^i\mathring x^a \, ds - \oint_{\partial M}g_{jk}w^k\tau^{ij}b_i\, ds.
\end{align}
After substituting \(w^k = -g^{kj}n_a \partial_j\mathring x^a\) and \(P_a^{i} = \delta_{ab}F_k^b(2\sigma^{ki}+S_j^k\tau^{ij})+n_a\nabla_j\tau^{ij}\),
we get
\begin{align*}
    \llangle -f,\mathring x\rrangle &= 
    \llangle \sigma,\mathring g\rrangle + \llangle\tau,\mathring\SFF\rrangle - \oint_{\partial M} b_i\Big(\mathring x^a\delta_{ab} F_k^b(2\sigma^{ki}+S_j^k\tau^{ij}) + n_a\mathring x^a\nabla_j\tau^{ij} - \underbrace{n_a\partial_j\mathring x^a}_{\substack{\partial_j(n_a\mathring x^a)\\ -\delta_{ab}F_k^bS^k_j\mathring x^a}}\tau^{ij}\Big)\\
    &= 
    \llangle \sigma,\mathring g\rrangle + \llangle\tau,\mathring\SFF\rrangle - \oint_{\partial M} b_i\Big(\mathring x^a\delta_{ab} 2 F_k^b(\sigma^{ki}+S_j^k\tau^{ij})+n_a\mathring x^a\nabla_j\tau^{ij} - \partial_j(n_a\mathring x^a)\tau^{ij}\Big)\, ds.
\end{align*}
Letting \(\mathring x = F\mathring x^\parallel + \mathring x^\bot n \) be the tangential and normal decomposition of \(\mathring x\) (with \(\mathring x^\bot n = \langle n,\mathring x\rangle\)), we arrive at
\begin{align}
\label{eq:GreenForDSemiFinal}
    \llangle-f,\mathring x\rrangle
    =
    \llangle \sigma,\mathring g\rrangle + \llangle \tau,\mathring\SFF\rrangle
    - \oint_{\partial M}2\mathring x^\parallel\cdot(\sigma+S\tau)b\, ds - \oint_{\partial M}\mathring x^
    \bot(\nabla\cdot\tau)b\, ds + \oint_{\partial M} d\mathring x^\bot\cdot\tau b\, ds.
\end{align}

We further expand the last term \(d\mathring x^\bot\cdot\tau b = \partial_t\mathring x^\bot \tau_{tb} + \partial_b\mathring x^\bot \tau_{bb}\) and apply an integration by parts for the tangential derivative \(\partial_t\) in the boundary integral \(\oint_{\partial M}\).  This step reorganizes \eqref{eq:GreenForDSemiFinal} into
\begin{align}
\label{eq:GreenForDAlmostFinal}
\nonumber
    \llangle-f,\mathring x\rrangle
    =
    \llangle \sigma,\mathring g\rrangle + \llangle \tau,\mathring\SFF\rrangle
    &- \oint_{\partial M}2\mathring x^\parallel\cdot(\sigma+S\tau)b\, ds\\
    &- \oint_{\partial M}\mathring x^
    \bot[(\nabla\cdot\tau)b + \partial_t\tau_{tb}]\, ds + \oint_{\partial M} (\partial_b\mathring x^\bot)\tau_{bb}\, ds.
\end{align}
Expanding the divergence term yields
\begin{align}
\nonumber
    (\nabla\cdot\tau)b &= 
(\nabla_t\tau)(t,b) + (\nabla_b\tau)(b,b)\\ 
&= \partial_t\tau_{tb}  - 
\tau(\underbrace{\nabla_tt}_{-\kappa_{\rm g}b},b) - \tau(t,\underbrace{\nabla_tb}_{\kappa_{\rm g}t})+\partial_b \tau_{bb} - 2\tau(b,\nabla_bb)
\end{align}
where \(\kappa_{\rm g}\) is the geodesic curvature.  The last two terms \(\partial_b\tau_{bb}-2\tau(b,\nabla_bb)\) both depend on how \(b\) extends away from the boundary.  The original expression \((\nabla_b\tau)(b,b)\) is a more invariant expression independent of the extension.  Therefore, \eqref{eq:GreenForDAlmostFinal} becomes
\begin{align}
\nonumber
    \llangle-f,\mathring x\rrangle
    =
    &\llangle \sigma,\mathring g\rrangle + \llangle \tau,\mathring\SFF\rrangle
    - \oint_{\partial M}2\mathring x^\parallel\cdot(\sigma+S\tau)b\, ds\\
    &- \oint_{\partial M}\mathring x^
    \bot[2\partial_t(\tau_{tb}) + (\nabla_b\tau)_{bb} + \kappa_{\rm g}(\tau_{bb}-\tau_{tt})]\, ds + \oint_{\partial M} (\partial_b\mathring x^\bot)\tau_{bb}\, ds.
\end{align}
This completes the derivation for \Cref{thm:GreenIdentityD}.

\subsubsection{Green's Identity for \(\GC\)}

\begin{lemma}[\({\delta\Gamma\over\delta g}\)]
    The variation \({\delta\Gamma\over\delta g}\) of the Levi-Civita connection with respect to metric is given by
    \begin{align}
        \mathring\Gamma = g^{-1}\kos(\nabla\mathring g);
    \end{align}
    that is, \(\mathring\Gamma_{ij}^k={1\over 2}g^{km}(\nabla_i\mathring g_{mj}+\nabla_{j}\mathring g_{mi} - \nabla_m\mathring g_{ij})\).
\end{lemma}

For each connection strain \(\mathring\Gamma\) we define the its fully covariant representation \(\gamma\) and the associated dual.
\begin{definition}[\(\gamma\) and \(\lambda\)]
\label{def:gammaAndLambda}
    Let
    \begin{align}
    \label{eq:gammaConnection}
        \gamma_{m|ij}\coloneqq g_{mk}\mathring\Gamma_{ij}^k,\quad\text{that is,}\quad
        \gamma = \kos(\nabla\mathring g).
    \end{align}
    Similarly, for each connection stress \(L\), define 
    \begin{align}
        \label{eq:lambdaStress}
        \lambda^{m|ij}\coloneqq g^{mk}L_k^{ij}.
    \end{align}
\end{definition}
Note that the types for \(\gamma\) and \(\lambda\) are
\begin{align}
    \gamma\in\Gamma(T^*M\otimes \Sym T^*M),\quad
    \lambda\in\Gamma(TM\otimes\Sym TM).
\end{align}
The dual pairing remains the same \(L_k^{ij}\mathring\Gamma_{ij}^k = \lambda^{m|ij}\gamma_{m|ij}\).

\begin{lemma}[\(({\delta\Gamma\over\delta g})^*\)]
    The component of the stretching stress \(\sigma_{\nabla}\coloneqq ({\delta\Gamma\over\delta g})^*L\) arising from a connection stress \(L\) is given by
    \begin{align}
        \sigma_\nabla = -\nabla\cdot\kos\lambda,\quad\text{or in indices,}\quad
        (\sigma_\nabla)^{ij} = -{1\over 2}\nabla_m(\lambda^{i|mj}+\lambda^{j|mi} - \lambda^{m|ij}),
    \end{align}
    with
    \begin{align}
        \llangle L,\mathring\Gamma\rrangle = \llangle\lambda,\gamma\rrangle = \llangle\sigma_\nabla,\mathring g\rrangle + \oint_{\partial M} b_m (\kos\lambda)^{m|ij}\mathring g_{ij}\,ds
    \end{align}
\end{lemma}
\begin{proof}
    \(\llangle\lambda,\gamma\rrangle = \int_M\langle \lambda|\kos\nabla\mathring g\rangle\, dA = \int_M\langle\kos\lambda|\nabla\mathring g\rangle\,dA = -\int_M\nabla\cdot(\kos\lambda):\mathring g\,dA + \oint_{\partial M}b\cdot(\kos \lambda:\mathring g)\, ds\)
\end{proof}

Next, we study the variation of the Gauss equation residual \(G = gR^\nabla + {1\over 2}\SFF\owedge\SFF\). 
Following the dependency network \eqref{eq:LinearizedNetwork}, we find that it has three contributions:
\begin{align}
\label{eq:GRingDecomposition}
    \mathring G = \underbrace{\mathring g R^\nabla}_{\mathring G_g} + \underbrace{g\left({\delta R^\nabla\over \delta\Gamma}\right)\mathring\Gamma}_{\mathring G_\nabla}
    + \underbrace{({\delta G\over\delta\SFF})\mathring\SFF}_{\mathring G_{\SFF}}.
\end{align}
Note that \(\nabla\) and \(g\) are ultimately dependent with each other, in the decomposition \eqref{eq:GRingDecomposition} we do not expect \(\mathring G_g\) or \(\mathring G_\nabla\) individually to be of the algebraic type of an algebraic curvature, while expecting that \(\mathring G_g + \mathring G_\nabla\) is an algebraic curvature.

\begin{lemma}[\({\delta G\over\delta\Gamma}\)]
    The component \(\mathring G_\nabla = ({\delta G\over\delta\Gamma})\mathring\Gamma\) of \(\mathring G\) arising from a variation of connection is given by
    \begin{align}
        \mathring G_\nabla = d^\nabla\gamma;\quad\text{in indices:}\quad
        (\mathring G_\nabla)_{ijk\ell} = \nabla_i\gamma_{\ell|jk} - \nabla_j\gamma_{\ell|ik}.
    \end{align}
\end{lemma}
\begin{proof}
    Since \(\mathring G_\nabla = g({\delta R^\nabla\over\delta \Gamma})\mathring\Gamma\), our goal is to show \(g\mathring R^\nabla = d^\nabla\gamma\) or equivalently \(\mathring R^\nabla = d^\nabla\mathring\Gamma\). For any fixed test vector field \(Z\), we have \(R^\nabla Z = d^\nabla \nabla Z\).  Taking variation on both sides yield 
    \(\mathring R^\nabla Z = \mathring\Gamma\wedge \nabla Z + d^\nabla(\Gamma Z) = \mathring\Gamma\wedge\nabla Z + d^\nabla\Gamma Z - \Gamma\wedge \nabla Z = d^\nabla \Gamma Z \).  Therefore, \(\mathring R^\nabla = d^\nabla\mathring\Gamma\).
\end{proof}

\begin{lemma}[\(({\delta G\over\delta\Gamma})^*\)]
    The component of the connection stress \(L_G = ({\delta G\over\delta \Gamma})^*\varphi\) arising from an Airy potential \(\varphi\), represented by \(\lambda_G\) via \eqref{eq:lambdaStress}, is given by
    \begin{align}
    \label{eq:lambda_G}
        (\lambda_G)^{m|ij} = -{1\over 4}\Big(
        \mu^{mi}(\sgrad\varphi)^j + 
        \mu^{mj}(\sgrad\varphi)^i \Big)
    \end{align}
    where \((\sgrad\varphi)^i = (-J\grad\varphi)^i = \mu^{ik}\partial_k\varphi\) is the symplectic gradient of \(\varphi\).  Moreover,
    \begin{align}
        \llangle \varphi,\mathring G_\nabla\rrangle = \llangle\lambda_G,\gamma\rrangle - {1\over 2}\oint_{\partial M}\varphi\gamma_{k|i\ell}\mu^{k\ell}t^i\,ds.
    \end{align}
\end{lemma}
\begin{proof}
\begin{align*}
    \llangle \varphi,\mathring G_\nabla\rrangle &= \int_M \varphi(\nabla_i\gamma_{\ell|jk} - \nabla_j\gamma_{\ell |ik}){\mu^{ij}\mu^{k\ell}\over 4}\,dA 
    =
    {1\over 4}\int_M -2\varphi\nabla_j\gamma_{\ell|ik}\mu^{ij}\mu^{k\ell}\, dA\\
    &
    =
    {1\over 2}\int_M (\partial_j\varphi)\gamma_{\ell|ik}\mu^{ij}\mu^{k\ell}\, dA - {1\over 2}\oint_{\partial M}\varphi\gamma_{\ell|ik}\mu^{ij}\mu^{k\ell}b_j\, ds\\
    &=
    {1\over 2}\int_M -(\sgrad\varphi)^i\mu^{\ell k}\gamma_{\ell|ik}\, dA - {1\over 2}\oint_{\partial M}\varphi\gamma_{\ell|ik}\mu^{\ell k}(Jb)^i\, ds.
\end{align*}
Note that \(t = Jb\).  We obtain \eqref{eq:lambda_G} by relabeling the indices and symmetrization for \(\lambda_G\) accounting for the symmetry in \(\gamma\).
\end{proof}

\begin{lemma}[\({\delta G\over\delta g}\)]
    The component \(\mathring G_g = ({\delta G\over\delta g})\mathring g \) in \eqref{eq:GRingDecomposition} is given by
    \begin{align}
        \mathring G_g = \mathring gR^\nabla,\quad (\mathring G_g)_{ijk\ell} = {(R^\nabla)_{ij}}_k^m\mathring g_{m\ell}
    \end{align}
\end{lemma}
\begin{lemma}[\(({\delta G\over\delta g})^*\)]
    The adjoint of \({\delta G\over\delta g}\) that maps an Airy potential \(\varphi\) to the component \(\sigma_G = ({\delta G\over\delta g})^*\varphi\) of the stretching stress is given by
    \begin{align}
        \sigma_G = -{1\over 2}K\varphi g^{-1}
    \end{align}
    with \(\llangle \varphi,\mathring G_g\rrangle = \llangle\sigma_G,\mathring g\rrangle\).
\end{lemma}
\begin{proof}
Using the Gauss equation \(R^\nabla = -KJ\mu\), we have
    \(\llangle \varphi,\mathring G_g\rrangle = \int_{M}\varphi {R^\nabla_{ij}}_k^m\mathring g_{m\ell}{\mu^{ij}\mu^{k\ell}\over 4} = \int_M-K\varphi J_k^m\mathring g_{m\ell}{\mu^{k\ell}\over 2} = \int_M -{1\over 2}K\varphi g^{m\ell}\mathring g_{m\ell}\).
\end{proof}

\begin{lemma}[\({\delta G\over\delta\SFF}\)]
    The component \(\mathring G_{\SFF} = ({\delta G\over\delta\SFF})\mathring\SFF\) in \eqref{eq:GRingDecomposition} is given by
    \begin{align}
        \mathring G_{\SFF} = \SFF\owedge\mathring\SFF,\quad(\mathring G_{\SFF})_{ijk\ell} = \SFF_{ik}\mathring\SFF_{j\ell}+\SFF_{j\ell}\mathring\SFF_{ik} - \SFF_{i\ell}\mathring\SFF_{jk} - \SFF_{jk}\mathring\SFF_{i\ell}.
    \end{align}
\end{lemma}
\begin{lemma}[\(({\delta G\over\delta\SFF})^*\)]
    The component \(\tau_{G} = ({\delta G\over\delta\SFF})^*\varphi\) of the bending stress, with \(\llangle\varphi,\mathring G_{\SFF}\rrangle = \llangle \tau_G,\mathring\SFF \rrangle\), is given by
    \begin{align}
        \tau_G = {\cof(\SFF)\over\det(g)}\varphi,\quad(\tau_G)^{ij} = \mu^{ik}\mu^{j\ell}\SFF_{k\ell}\varphi.
    \end{align}
\end{lemma}
\begin{proof}
    \(\llangle\varphi,\mathring G_{\SFF}\rrangle = \int_M \varphi (\SFF\owedge\mathring\SFF)_{ijk\ell}{\mu^{ij}\mu^{k\ell}\over 4}\, dA \), and comparing it with \(\llangle\tau_G,\mathring\SFF\rrangle = \int_M \tau_G^{ij}\mathring\SFF_{ij}\, dA\) leads to 
    \((\tau_G)^{ij} = \mu^{ik}\mu^{j\ell}\SFF_{k\ell}\varphi\).  Using \(\mu_{ij} = \sqrt{\det(g)}\epsilon_{ij}\),  we get \(\mu^{ik}\mu^{j\ell}\SFF_{k\ell} = {1\over\det(g)}\epsilon^{ik}\epsilon^{j\ell}\SFF_{k\ell} = {1\over\det(g)}\cof(\SFF)^{ij}\).
\end{proof}
Next, we expand the variation of
\begin{align}
    Q = d^\nabla\SFF,\quad Q_{ij|k} = \partial_i\SFF_{jk} - \partial_j\SFF_{ik} - \Gamma_{ik}^\ell\SFF_{j\ell} + \Gamma_{jk}^\ell\SFF_{i\ell}.
\end{align}
(Note that the connection \(\nabla\) in \(d^\nabla\) is on the cotangent bundle, leading to the minus sign in front of the Christoffel symbol.)
\begin{align}
    \mathring Q_{ij|k} = \underbrace{
    \partial_i\mathring\SFF_{jk} - \partial_j\mathring\SFF_{ik} - \Gamma_{ik}^\ell\mathring\SFF_{j\ell}+\Gamma_{jk}^\ell\mathring\SFF_{i\ell}
    }_{(\mathring Q_{\SFF})_{ij|k}}  \underbrace{
    -\mathring\Gamma_{ik}^\ell \SFF_{j\ell} + \mathring\Gamma_{jk}^\ell\SFF_{i\ell}
    }_{(\mathring Q_{\nabla})_{ij|k}}.
\end{align}

\begin{lemma}[\({\delta Q\over\delta\SFF}\)]
    The variation \(\mathring Q_{\SFF} = ({\delta Q\over\delta\SFF})\mathring\SFF\) with respect to variation in \(\SFF\) is given by 
    \begin{align}
        \mathring Q_{\SFF} = d^\nabla\mathring\SFF,\quad (\mathring Q_{\SFF})_{ij|k} = \nabla_i\mathring\SFF_{jk} - \nabla_j\mathring\SFF_{ik}.
    \end{align}
\end{lemma}
\begin{lemma}[\(({\delta Q\over\delta\SFF})^*\)]
    The component \(\tau_{Q} = ({\delta Q\over\delta\SFF})^*\psi\) of the bending stress is given by
    \begin{align}
        \tau_Q = -\sym(\sgrad^\nabla\psi),\quad 
        (\tau_Q)^{ij} = -{1\over 2}\left(\mu^{ik}\nabla_k\psi^j + \mu^{jk}\nabla_k\psi^i\right),
    \end{align}
    with
    \begin{align}
        \llangle \psi,\mathring Q_{\SFF}\rrangle = \llangle\tau_Q,\mathring\SFF\rrangle + \oint_{\partial M}\mathring\SFF(\psi,t)\, ds
    \end{align}
\end{lemma}
\begin{proof}
    \(\llangle \psi,\mathring Q_{\SFF}\rrangle = \int_M \langle d^\nabla\mathring\SFF,\psi\rangle = \int_M\langle\mathring\SFF \wedge\nabla\psi\rangle + \oint_{\partial M}\mathring\SFF(\psi,\cdot) = \int_M\mathring\SFF : (-\sgrad^\nabla\psi)\, dA + \oint_{\partial M}\mathring\SFF(\psi,t)\, ds\).
\end{proof}
\begin{lemma}[\({\delta Q\over\delta\Gamma}\)]
    The variation \(\mathring Q_{\nabla} = ({\delta Q\over\delta\Gamma})\mathring\Gamma\) with respect to variations in the connection is given by
    \begin{align}
        \mathring Q_\nabla = -\gamma\wedge S,\quad
        (\mathring Q_\nabla)_{ij|k} = -\gamma_{m|ki} S_j^m + \gamma_{m|kj} S_i^m
    \end{align}
    where \(\gamma = g\mathring\Gamma\) (\Cref{def:gammaAndLambda}) and \(S = g^{-1}\SFF\).
\end{lemma}

For the next expression of the adjoint of \({\delta Q\over\delta\Gamma}\), it is useful to decompose the second fundamental form into its trace and trace-free parts.  Recall \eqref{eq:GeodesicTorsionForm} and \eqref{eq:GeodesicTorsionBivector}.
Let \(H = {1\over 2}\tr(S) = {1\over 2}\tr_g(\SFF)\) be the mean curvature, \(q = \SFF - Hg\) be the trace-free part of \(\SFF\), and \(\tau_{\rm g}(X,Y)\coloneqq q(JX,Y)\) be the geodesic torsion symmetric bilinear form.  Let \(T_{\rm g} \coloneqq qJ^{\sharp\sharp}\) be the associated symmetric bivector \(T_{\rm g}^{ij}=(qJ^{\sharp\sharp})^{ij} = (qJ)_{i'j'}g^{i'i}g^{j'j}\).
A useful identity is
\begin{align}
\label{eq:muS}
    \mu^{ik}S_k^j = H\mu^{ij} + T_{\rm g}^{ij}.
\end{align}

\begin{lemma}[\(({\delta Q\over\delta\Gamma})^*\)]
    The component \(L_Q = ({\delta Q\over\delta \Gamma})^*\psi\) of the connection stress, represented by \(\lambda_Q\) via \Cref{def:gammaAndLambda}, is given by
    \begin{align}
        (\lambda_Q)^{m|ij} = {1\over 2}
        \left(\psi^i\mu^{j\ell}+\psi^j\mu^{i\ell}\right)S_\ell^m
    \end{align}
    with \(\llangle \psi,\mathring Q_\nabla\rrangle = \llangle L_Q,\mathring\Gamma\rrangle =  \llangle \lambda_Q,\gamma\rrangle\).  Via \eqref{eq:muS}, an alternative expression for \(\lambda_Q\) is
    \begin{align}
    \label{eq:lambda_Q}
        (\lambda_Q)^{m|ij} = {H\over 2}(\psi^i\mu^{jm}+\psi^{j}\mu^{im})
        + {1\over 2} \left(\psi^iT_{\rm g}^{mj}+\psi^jT_{\rm g}^{mi}\right).
    \end{align}
\end{lemma}
\begin{proof}
    \(\llangle\psi|\mathring Q_\nabla\rrangle = \int_M\psi^k(-\gamma_{m|ki}S_j^m + \gamma_{m|kj}S_i^m){\mu^{ji}\over 2}\, dA = \int_M \psi^k\gamma_{m|ki}\mu^{ij}S_j^m\, dA\).  Compare this expression to \(\llangle\lambda_Q,\gamma\rrangle = \int_M (\lambda_Q)^{m|ki}\gamma_{m|ki}\) yields
    \((\lambda_Q)^{m|ki} = {1\over 2}(\psi^k\mu^{ij}S_j^m + \psi^i\mu^{kj}S_j^m).\)
\end{proof}

\begin{remark}
\label{rmk:EigenPropertyOfLambda}
    An important property about 
    the two terms in \eqref{eq:lambda_Q}, as well as the expression \eqref{eq:lambda_G} for \(\lambda_G\), is that they work much better with the Koszul symmetrization operator \(\kos\). Both the first term of \eqref{eq:lambda_Q} and \(\lambda_G\) are eigentensors of \(\kos\) with eigenvalue \(-1\).  The second term of \eqref{eq:lambda_Q} is \({1\over 2}\kos^{-1}(\psi\otimes T_{\rm g})\).
\end{remark}

We now assemble the above lemmas to compute \(\GC\).

The collection of terms in \(\mathring G\) dependent on \(\mathring g\) are \(\mathring G_\nabla + \mathring G_g\):
\begin{align}
    \mathring G_\nabla + \mathring G_g &= d^\nabla\gamma + \mathring g R^\nabla = d^\nabla(\kos(\nabla\mathring g)) +\mathring R^\nabla.
\end{align}
Expanding \(d^\nabla(\kos(\nabla\mathring g))\) yields
\begin{align*}
    (d^\nabla(\kos(\nabla \mathring g)))_{ijk\ell} &= {1\over 2}\Big(\nabla_i(\nabla_j\mathring g_{\ell k} + \nabla_k \mathring g_{\ell j} - \nabla_\ell\mathring g_{jk})\\
    &\qquad -\nabla_j(\nabla_i\mathring g_{\ell k} + \nabla_k \mathring g_{\ell i} - \nabla_\ell\mathring g_{ik})\Big)\\
    &={1\over 2}(\nabla_i\nabla_k\mathring g_{j\ell} +\nabla_j\nabla_\ell \mathring g_{ik} - \nabla_{i}\nabla_\ell\mathring g_{jk} - \nabla_{j}\nabla_k\mathring g_{i\ell}) +{1\over 2}[\nabla_i,\nabla_j]\mathring g_{k\ell}\\
    &={1\over 2}(\nabla\nabla\owedge\mathring g)_{ijk\ell} - {1\over 2}{(R^\nabla)_{ij}}_k^m\mathring g_{m\ell}-{1\over 2}{(R^\nabla)_{ij}}_\ell^m\mathring g_{mk}
\end{align*}
using the fact that \([\nabla_i,\nabla_j]\) gives us the curvature 2-form, whose action on \(\mathring g\in\Gamma(\Sym T^*M)\) is the negative adjoint of the endomorphism-valued 2-form \(R^\nabla\) acting on both covectors in \(\mathring g\).  Therefore,
\begin{align}
\label{eq:GRingWithBothNablaAndMetric}
    (\mathring G_\nabla + \mathring G_g)_{ijk\ell} = {1\over 2}(\nabla\nabla\owedge\mathring g)_{ijk\ell} + {1\over 2}
    \left(\mathring g_{\ell m}{(R^\nabla)_{ij}}_k^m-\mathring g_{km}{(R^\nabla)_{ij}}_\ell^m\right).
\end{align}
Note that after combining \(\mathring G_\nabla\) and \(\mathring G_g\) the tensor has the correct index symmetry as an algebraic curvature.  
Now, the curvature form \(R^\nabla\) without variation satisfies the Gauss equation \(R^\nabla = -KJ\mu\).  
Hence,
\begin{align*}
    \mathring g_{\ell m}{(R^\nabla)_{ij}}_k^m-\mathring g_{km}{(R^\nabla)_{ij}}_\ell^m &= 
-K\mu_{ij}(\mathring g_{\ell m}J_k^m - \mathring g_{km}J_\ell^m) \\
&= -2K\mu_{ij}\mathring\mu_{k\ell} = -K\mu_{ij}\mu_{k\ell}\tr_g(\mathring g) = -{1\over 2}K(g\owedge g)_{ijk\ell}\tr_g(\mathring g),
\end{align*}
simplifying the second term of \eqref{eq:GRingWithBothNablaAndMetric}.  Therefore, the final expression of \(\mathring G\) in terms of \(\mathring g,\mathring\SFF\) is given by
\begin{align}
    &\mathring G = \mathring G_\nabla + \mathring G_g + \mathring G_{\SFF}
    =
    {1\over 2}(\nabla\nabla\owedge \mathring g) - {K\over 4}(g\owedge g)\tr_g(\mathring g) + \SFF\owedge\mathring\SFF.
\end{align}

The variation of the Codazzi residual \(\mathring Q =  \mathring Q_{\SFF}+\mathring Q_\nabla\)
is straightforward:
\begin{align}
    &\mathring Q = d^\nabla\mathring\SFF - \gamma\wedge S,\quad\mathring Q_{ij|k} = 
    \nabla_i\mathring\SFF_{jk}-\nabla_j\mathring\SFF_{ik} - \gamma_{m|ki}S_j^m + \gamma_{m|kj}S_i^m.
\end{align}

Next, we assemble the terms for the adjoint \((\sigma,\tau) = \GC^*(\varphi,\psi)\).

Collecting the terms in \(\sigma = \sigma_\nabla + \sigma_G\) according to the  graph \eqref{eq:AdjointNetwork} yields
\begin{align}
    \sigma = \sigma_{\nabla} + \sigma_G = -\nabla\cdot\kos(\lambda_G+\lambda_Q) + \sigma_G.
\end{align}
Recall \Cref{rmk:EigenPropertyOfLambda} that 
\begin{align}
    \kos\lambda_G = -\lambda_G
\end{align}
 and 
\begin{align}
    \kos(\lambda_Q)^{m|ij} = -{H\over 2}(\psi^i\mu^{jm}+\psi^j\mu^{im}) + {1\over 2}\psi^m T_{\rm g}^{ij}.
\end{align}
Therefore,
\begin{align*}
    \sigma^{ij} = -\nabla_m\bigg[
    &{\mu^{mi}(\sgrad\varphi)^j + \mu^{mj}(\sgrad\varphi)^i\over 4}\\
    &
    -{H\over 2}(\psi^i\mu^{jm}+\psi^j\mu^{im}) + {1\over 2}\psi^mT_{\rm g}^{ij}
    \bigg]-{1\over 2}K\varphi g^{ij}.
\end{align*}
Passing the covariant derivative through the metric tensors and area forms, we have \(-\nabla_m[\mu^{mi}\sgrad(\varphi)^j] = -\nabla_m(\mu^{mi}\mu^{jn}\partial_n\varphi) = \mu^{im}\mu^{jn}\partial_{m}\partial_n\varphi\), which is the classical planar Airy representation.  Performing the similar calculation to the \((H\psi^i\mu^{jm})\) term yields \(\nabla_m(H\psi^i\mu^{jm}) = \mu^{jm}\nabla_m(H\psi^i) = (\sgrad^\nabla(H\psi))^{ji}\).  
Hence, 
\begin{align}
    \sigma = {1\over 2}(\mu^{-1}\otimes\mu^{-1})\Hess\varphi + \sym\left(\sgrad^{\nabla}(H\psi)\right)
    -{(\nabla\cdot\psi)T_{\rm g} + \nabla_\psi T_{\rm g}\over 2}-{1\over 2}K\varphi g^{-1}.
\end{align}

The expression for the bending stress \(\tau = \tau_{G} + \tau_{Q}\) is more straightforward:
\begin{align}
    \tau = {\cof(\SFF)\over\det(g)}\varphi - \sym(\sgrad^\nabla \psi).
\end{align}

Finally, we collect the boundary terms:
\begin{align}
    \llangle\varphi,\mathring G\rrangle + \llangle\psi,\mathring Q\rrangle &=
    \llangle \varphi,\mathring G_\nabla\rrangle + \llangle\varphi,\mathring G_g\rrangle + \llangle \varphi,\mathring G_{\SFF}\rrangle + \llangle\psi,\mathring Q_\nabla\rrangle + \llangle \psi,\mathring Q_{\SFF}\rrangle\\
    \nonumber
    &=
    \llangle \lambda_G,\gamma\rrangle + \llangle\sigma_G,\mathring g\rrangle + \llangle\tau_G,\mathring\SFF\rrangle + \llangle \lambda_Q,\gamma\rrangle + \llangle\tau_Q,\mathring\SFF\rrangle\\
    &\quad
    -{1\over 2}\oint_{\partial M}\varphi\underbrace{\gamma_{k|i\ell}}_{\kos(\nabla\mathring g)_{k|i\ell}}\mu^{k\ell}t^i\, ds
    +\oint_{\partial M}\mathring\SFF(\psi,t)\,ds.
\end{align}
Note that \(\kos(\nabla\mathring g)_{k|i\ell}\mu^{k\ell} = -(\nabla\mathring g)_{k|i\ell}\mu^{k\ell}\) since the contraction with \(\mu^{k\ell}\) projects to the \(-1\) eigenspace of \(\kos\).  Next, combine \(\llangle\lambda_G,\gamma\rrangle + \llangle\lambda_Q,\gamma\rrangle = \llangle\lambda,\gamma\rrangle\) which becomes \(\llangle\sigma_\nabla,\mathring g\rrangle + \oint_{\partial M}b_m(\kos\lambda)^{m|ij}\mathring g_{ij}\, ds\).
Thus,
\begin{align}
    \nonumber
    \llangle\varphi,\mathring G\rrangle + \llangle\psi,\mathring Q\rrangle &=
    \llangle\sigma,\mathring g\rrangle + \llangle\tau,\mathring\SFF\rrangle +
    \underbrace{
    \oint_{\partial M}(\kos\lambda)^{m|ij}\mathring g_{ij}\, ds
    }_{\text{(i)}} \\
    &\quad+ \underbrace{
    \oint_{\partial M}{1\over 2}\varphi\nabla_k\mathring g_{i\ell}\mu^{k\ell}t^i\, ds
    }_{\text{(ii)}} + \underbrace{\oint_{\partial M}\mathring\SFF(\psi,t)\,ds}_{\text{(iii)}}.
    \label{eq:ThreeBoundaryTerms}
\end{align}

In boundary part (i), we have \((\kos\lambda)^{m|ij} = -\lambda_G^{m|ij} - {H\over 2}(\psi^i\mu^{jm}+\psi^j\mu^{im})+{1\over 2}\psi^m T_{\rm g}^{ij}\) and \(-\lambda_{G}^{m|ij}={1\over 4}(\mu^{mi}(\sgrad\varphi)^j + \mu^{mj}(\sgrad\varphi)^i)\).  Therefore, written in the \((t,b)\) basis,
\begin{align}
    \text{(i)} &= \oint_{\partial M} b_m{\mu^{mi}(\sgrad \varphi)^j\over 2}\mathring g_{ij} + H\psi^it^j\mathring g_{ij} + {1\over 2} b_m\psi^m T_{\rm g}^{ij}\mathring g_{ij}\\
    \nonumber
    &=\oint_{\partial M}
    \left(
    \begin{bmatrix}
        -{\partial_b\varphi\over 2} & {\partial_t\varphi\over 4}\\
        {\partial_t\varphi\over 4}
    \end{bmatrix}
    +
    H
    \begin{bmatrix}
        \langle\psi,t\rangle & {\langle\psi,b\rangle\over 2}\\
        {\langle\psi,b\rangle\over 2} 
    \end{bmatrix}
    \right.
    \\
    &\qquad\qquad\qquad \left.
    {}+
    {1\over 2}\langle b,\psi\rangle
    \begin{bmatrix}
            T_{\rm g}(t,t) & T_{\rm g}(t,b)\\
            T_{\rm g}(b,t) & T_{\rm g}(b,b)
        \end{bmatrix}
    \right):
    \begin{bmatrix}
        \mathring g_{tt} & \mathring g_{tb}\\
        \mathring g_{bt}&
        \mathring g_{bb}
    \end{bmatrix}
\end{align}
Note that \(T_{\rm g}(t,t) = -\SFF(t,b)\) is the geodesic torsion \(\tau_{\rm g}(t)\) along the \(t\) direction whereas \(T_{\rm g}(b,b) = -\tau_{\rm g}(t)\).  On the other hand, \(T_{\rm g}(t,b) = q(t,t) = \SFF(t,t)-H = \kappa_{\rm n}(t)\) is the normal curvature along the \(t\) direction.  They simplify the expression:
\begin{align}
    \text{(i)} &=
    \oint_{\partial M}
    \left(
    \begin{bmatrix}
        -{\partial_b\varphi\over 2} & {\partial_t\varphi\over 4}\\
        {\partial_t\varphi\over 4}
    \end{bmatrix}
    +
    \begin{bmatrix}
        H\langle t,\psi\rangle + {1\over 2}\tau_{\rm g}(t)\langle b, \psi\rangle & {1\over 2}\kappa_{\rm n}(t)\langle b,\psi\rangle\\
        {1\over 2}\kappa_{\rm n}(t)\langle b,\psi\rangle & -{1\over 2}\tau_{\rm g}(t)\langle b,\psi\rangle
    \end{bmatrix}
    \right)
    :
    \begin{bmatrix}
        \mathring g_{tt} & \mathring g_{tb}\\
        \mathring g_{bt}&
        \mathring g_{bb}
    \end{bmatrix}.
\end{align}

For boundary part (ii), writing in \((t,b)\) basis yields
\begin{align}
    \text{(ii)} &= {1\over 2}\oint_{\partial M}
    \varphi\Big(
    -(\nabla_b\mathring g)(t,t) + (\nabla_t\mathring g)(b,t)
    \Big)\, ds.
\end{align}
The second term involving \(\nabla_t\) can be applied with an integration by parts since the integral is along \(t\). Hence,
\begin{align}
    \text{(ii)} &= 
    -{1\over 2}\oint_{\partial M}
    (\nabla_b\mathring g)(t,t)\varphi + (\partial_t\varphi)\mathring g_{bt} + \varphi\mathring g(\underbrace{\nabla_tb}_{\kappa_{\rm g}t},t) + \varphi\mathring g(b,\underbrace{\nabla_tt}_{-\kappa_{\rm g}b})\\
    &=-{1\over 2}
    \oint_{\partial M}
    (\nabla_b\mathring g)(t,t)\varphi
    -{1\over 2}\oint_{\partial M}
    \begin{bmatrix}
        \kappa_{\rm g}\varphi & {1\over 2}\partial_t\varphi\\
        {1\over 2}\partial_t\varphi & -\kappa_{\rm g}\varphi
    \end{bmatrix}
    :
    \begin{bmatrix}
        \mathring g_{tt} & \mathring g_{tb}\\
        \mathring g_{bt}&
        \mathring g_{bb}
    \end{bmatrix}
\end{align}
where \(\kappa_{\rm g}\) is the geodesic curvature of \(\partial M\).

Part (iii) is more straightforward:
\begin{align}
    \text{(iii)} &=
    \oint_{\partial M}\mathring\SFF(\psi,t) = \oint_{\partial M}
    \begin{bmatrix}
        \langle t,\psi\rangle & {1\over 2}{\langle b,\psi\rangle}\\
        {1\over 2}{\langle b,\psi\rangle}
    \end{bmatrix}
    :
    \begin{bmatrix}
        \mathring \SFF_{tt} & \mathring \SFF_{tb}\\
        \mathring \SFF_{bt}&
        \mathring \SFF_{bb}
    \end{bmatrix}.
\end{align}

Combining all three boundary terms for \eqref{eq:ThreeBoundaryTerms}, we get
\begin{align}
    \nonumber
    \llangle\varphi,\mathring G\rrangle + \llangle\psi,\mathring Q\rrangle &=
    \llangle\sigma,\mathring g\rrangle + \llangle\tau,\mathring\SFF\rrangle \\
    \nonumber
    &\quad
    -{1\over 2}\oint_{\partial M}(\nabla_b\mathring g)(t,t)\varphi\, ds\\
    \nonumber
    &\quad+
    \oint_{\partial M}
    \left(
    \begin{bmatrix}
        -{\partial_b\varphi\over 2}-{\kappa_{\rm g}\over 2}\varphi\\
        &{\kappa_{\rm g}\over 2}\varphi
    \end{bmatrix}
    +
    \begin{bmatrix}
        H\psi_t +{\tau_{\rm g}\over 2}\psi_b & {\kappa_{\rm n}\over 2}\psi_b\\
        {\kappa_{\rm n}\over 2}\psi_b &
        -{\tau_{\rm g}\over 2}\psi_b
    \end{bmatrix}
    \right)
    :
    \begin{bmatrix}
        \mathring g_{tt} & \mathring g_{tb}\\
        \mathring g_{bt}&
        \mathring g_{bb}
    \end{bmatrix}\\
    &\quad
    +\oint_{\partial M}
    \begin{bmatrix}
        \psi_t & {1\over 2}\psi_b\\
        {1\over 2}\psi_b
    \end{bmatrix}
    :
    \begin{bmatrix}
        \mathring \SFF_{tt} & \mathring \SFF_{tb}\\
        \mathring \SFF_{bt}&
        \mathring \SFF_{bb}
    \end{bmatrix}.
\end{align}

\section{Cohomology of the Shell Complex}

Putting the kinematic complex \eqref{eq:KinematicShellComplex} and its dual complex \eqref{eq:DynamicShellComplex} together, we obtain the full shell complex.
Let us assume \(M\) is connected, so that the space of global rigid motions \(\underline{\se(3)}(M)\) and the space of global residual class \(H^2(M;\se(3))\) are both just the six dimensional vector space \(\se(3)\).
\begin{align}
\label{eq:FullShellComplex}
    \begin{aligned}[c]
\xymatrix@C=1.5em{
\se(3)
\ar@{^(->}[r]
\ar@{.>}[d]^{\bM}
\ar@{}[dr]|{\displaystyle\circlearrowright}
& 
\mathring x\in\fX^0(M)
\ar[r]^-{D}
\ar@{.>}[d]^{\star_0}
& 
(\mathring g,\mathring\SFF)\in
\fX^1(M)
\ar[r]^-{\GC}
\ar@{.>}[d]^{\star_1}
&
(\mathring G,\mathring Q)\in\fX^2(M)
\ar@{->>}[r]
\ar@{.>}[d]^{\star_2}
&
\se(3)\\
\se(3)^*
&
-f\in\fX^0(M)^*
\ar@{->>}[l]
&
(\sigma,\tau)\in\fX^1(M)^*
\ar@{->}_-{D^*}[l]
&
(\varphi,\psi)\in\fX^2(M)^*
\ar@{->}_-{\GC^*}[l]
&
\se(3)^*
\ar@{_(->}[l]
\ar@{.>}[u]^-{\bW}
\ar@{}[ul]|{\displaystyle\circlearrowright}
}
    \end{aligned}
\end{align}
Here, each \(\star_k\colon\fX^k(M)\to \fX^k(M)^*\) is any fiberwise symmetric positive definite map encoding constitutive material properties.  Specifically, \(\star_0\) is the mass density and \(\star_1\) is the stress--strain relation.  The operator \(\star_2\) is auxiliary that does not have a clear material interpretation.  
The \(6\times 6\) matrices \(\bM\) and \(\bW\) are induced from \(\star_0\) and \(\star_2^{-1}\) so that the squares on the ends of the diagram commute.
The matrix \(\bM\) is the global inertia tensor (consisting of mass and moment of inertia) of the shell as a rigid body, induced from the mass distribution \(\star_0\). Analogously, the matrix \(\bW\) is the global weight for the gauge fields in the Airy potentials induced from the weights \(\star_2^{-1}\).

Without imposing boundary conditions, we have \(\ker(D) \cong\se(3)\) and \(\fX^2(M)/\im(\GC)\cong\se(3)\).  That is, the sequence  \eqref{eq:FullShellComplex} is exact at both ends, explicitly expressing the cohomology groups \(H^0(M;\se(3))\) and \(H^2(M;\se(3))\). The remaining part of \eqref{eq:FullShellComplex} is a chain complex whose cohomology is \( \ker(\GC)/\im(D)\cong H^1(M;\se(3))\) given by the \(\se(3)\)-sheaf cohomology.  
The constitutive isomorphisms \(\star_k\) play a role analogous to that of the Hodge stars in the de Rham complex, allowing us to select harmonic representatives of the cohomology classes in \(H^\bullet(M;\se(3))\).  This resulting cohomological splitting is known as the \emph{Hodge decomposition}.

In the presence of boundaries, the Hodge decomposition must account for boundary conditions.  The boundary terms of \Cref{thm:GreenIdentityD,thm:GreenIdentityGC} provide guidance for identifying the natural boundary conditions. Incorporating the boundary conditions finally leads to a \emph{Hodge--Morrey--Friedrichs decomposition} for the shell complex, representing both absolute and relative cohomology.

\subsection{Hilbert Structures}
Any positive-definite stress--strain relation \(\star_1\colon\fX^1(M)\to\fX^1(M)^*\) defines an \(L^2\) inner product on \(\fX^1(M)\) (and dually on \(\fX^1(M)^*\)) by \(\llangle (\dot g,\dot\SFF),(\mathring g,\mathring\SFF)\rrangle_{\star_1} = \llangle (\dot g,\dot\SFF),\star_1(\mathring g,\mathring\SFF)\rrangle\) where the pairing on the right-hand side is the natural dual pairing.

The Hilbert space \(L^2\fX^1\) induces, through the deformation map \(D\), a ``Dirichlet energy'' inner product and thereby a Hilbert structure on the displacement space \(\fX^0\) modulo infinitesimal rigid motions.  Unlike the exterior derivative in the de Rham complex, the deformation map \(D\) involves both first- and second-order derivatives (see \Cref{thm:DeformationMap}).  More precisely, for a smooth immersion \(x\colon M\to\RR^3\), we consider displacements whose tangential components belong to \(W^{1,2}\) and whose normal components belong to \(W^{2,2}\):
\begin{align}
    W^{1,2;2,2}\fX^0(M)\coloneqq \left\{\mathring x\colon M\to \RR^3\,\middle|\,\mathring x^\parallel\in W^{1,2}\Gamma(TM), \mathring x^\bot \in W^{2,2}(M;\RR)\right\}
\end{align}
where \(\mathring x^\parallel\) and \(\mathring x^\bot\) denote the tangential and normal components of the \(\RR^3\)-valued displacement field \(\mathring x = dx(\mathring x^\parallel) + \mathring x^\bot n\). 
This choice of function spaces is consistent with the foundational work of \cite{ciarlet2000mathematical} on the mathematical analysis and well-posedness of shell deformation problems.

The associated Dirichlet bilinear form
\begin{align}
    \cB_D(\dot x,\mathring x)\coloneqq \llangle D\dot x, D\mathring x\rrangle_{\star_1}
\end{align}
is continuous on \(W^{1,2;2,2}\fX^0(M)\) and coercive on the quotient  \(W^{1,2;2,2}\fX^0(M)/\ker(D)\).%
\footnote{One checks that there exists a constant \(C\) such that \(\inf_{r\in\ker D}\|\mathring x-r\|_{W^{1,2;2,2}}^2
\le C\|D\mathring x\|_{\star_1}^2.\)  The related \emph{Korn's inequality} is established by \cite{ciarlet2000mathematical}.}
That is, the mixed-order Sobolev space \(W^{1,2;2,2}\fX^0(M)\) is equivalent to the graph space of the \(D\) operator \(\{\mathring x\in L^2\fX^0(M)\,|\,D\mathring x\in L^2\fX^1(M)\}\). 
In fact, the constitutive elastic potential energy of a displacement \(\mathring x\) is precisely \({1\over 2}\cB_D(\mathring x,\mathring x)\).  Thus, \(W^{1,2;2,2}\fX^0(M)\) is the natural functional space for finite-energy displacements.

This Hilbert space setting allows us to establish the following well-posedness result by a direct application of the Lax--Milgram theorem.
\begin{proposition}
    Given \((\mathring g,\mathring\SFF)\in L^2\fX^1(M)\), there exists \(\mathring x\in W^{1,2;2,2}\fX^0(M)\) satisfying
    \begin{align}
        \llangle D\dot x, D\mathring x\rrangle_{\star_1} = \llangle D\dot x,(\mathring g,\mathring\SFF)\rrangle_{\star_1},\quad\text{for all \(\dot x\in W^{1,2;2,2}\fX^0(M)\).}
    \end{align}
    The  solution is unique up to \(\ker(D)\).
\end{proposition}
Equivalently, the PDE
\begin{align}
    (D^*\star_1 D)\mathring x  = D^*\star_1(\mathring g,\mathring\SFF)
\end{align}
together with the corresponding natural boundary conditions, admits a unique weak solution modulo \(\ker(D)\).  This solution is the pseudoinverse, or least-squares, solution of \(D\mathring x = (\mathring g,\mathring\SFF)\).  The analogous construction for the de Rham complex is also known as ``Poisson reconstruction.''

Similarly, we equip the space of Airy potentials with the \(W^{2,2;1,2}\) Hilbert structure, \(W^{2,2;1,2}\fX^2(M)^* = \{(\varphi,\psi)\,|\,\varphi\in W^{2,2}(M;\RR), \psi\in W^{1,2}\Gamma(TM)\}\).  The corresponding spaces of forces and curvatures are \(W^{-1,2;-2,2}\fX^0(M)^*\) and \(W^{-2,2;-1,2}\fX^2(M)\)  defined as the dual spaces of displacements and Airy potentials respectively.  With these Sobolev structures, the operators \(D,D^*,\GC,\GC^*\) are continuous linear maps.  

We further consider the graph space \(W^{\GC,2}\fX^1(M)\coloneqq\{(\mathring g,\mathring\SFF)\in L^2\fX^1(M)\,|\,\GC(\mathring g,\mathring\SFF)\in L^2\fX^{2}(M)\}\) and analogously define \(W^{D^*,2}\fX^1(M)^*\) for stress fields.  These spaces are the shell analogues of the familiar \(H(\div), H(\curl)\) spaces in the de Rham complex. They provide the natural setting for defining weak boundary traces of strain and stress fields through the Green identities (\Cref{thm:GreenIdentityD,thm:GreenIdentityGC}) associated with \(\GC\) and \(D\).

In the following discussion of Hodge decompositions, we proceed formally with Hilbert space operations, focusing on the algebraic structure of the decompositions while leaving the functional-analytic details implicit. 
 

\subsection{Hodge Decomposition for Shells without Boundary}
The following Hodge decomposition for a closed shell states that every strain admits an orthogonal decomposition into three components: one induced by a displacement, one corresponding to a stress represented by Airy potentials, and a harmonic strain that satisfies the linearized Gauss--Codazzi equations and corresponds to an equilibrium stress.
\begin{theorem}
If \(\partial M = \varnothing\), then
the space of strains admits a \(\star_1\)-orthogonal decomposition
    \begin{align}
    \label{eq:StrainHodgeDecompositionNoBoundary}
        \fX^1(M) = \im(D)\oplus \im(\star_1^{-1}\GC^*) \oplus \cH\fX^1(M)
    \end{align}
    where
    \begin{align}
        \cH\fX^1(M) \coloneqq \ker(\GC)\cap \ker(D^*\star_1) \cong H^1(M;\se(3)).
    \end{align}
\end{theorem}
\begin{proof}
    The space \(\fX^1(M)\) splits into \(\im(D)\) and its orthogonal complement \(\im(D)^\bot = \ker(D^*\star_1)\).  The dual chain complex \(\im(\GC^*)\subset\ker(D^*)\) further splits \(\ker(D^*\star_1)\) orthogonally into \(\im(\star_1^{-1}\GC^*)\) and \(\im(\star_1^{-1}\GC^*)^\bot\cap\ker(D^*\star_1)\). Note that \(\im(\star_1^{-1}\GC^*)^\bot = \ker(\GC)\). Therefore, \(\fX^1(M) = \im(D)\oplus \im(\star_1^{-1}\GC^*)\oplus (\ker(\GC)\cap\ker(D^*\star_1))\).  To show that \(\cH\fX^1(M)\cong H^1(M;\se(3))\), observe that \(\im(D)\oplus\cH\fX^1(M)=\im(\star_1^{-1}\GC^*)^\bot = \ker(\GC)\).  Therefore, \(\cH\fX^1(M)\cong \ker(\GC)/\im(D) \cong H^1(M;\se(3))\).
\end{proof}
By applying \(\star_1\) to \eqref{eq:StrainHodgeDecompositionNoBoundary}, we obtain the Hodge decomposition for the space of stress fields.
\begin{corollary}
    If \(\partial M = \varnothing\), then
the space of stress fields admits a \(\star_1^{-1}\)-orthogonal decomposition
    \begin{align}
    \label{eq:StrainHodgeDecompositionNoBoundary}
        \fX^1(M)^* = \im(\GC^*)\oplus \im(\star_1 D^*) \oplus \cH\fX^1(M)^*
    \end{align}
    where
    \begin{align}
        \cH\fX^1(M)^* = \ker(D^*)\cap \ker(\GC\star_1^{-1}) \cong H^1(M;\se(3)).
    \end{align}
\end{corollary}

In other words, every stress field admits an orthogonal decomposition into three components: one represented by Airy potentials, one arising from a displacement through the constitutive relation, and a harmonic stress field that is in equilibrium and whose constitutive inverse satisfies the linearized Gauss--Codazzi equations.

\subsection{Boundary Conditions}

When \(M\) has a boundary, the Hodge decomposition requires specifying suitable boundary conditions. The Green identities in \Cref{thm:GreenIdentityD,thm:GreenIdentityGC} provide the expressions for the boundary pairings.  We first consider two complementary boundary conditions for displacements and stresses.

\begin{definition}
    A displacement field \(\mathring x\in\fX^0(M)\) is said to have \emph{free boundary} if it is not constrained by any boundary condition.
\end{definition}

\begin{definition}[Clamped displacement]
    A displacement field \(\mathring x\in W^{1,2;2,2}\fX^0(M)\) is said to satisfy the \emph{clamped boundary condition} if
    \begin{align}
    \label{eq:ClampedDisplacement}
        \mathring x = 0, \quad \partial_b\mathring x^\bot = 0\quad\text{on \(\partial M\)}.
    \end{align}
\end{definition}

To define the complementary stress boundary condition, recall that the boundary terms in \Cref{thm:GreenIdentityD} can be expressed in terms of three independent displacement traces: \(\mathring x^\parallel\), \(\mathring x^\bot\), and \(\partial_b\mathring x^\bot\).  Their conjugate stress quantities are the tangential force traction, effective transverse shear traction, and normal bending moment respectively.

\begin{definition}[Traction-free stress]
    A stress field \((\sigma,\tau)\in W^{D^*,2}\fX^1(M)^*\) is said to satisfy the \emph{traction-free boundary condition} if on \(\partial M\)
    \begin{subequations}
        \begin{align}
        & 2(\sigma + S\tau)b = 0,\\
        & 2\partial_t(\tau_{tb})+(\nabla_b\tau)_{bb} + \kappa_{\rm g}(\tau_{bb} - \tau_{tt}) = 0\\
        & \tau_{bb} = 0.
    \end{align}
    \end{subequations}
\end{definition}
The first two conditions express the vanishing of the tangential and effective transverse force tractions, while the third requires the normal bending moment to vanish. These are the natural free-edge conditions in shell theory. In particular, when the surface is flat and the edge is straight (\(\kappa_{\rm g} = 0\)) the vanishing effective transverse shear traction reduces to \(2\partial_t\tau_{tb} + \partial_b\tau_{bb}=0\) in the classical Kirchhoff--Love plate theory.

\begin{definition}
    A strain field \((\mathring g,\mathring\SFF)\in\fX^1(M)\) is said to have traction-free boundary if its corresponding stress \(\star_1(\mathring g,\mathring\SFF)\in\fX^1(M)^*\) has traction-free boundary.
\end{definition}

Let \(\fX^0_{\rm clamped}\) denote the subspaces of clamped displacements, and \(D_\clamped\) be \(D\) restricted to \(\fX^0_\clamped\).

\begin{proposition}\label{prop:StrainOrthogonalComplementWithBC}
    We have the orthogonal complement relationships in \(\fX^1(M)\)
    \begin{align}
        \im(D_{\rm clamped})^\bot = \ker(D^*\star_1)\quad\text{and}\quad
        \im(D)^\bot = \ker(D^*\star_1)|_{\substack{{\rm traction}\\{\rm free}}}
    \end{align}
\end{proposition}
\begin{proof}
    These orthogonal complements follow from the Green identity  (\Cref{thm:GreenIdentityD}).
\end{proof}

Next, we consider the complementary boundary conditions inferred from \Cref{thm:GreenIdentityGC}.

\begin{definition}[Clamped Airy potential]
    An Airy potential \((\varphi,\psi)\in W^{2,2;1,2}\fX^2(M)^*\) is said to satisfy the clamped boundary condition if 
    \begin{align}
        \varphi = 0,\quad \partial_b\varphi = 0,\quad \psi = 0\quad \text{on $\partial M$.}
    \end{align}
\end{definition}

\begin{definition}[Gauss--Codazzi relative boundary condition]
    A strain field \((\mathring g,\mathring\SFF)\in W^{\GC,2}\fX^1(M)\) is said to satisfy the \emph{Gauss--Codazzi relative boundary condition} (GCBC) if its boundary pairing in \Cref{thm:GreenIdentityGC} vanishes for all Airy potentials.  That is, on \(\partial M\),
    \begin{subequations}
    \label{eq:GCBC}
        \begin{align}
            &\mathring g_{tt} = 0,\qquad (\nabla_b\mathring g)_{tt} - \kappa_{\rm g}\mathring g_{bb} = 0,\\&\mathring\SFF_{tt} = 0,\qquad 2\mathring\SFF_{tb}+ 2\kappa_{\rm n}\mathring g_{tb} - \tau_{\rm g}\mathring g_{bb} = 0.
        \end{align}
    \end{subequations}
\end{definition}
Geometrically, \eqref{eq:GCBC} constrains stretching along the boundary curve (\(\mathring g_{tt}=0\)), the variation of tangential stretching across a boundary collar (\((\nabla_b\mathring g)_{tt} - \kappa_{\rm g}\mathring g_{bb} = 0\)), and variations of bending and twisting of the collar.  These conditions impose constraints on the strains of a collar neighborhood, analogous to how clamped boundary conditions constrain the position and orientation of a shell boundary.  One may think of GCBC as a geometric clamping at the level of strains \((\mathring g,\mathring\SFF)\in \fX^1(M)\) without assuming \((\mathring g,\mathring\SFF)\) satisfies Gauss--Codazzi equation or comes from a displacement field.
\begin{definition}
    A stress field \((\sigma,\tau)\in\fX^1(M)^{*}\) is said to satisfy the GCBC if its corresponding strain field \(\star_1^{-1}(\sigma,\tau)\) satisfies the GCBC.
\end{definition}
\begin{proposition}\label{prop:StressOrthogonalComplementWithBC}
    We have the orthogonal complement relationships in and \(\fX^1(M)^*\)
    \begin{align}
        \im(\GC_\clamped^*)^\bot = \ker(\GC\star_1^{-1})\quad\text{and}\quad
        \im(\GC^*)^\bot = \ker(\GC\star^{-1})|_{\GCBC}.
    \end{align}
\end{proposition}
\begin{proposition}\label{prop:SubChainComplex}
The strain field of a clamped displacement satisfies GCBC, and the stress field represented by a clamped Airy potential has traction free boundary:
    \begin{align}
        \im(D_\clamped)\subset\ker(\GC)|_{\GCBC},\quad 
        \im(\GC_\clamped^*)\subset\ker(D^*)|_\tractionfree.
    \end{align}
\end{proposition}
\Cref{prop:SubChainComplex} implies that we have relative chain complexes 
    \begin{align}
        \fX^0_{\clamped}(M)\xrightarrow{D_{\clamped}}\fX^1_{\GCBC}(M)\xrightarrow{\GC_{\GCBC}}\fX^2(M) 
    \end{align}
    and
    \begin{align}
        \fX^{0*}(M)\xleftarrow{D^*_{\tractionfree}}\fX^{1*}_{\tractionfree}(M)\xleftarrow{\GC^*_{\clamped}}\fX^{2*}_\clamped(M).
    \end{align}
\subsection{Hodge--Morrey--Friedrichs Decomposition for Shells with Boundaries}
For shell with boundary, define the space of harmonic stress and strain fields with various boundary conditions as
\begin{align}
    &\cH\fX^1(M) \coloneqq \ker(\GC)\cap\ker(D^*\star_1)\\
    &\cH\fX^{1*}(M) \coloneqq \ker(\star_1^{-1}\GC)\cap\ker(D^*) =  \star_1\cH\fX^1(M)\\
    &\cH_{\GCBC}\fX^1(M) \coloneqq \cH\fX^1(M)|_{\GCBC}\\
    &\cH_{\tractionfree}\fX^{1*}(M)\coloneqq \cH\fX^{1*}(M)|_{\tractionfree}\\
    &\cH_{\GCBC}\fX^{1*} \coloneqq \star_1\cH_{\GCBC}\fX^1(M),\\
    &\cH_{\tractionfree}\fX^{1}(M) \coloneqq \star^{-1}\cH_{\tractionfree}\fX^{1*}(M).
\end{align}
Like the Hodge decompositions for the de Rham complex, the harmonic fields without boundary conditions are generally infinite dimensional, and are finite dimensional once the natural boundary conditions are imposed.
\begin{theorem}[Morrey decomposition for strain fields on shells]
    The space \(\fX^1(M)\) of strain fields is \(\star_1\)-orthogonally decomposed into 
    \begin{align}
        \fX^1(M) = \im(D_{\clamped})\oplus\im(\star_1\GC_{\clamped}^*)\oplus \cH\fX^1(M).
    \end{align}
\end{theorem}
\begin{proof}
    By \Cref{prop:StrainOrthogonalComplementWithBC} we have \(\fX^1(M) = \im(D_{\clamped})\oplus\ker(D^*\star_1)\).  Observe that \(\im(\star_1^{-1}\GC_\clamped^*)\subset\ker(D^*\star_1)\) and \Cref{prop:StressOrthogonalComplementWithBC}.  Therefore, \(\ker(D^*\star_1) = \im(\star_1^{-1}\GC_\clamped^*)\oplus \ker(\GC)\cap\ker(D^*\star_1)\).
\end{proof}
\begin{corollary}
    The space \(\fX^{1*}(M)\) of stress fields is \(\star_1^{-1}\)-orthogonally decomposed into
    \begin{align}
        \fX^{1*}(M) = \im(\star_1 D_\clamped)\oplus \im(\GC^*_\clamped)\oplus \cH\fX^{1*}(M).
    \end{align}
\end{corollary}
\begin{theorem}[Friedrichs decomposition for harmonic strain fields]
    The space \(\cH\fX^1(M)\) of harmonic strain fields admits the following orthogonal decompositions
    \begin{align}
    \label{eq:Friedrichs1}
        \cH\fX^1(M) &= \im(D)\cap\ker(D^*\star_1) \oplus \cH_{\tractionfree}\fX^1(M)\\
        \label{eq:Friedrichs2}
        &= \im(\star^{-1}\GC^*)\cap(\GC) \oplus \cH_{\GCBC}\fX^1(M).
    \end{align}
\end{theorem}
\begin{proof}
    Applying \(\im(D)\subset\ker(\GC)\) in \(\cH\fX^1(M) = \ker(\GC)\cap\ker(D^*\star_1)\) and \(\im(D)^\bot = \ker(D^*\star_1)_{\tractionfree}\) (\Cref{prop:StrainOrthogonalComplementWithBC}), we have \(\cH\fX^1(M) = \im(D)\cap\ker(D^*\star_1)\oplus\ker(D^*\star_1)_{\tractionfree}\cap\ker(\GC)\), showing \eqref{eq:Friedrichs1}.  The decomposition \eqref{eq:Friedrichs2} is derived analogously starting with \(\im(\star_1^{-1}\GC^*)\subset\ker(D^*\star_1)\) in \(\cH\fX^1(M) = \ker(\GC)\cap\ker(D^*\star_1)\).
\end{proof}

\begin{corollary}[Hodge--Morrey--Friedrichs decomposition for shells]
The space \(\fX^1(M)\) of strain fields on a shell can be orthogonally decomposed in two ways:
    \begin{align}
    \label{eq:HodgeMorreyFriedrichs1}
    \begin{aligned}
        &\overbrace{
        \overbrace{
        \phantom{\im(D_\clamped)\oplus \im(D)\cap\ker(D^*\star_1)}
        }^{\im(D)\,\,\text{from displacement}}
        \phantom{
        \oplus}
        \overbrace{
        \phantom{
        \cH_{\tractionfree}\fX^1(M)
        }}
        ^{\mathclap{\substack{\text{traction-free}\\\text{harmonic strain}}}}
        }^{\ker(\GC)\,\,\text{Gauss--Codazzi satisfying}}
        \phantom{\oplus}
        \overbrace{
        \phantom{\im(\star_1^{-1}\GC_\clamped^*)}
        }^{\mathclap{\substack{\text{from clamped}\\\text{Airy potential}}}}
        \\[-2em]
        \fX^1(M) = & 
        \underbrace{\im(D_\clamped)}_{\mathclap{\substack{\text{from clamped}\\\text{displacement}}}}
        \oplus
        \underbrace{\im(D)\cap\ker(D^*\star_1)}_{\mathclap{\text{harmonic displacement}}}
        \oplus
        \underbrace{\cH_{\tractionfree}\fX^1(M) \oplus \im(\star_1^{-1}\GC^*_{\clamped})}_{\mathclap{\substack{\text{traction-free }\\\text{equilibrial stress}}\,\ker(D^*\star_1)_{\tractionfree}}}
        \\[-2em]
        &
        \phantom{\im(D_\clamped)}
        \phantom{\oplus}
        \underbrace{
        \underbrace{\phantom{\im(D)\cap\ker(D^*\star_1)\oplus\cH_\tractionfree\fX^1(M)}}_{\mathclap{\cH\fX^1(M)\,\,\text{harmonic strain}}}
        \phantom{{}\oplus\im(\star_1^{-1}\GC_\clamped^*)}
        }_{\ker(D^*\star_1)\,\,\text{corresponding stress is in equilibrium}}
    \end{aligned}
    \end{align}
    \begin{align}
    \label{eq:HodgeMorreyFriedrichs2}
    \begin{aligned}
        &\overbrace{
        \overbrace{
        \phantom{\im(D_\clamped)\oplus \cH_{\GCBC}\fX^1(M)}
        }^{\mathclap{\ker(\GC)_{\GCBC}\,\,\substack{\text{Gauss--Codazzi}\\\text{satisfying with GCBC}}}}
        \phantom{
        \oplus}
        \overbrace{
        \phantom{
        \im(\star_1^{-1}\GC^*)\cap\ker(\GC)
        }}
        ^{\mathclap{\text{harmonic Airy potential}}}
        }^{\ker(\GC)\,\,\text{Gauss--Codazzi satisfying}}
        \phantom{\oplus}
        \overbrace{
        \phantom{\im(\star_1^{-1}\GC_\clamped^*)}
        }^{\mathclap{\substack{\text{from clamped}\\\text{Airy potential}}}}
        \\[-1.5em]
        \fX^1(M) = & 
        \underbrace{\im(D_\clamped)}_{\mathclap{\substack{\text{from clamped}\\\text{displacement}}}}
        \oplus
        \underbrace{\cH_{\GCBC}\fX^1(M)}_{\mathclap{\substack{\text{harmonic strain}\\\text{with GCBC}}}}
        \oplus
        \underbrace{\im(\star_1^{-1}\GC^*)\cap\ker(\GC) \oplus \im(\star_1^{-1}\GC^*_{\clamped})}_{\mathclap{\text{from Airy portential}\,\, \im(\star_1^{-1}\GC^*)}}
        \\[-1.5em]
        &
        \phantom{\im(D_\clamped)}
        \phantom{\oplus}
        \underbrace{
        \underbrace{\phantom{\cH_\GCBC\fX^1(M)\oplus\im(\star_1^{-1}\GC^*)\cap\ker(\GC)}}_{\mathclap{\cH\fX^1(M)\,\,\text{harmonic strain}}}
        \phantom{{}\oplus\im(\star_1^{-1}\GC_\clamped^*)}
        }_{\ker(D^*\star_1)\,\,\text{corresponding stress is in equilibrium}}
        \end{aligned}
    \end{align}
    The space \(\fX^1(M)^*\) of stress fields share the same decompositions via the \(\star_1\) isomorphism.
\end{corollary}

\begin{corollary}
    Harmonic strains with traction-free/GCBC boundary conditions represent absolute and relative cohomology:
    \begin{align}
        &\cH_\tractionfree\fX^1(M)\cong H^1(M;\se(3)) = {\ker(\GC)\over\im(D)}\\
        &\cH_\GCBC\fX^1(M)\cong H^1(M,\partial M;\se(3)) = {\ker(\GC)_\GCBC\over\im(D_\clamped)}\\
        &\cH_{\GCBC}\fX^1(M)^*\cong H^1(M;\se(3)^*) = {\ker(D^*)\over\im(\GC^*)}\\
        &\cH_{\tractionfree}\fX^1(M)^*\cong H^1(M,\partial M;\se(3)^*) = {\ker(D^*)_\tractionfree\over\im(\GC^*_\clamped)}
    \end{align}
\end{corollary}
Each of these absolute or relative cohomology groups are given by the tensor product of the six-dimensional space \(\se(3)\) with the corresponding absolute/relative de Rham cohomology of \(M\).  In particular, for an oriented surface with a given genus and a nonzero number of boundary components:
\begin{align}
    \dim(H^1(M;\se(3))= \dim(H^1(M,\partial M;\se(3)) = 6(2{\rm genus} + \#{\rm boundaries} - 1)
\end{align}

While the absolute 0th cohomology \(H^0(M;\se(3))\) is \(\se(3)\), the relative cohomology is different.  For a connected \(M\) with \(\partial M\neq\varnothing\), we have 
\begin{align}
    H^0(M,\partial M;\se(3))=\ker(D_\clamped)=\{0\}
\end{align}
as the clamped boundary condition removes the rigid body kernel.  Similarly, \(H^0(M,\partial M;\se(3)^*) = \ker(\GC_\clamped^*) = \{0\}\).

\begin{remark}
    The clamped boundary for \(\fX^0\) and GCBC for \(\fX^1\) can be applied to only a portion \(\partial M_{\clamped}\subset\partial M\) of the boundary, and leaving the complement \(\partial M_\free = \partial M\setminus \partial M_\clamped\) free.  The above machinery continues to work for such mixed boundary conditions.  The complementary boundary conditions, namely clamped condition for Airy potential and traction-free condition for stress field, are imposed on \(\partial M_\free\).  We expect the relative cohomology groups are replaced by \(H^\bullet(M,\partial M_\clamped;\se(3))\). 
\end{remark}

\section{Conclusion}
We developed an elasticity complex for Kirchhoff--Love shell theory from a geometric and sheaf-theoretic perspective. Building on Bonnet's fundamental theorem for surfaces, we showed that the deformation map and the linearized Gauss--Codazzi operator form a fine resolution of the sheaf of infinitesimal rigid motions. Taking the adjoint of this sequence yields a dual complex, revealing an Airy potential representation of equilibrium stress fields. The Green identities associated with these operators further provide a systematic derivation of natural boundary conditions for displacements, strains, stresses, and Airy potentials.

The resulting framework establishes a Hodge--Morrey--Friedrichs decomposition for shells with boundary, revealing a concrete interplay between kinematic compatibility, static equilibrium, boundary conditions, and the topology of the shell. In particular, the decomposition identifies orthogonal components arising from displacements, Airy potentials, and harmonic fields associated with the absolute and relative cohomology of the shell.

This geometric framework also extends naturally to a discrete shell complex that preserves the underlying geometric and cohomological structures. The resulting discrete Airy representation generalizes the classical Maxwell--Cremona theory in graphic statics and provides a versatile toolbox for shell simulation. We derive the discrete shell complex in the appendix \ref{appendix-discrete-closed} and \ref{appendix_discrete-boundary}.

\bibliographystyle{plainnat}
\bibliography{Reference}

\appendix

\section{Discrete Elastic Complex on Closed Surfaces}
\label{appendix-discrete-closed}

In the discrete setting, we discretize the surface as a triangle mesh. The space of all
deformations is a 3-vector per triangle vertex. Then we adapt Discrete Elastic Shells as
invented by \cite{Grinspun:2003:DS}, where the discrete first fundamental form is represented by edge lengths, and the discrete second fundamental form is given by the dihedral angle between two triangles.
Then the stretching strain is the change in edge length and the bending strain is the
change of dihedral angle. 

A shell is discretized as a triangle mesh with vertices $V$, edges $E$, and faces $F$.
Each edge $e$ has an orientation $(i, j) = (\mathrm{src}(e), \mathrm{dst}(e))$ where $i, j$ are vertices. The vertex
positions are denoted $\vec{x}_i$, and the edge lengths are denoted $\ell_e$. We denote the
triangle corner angle $\alpha_i^f$ inside triangle $f \in F$ at $i \in V$, and the dihedral
angle on edges as $\beta_e$ for $e \in E$. Face normals are denoted $\vec{n}_f$.


The variation of the positions is encoded as a 3-vector per triangle vertex $\mathring x$.
The variation of the first fundamental form is the change of triangle edge lengths, denoted
$\mathring \ell$. The variation of the second fundamental form is the change of the bending
(dihedral) angle between two neighbouring triangles, denoted $\mathring \beta$.

The discrete differential operator $\rm D$ maps $\mathring x$ to $(\mathring \ell,
\mathring \beta)$. The second operator $\rm GC$ is the discrete linearized Gauss--Codazzi
operator, mapping the space of $(\mathring \ell, \mathring \beta)$ to the Gauss--Codazzi
residual $\mathring r$, which is a 3-vector per triangle vertex.

We use first-order elements. 
The cochain spaces are
$$C^0 = \mathbb{R}^{3 |V|} \ni \mathring \bx,
\qquad
C^1 = \mathbb{R}^{|E|} \oplus \mathbb{R}^{|E|} \ni \vcat{\mathring \bl}{\mathring \bbeta},
\qquad
C^2 = \mathbb{R}^{3 |V|} \ni \mathring \br,$$
$$
C^0 \xrightarrow{\rm D\ } C^1 \xrightarrow{\rm GC\ } C^2,
\qquad
{\rm D} = \begin{bmatrix} \rm D_L \\ \rm D_B \end{bmatrix},
\qquad
{\rm GC} = \begin{bmatrix} \rm GC_L & \rm GC_B \end{bmatrix},
$$
where the aggregated data are $\mathring \bx \in \mathbb{R}^{3|V|}$,
$\mathring \bl \in \mathbb{R}^{|E|}$, $\mathring \bbeta \in \mathbb{R}^{|E|}$, and
$\mathring \br \in \mathbb{R}^{3|V|}$. Sometimes it is helpful to stack the stretching and
bending strain vectors together as the total strain $\vcat{\mathring \bl}{\mathring \bbeta}$.

\subsection{Discrete Differential Operator}

The discrete deformation map $\rm D$  consists of two parts,
\begin{align}
    {\rm D} \coloneqq
    \begin{bmatrix}
        {\rm D_L}\\
        {\rm D_B}
\end{bmatrix} \in \RR^{(2\sfE) \times (3\sfV)} .
\end{align}
$\rm D_L$ represents the linear map that takes $\mathring{\bx}$ to $\mathring{\bl}$, and $\rm D_B$
maps $\mathring{\bx}$ to $\mathring{\bbeta}$.

\subsubsection{Stretching Block}

$\rm D_L$ is the differential of edge length,
$$
({\rm D_L} \mathring\bx)_e \;=\; \mathring{\ell}_e \;=\; \hat{e}_e \cdot
\bigl( \mathring x_{v_j} - \mathring x_{v_i} \bigr),
$$
where $\hat{e}_e = (X_j - X_i)/\ell_e$ is the unit edge vector of $e$, and
$i = \mathrm{src}(e)$, $j = \mathrm{dst}(e)$. The only nonzero blocks of row $e$ are
$$
{\rm D_L}(e, i) = -\,\hat{e}_e^{\top}, \qquad {\rm D_L}(e, j) = \hat{e}_e^{\top}.
$$

\subsubsection{Bending Block}

For an edge $e$,
$({\rm D_B} \mathring\bx)_e = \mathring \beta_e =
\sum_{v \in \{i,j,k,l\}} \nabla_{\!v}\beta_e \cdot \mathring x_v$,
where $i, j, k, l$ are the vertices of the two triangles sharing the edge $e$.

\begin{wrapfigure}[5]{r}{0.2\textwidth}
    \centering
\includegraphics[scale=0.5]{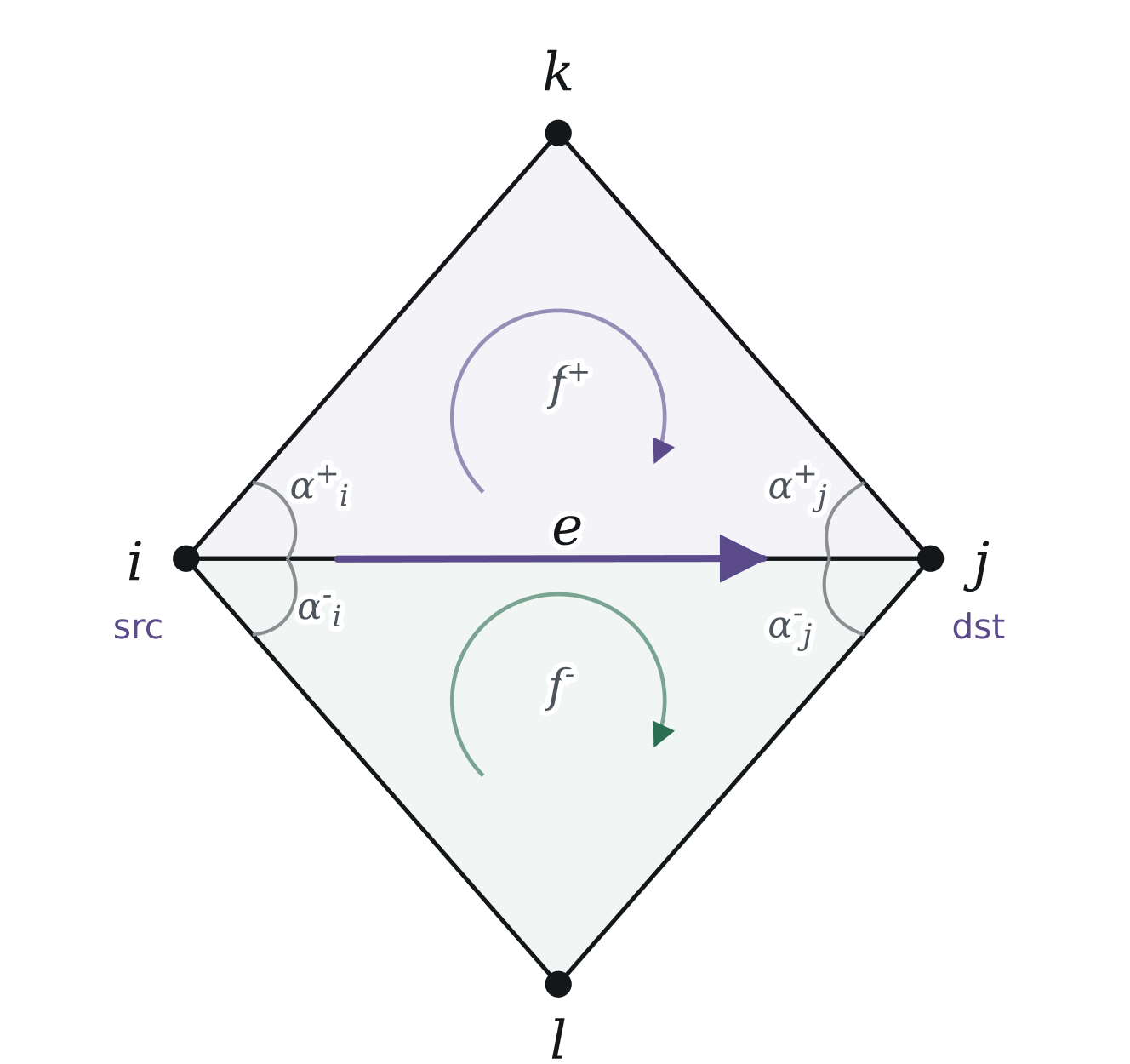}
    \label{fig:bending}
\end{wrapfigure}

We identify $f^+$ as the triangle positively oriented with respect to the edge $e$'s
orientation --- the one whose cyclic order contains $i \to j$ --- and $f^-$ as the triangle
containing $j \to i$. Then at each row $e$, the only non-zero entries are at columns
$i, j, k, l$:
$$
\begin{aligned}
\nabla_{\!i}\beta_e &=
  \frac{1}{\ell_e}\Bigl( \cot\alpha^{f^{+}}_{j}\, \vec{n}_{f^{+}}
                       + \cot\alpha^{f^{-}}_{j}\, \vec{n}_{f^{-}} \Bigr), \\[2pt]
\nabla_{\!j}\beta_e &=
  \frac{1}{\ell_e}\Bigl( \cot\alpha^{f^{+}}_{i}\, \vec{n}_{f^{+}}
                       + \cot\alpha^{f^{-}}_{i}\, \vec{n}_{f^{-}} \Bigr), \\[2pt]
\nabla_{\!k}\beta_e &=
  -\frac{1}{\ell_e}\bigl( \cot\alpha^{f^{+}}_{i} + \cot\alpha^{f^{+}}_{j} \bigr)
  \vec{n}_{f^{+}}, \\[2pt]
\nabla_{\!l}\beta_e &=
  -\frac{1}{\ell_e}\bigl( \cot\alpha^{f^{-}}_{i} + \cot\alpha^{f^{-}}_{j} \bigr)
  \vec{n}_{f^{-}} .
\end{aligned}
$$
The four gradients sum to zero, as they must for translation invariance. Using
$\cot\alpha_i^{f} + \cot\alpha_j^{f} = \ell_e^2 / (2A_f)$, the last two collapse to
$\nabla_{\!k}\beta_e = -\vec{n}_{f^+} / h_k$ and
$\nabla_{\!l}\beta_e = -\vec{n}_{f^-} / h_l$, where $h_k$ is the height of $k$ above $e$.

\subsection{Discrete Linearized Gauss--Codazzi Equation}

The discrete version of the Gauss--Codazzi equation has been previously adopted in the
graphics community as a ``non-linear'' rotation-invariant coordinate
\cite{wang2012GC, Sassen19thesis, Sassen2020} for reconstruction and shape space
representation. On a discrete triangle mesh, the Gauss--Codazzi equation can be expressed
as a statement on whether an edge vector rotated across the 1-ring neighbourhood comes back
to its original position.

Using unit quaternions in global coordinates, it reads
$$
e^{\frac{\alpha_1}{2}\vec{n}_1}\,
e^{\frac{\beta_1}{2}\vec{e}_1}\,
e^{\frac{\alpha_2}{2}\vec{n}_2} \cdots
e^{\frac{\beta_k}{2}\vec{e}_k} = -1.
$$
Here, the right-hand side is \(-1\) instead of \(1\) by the discrete spin theory of \cite{Chern2018}.
Taking the variation with respect to $\mathring{\alpha}$ and $\mathring{\beta}$, and
dropping the common factor $\tfrac12$,
$$
\mathring{\alpha}_1 \vec{n}_1 + \mathring{\beta}_1 \vec{e}_1
+ \mathring{\alpha}_2 \vec{n}_2 + \cdots + \mathring{\beta}_k \vec{e}_k = 0 .
$$
Then our discrete linearized Gauss--Codazzi operator is
\begin{align}
    \GC \coloneqq
    \begin{bmatrix}
        {\rm GC_L} & {\rm GC_B}
    \end{bmatrix}
    \in\RR^{(3\sfV) \times (2\sfE)} .
\end{align}
Essentially, at vertex $v$ we aggregate all the triangle interior angles at corner $v$
multiplied by the triangle normal, together with all the dihedral angles on edges incident
to $v$ multiplied by the unit edge vector.

\begin{figure}[h]
\centering
\includegraphics[scale=0.3]{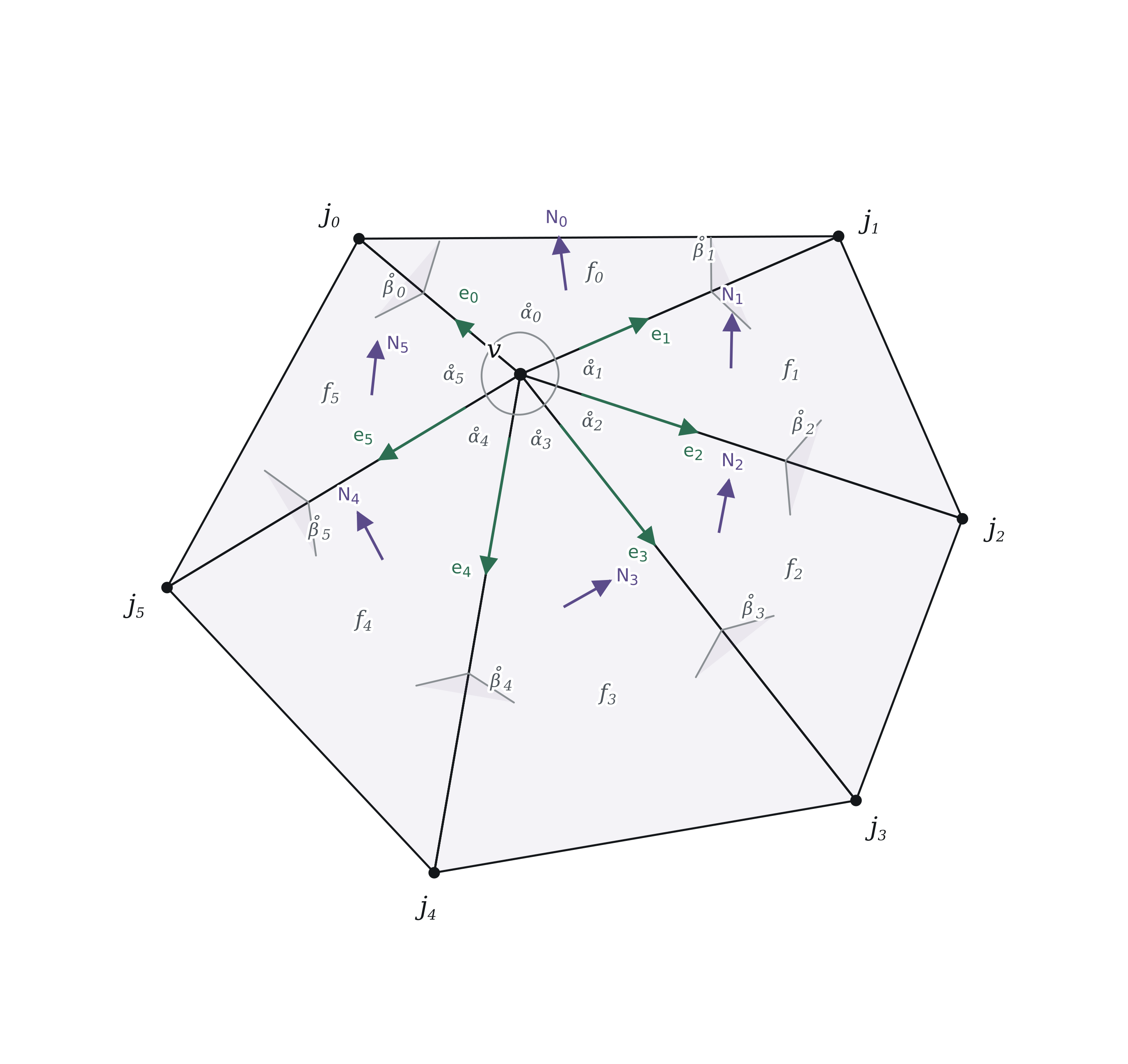}
\caption{A illustration of the quantities to compute the linearized Gauss-Codazzi residual around a vertex $v$.}
\label{fig:gc_3d}
\end{figure}

\subsubsection{Stretching Block}

Varying the three edge lengths of a triangle $f = ijk$ changes its angle at $i$ by the
variation of the law of cosines,
\begin{equation}
  \mathring \alpha_i^{f} :=
    \frac{\ell_{jk}}{2 A_{f}}\, \mathring \ell_{jk}
    \;-\; \frac{c_j^{f}}{\ell_{ij}}\, \mathring \ell_{ij}
    \;-\; \frac{c_k^{f}}{\ell_{ki}}\, \mathring \ell_{ki} ,
  \label{eq:dalpha}
\end{equation}
where $c_j^f = \cot \alpha_j^f$, so that the edge opposite the corner enters with weight
$\ell / 2A_{f}$, and each adjacent edge with minus the cotangent of the angle at its far
end, divided by its own length. Summing these over the triangles at a vertex, weighted by
their normals, gives the block of $\rm GC_L$ at vertex $i$:
\begin{equation}
  \bigl( {\rm GC_L}\, \mathring\bl \bigr)_i := \sum_{f \ni i} \vec{n}_f\; \mathring\alpha_i^{f} .
  \label{eq:gca}
\end{equation}

\subsubsection{Bending Block}

The bending block simply multiplies $\mathring\beta$ on each edge by the unit edge vector
of that edge:
\begin{equation}
  \bigl( {\rm GC_B}\, \mathring\bbeta \bigr)_i :=
  \sum_{e \in N(i)} \mathrm{sign}(e, i)\;\hat{e}_e\; \mathring\beta_{e} ,
  \label{eq:gcb}
\end{equation}
where $\mathrm{sign}(e, i) = 1$ if $i = \mathrm{src}(e)$ and $-1$ if $i = \mathrm{dst}(e)$.

\subsection{Material Property as Hodge Star}

We adopt the energy model from dicrete elastic shells \cite{Grinspun:2003:DS}, where the total energy is $W = W_S + W_B$, where $W_S$ is the membrane or strecthing energy and $W_B$ is the bending energy
\begin{align}
    W_{\rm S} = \sum_{\sfe\in\sfE}{1\over 2}Y_{\rm S}A_{\sfe}\varepsilon_\sfe^2 = \sum_{\sfe\in\sfE}{1\over 4}Y_{\rm S}{h_\sfe\over \ell_\sfe}\mathring\ell_\sfe^2,
\end{align}
where \(A_{\sfe}\) is the sum of \(1/3\) of the triangle area from both side (or from one side if it's a boundary edge).  That is,
$A_{\sfe} = {1\over 2}\ell_{\sfe}h_{\sfe}$, where $l_e$ is the edge length and $h_e$ is the length of the dual edge of $e$.

And the bending energy is 
\begin{align}
    W_{\rm B} = \sum_{\sfe\in\mathsf{intE}} {1\over 2}Y_{\rm B}A_{\sfe}\kappa_{\sfe}^2 = \sum_{\sfe\in\mathsf{intE}}{1\over 4}Y_{\rm B}{\ell_\sfe\over h_\sfe}\mathring\beta_\sfe^2.
\end{align}
This gives us diagonal ``Hodge star'' that maps from the strains to stresses on triangle edges
\begin{align}
    \star_\sfe = 
    \begin{bmatrix}
        {Y_{\rm S}\over 2}{h_\sfe\over\ell_\sfe}\\
        &{Y_{\rm B}\over 2}{\ell_\sfe\over h_\sfe}
    \end{bmatrix}.
\end{align}
Given Young's modulus \(E\), Poisson ratio \(\nu\), and thickness \(t\):
\begin{align}
    Y_{\rm S}\approx {Et\over 1-\nu^2},\quad Y_{\rm B}\approx {Et^3\over 12(1-\nu^2)}.
\end{align}

\subsection{Cohomology}

\begin{theorem}
$C^0 \xrightarrow{\rm D\ } C^1 \xrightarrow{\rm GC\ } C^2$ is a chain complex.
\end{theorem}
Together with the dual complex, we have a discrete analogue of \eqref{eq:FullShellComplex}
\[
\begin{tikzcd}
\mathfrak{se}(3) \arrow[r]
  & \mathring \bx \arrow[r, "\rm D"] \arrow[d, dashed, "\star_0"'] 
  & \vcat{\mathring \bl}{ \mathring \bbeta} \arrow[r, "\rm GC"] \arrow[d, dashed, "\star_1"] 
  & \mathring \br \arrow[r]  \arrow[d, dashed, "\star_2"] 
  & \mathfrak{se}(3)\\
\mathfrak{se}(3)^* & \bold{f} \arrow[l] & \vcat{\btau}{\bsigma} \arrow[l, "-D^T"] & \brho \arrow[l, "GC^T"] & \mathfrak{se}(3)^* \arrow[l]\\
\end{tikzcd}
\]
This discrete shell complex has cohomology
$$H^0 = \ker(D),\quad H^1 = \ker(GC)/\im(D),\quad H^2 \cong\coker(GC)$$
with
$$(H^0, H^1, H^2) = (6, 12g , 6).$$
The discrete Atiyah--Singer index for the chain complex (in the sense of the associated Hodge--Dirac operator) is given by
$$h^0 - h^1 + h^2  =  3|V| - 2|E| + 3|V|  =  6|V| - 2|E| = 6\chi.$$

\subsection{Strain Field Decomposition on Closed Surfaces}

Any strain field $\vcat{\mathring \bl}{\mathring \bbeta}$ on a closed surface splits into
three parts,
$$
\vcat{\mathring \bl}{\mathring \bbeta}
\;=\; \underbrace{{\rm D}\, \mathring\bx}_{\text{exact}}
\;\oplus\; \underbrace{\bH}_{\text{harmonic}}
\;\oplus\; \underbrace{\star_1^{-1}\,{\rm GC}^{\top}\star_2\, \mathring \br}_{\text{coexact}}.
$$
This decomposition is orthogonal with respect to the inner product $\star_1$. Here
$\star_1^{-1}{\rm GC}^{\top}\star_2$ is the adjoint of $\rm GC$ under $\star_1$ and
$\star_2$. The exact part is the strain of an actual displacement. The coexact part is
the strain generated by a stress potential. The harmonic part is what is left: it is
closed, ${\rm GC}\,\bH = 0$, and co-closed, ${\rm D}^{\top}\star_1 \bH = 0$, but it is
neither the strain of a displacement nor the image of a potential.

We have the following algorithm for computing strain field decomposition on surfaces by solving two elliptic PDEs:

\begin{algorithm}
\caption{Strain field decomposition on a closed surface}
\label{alg:decomp}
\begin{algorithmic}[1]
\Require total strain $\vcat{\mathring \bl}{\mathring \bbeta}$; operators $\rm D$, $\rm GC$; Hodge stars $\star_1$, $\star_2$
\Ensure $\vcat{\mathring \bl}{\mathring \bbeta} = \vcat{\mathring \bl}{\mathring \bbeta}_{\rm exact} \oplus \bH \oplus \vcat{\mathring \bl}{\mathring \bbeta}_{\rm coexact}$,
        orthogonal with respect to $\star_1$
\State $\mathring\bx \;\gets\;
       \textsc{Solve}\bigl({\rm D}^{\top}\star_1 {\rm D},\;\;
       {\rm D}^{\top}\star_1 \vcat{\mathring \bl}{\mathring \bbeta}\bigr)$
       \Comment{pin $\mathfrak{se}(3)$}
       \label{line:solve0}
\State $\vcat{\mathring \bl}{\mathring \bbeta}_{\rm exact} \;\gets\; {\rm D}\,\mathring\bx$
\State $\mathring\br \;\gets\;
       \textsc{Solve}\bigl({\rm GC}\star_1^{-1}{\rm GC}^{\top}\star_2,\;\;
       {\rm GC}\,\vcat{\mathring \bl}{\mathring \bbeta}\bigr)$
       \Comment{pin $\mathfrak{se}(3)$}
       \label{line:solve2}
\State $\vcat{\mathring \bl}{\mathring \bbeta}_{\rm coexact} \;\gets\; \star_1^{-1}{\rm GC}^{\top}\star_2\,\mathring\br$
\State $\bH \;\gets\; \vcat{\mathring \bl}{\mathring \bbeta} - \vcat{\mathring \bl}{\mathring \bbeta}_{\rm exact} - \vcat{\mathring \bl}{\mathring \bbeta}_{\rm coexact}$
\State \Return $\vcat{\mathring \bl}{\mathring \bbeta}_{\rm exact},\; \bH,\; \vcat{\mathring \bl}{\mathring \bbeta}_{\rm coexact}$
\end{algorithmic}
\end{algorithm}

\begin{remark}
${\rm D}^{\top}\star_1 {\rm D}$ and ${\rm GC}\star_1^{-1}{\rm GC}^{\top}$ has a 2nd order part for membrane that behaves like the laplacian and 4th order part for bending that behaves like the bi-laplacian.
    Both systems are
symmetric positive semi-definite with $\mathfrak{se}(3)$ as its kernel.
In practice, we solve the least-norm solution for linear systems, effectively remove the \(\se(3)\) part in \(\mathring\bx\) and \(\mathring\br\).
\end{remark}

\subsection{Finding a Basis for Harmonic Strain and Stress}
To get a set of harmonic basis, we can employ a singular value decomposition (SVD). By definition
$\bH \in \ker({\rm GC}) \cap \ker({\rm D}^{\top}\star_1)$, so we need the intersection of
two null spaces, taken in the $\star_1$ metric. Absorbing the metric with
$\tilde\bH = \star_1^{1/2}\bH$ turns both conditions Euclidean, so they can be stacked and
handled by a single SVD of
$$
{\rm A} =
\begin{bmatrix}
    {\rm GC}\,\star_1^{-1/2}\\[2pt]
    \bigl(\star_1^{1/2}{\rm D}\bigr)^{\top}
\end{bmatrix},
\qquad
\bH = \star_1^{-1/2}\,\ker({\rm A}).
$$

Alternatively, 
since our cohomology is a sheaf cohomology of the resolution of $\se(3)$, we can adpot the classical tree-cotree algorithm for efficient computation. The tree-cotree algorithm identifies a set of homology generator, and assigns a locally exact 1-form on the zig-zag path on the homology generator. This locally exact 1-form is obtained by setting one side of the potential be 0 and the other side of the potential to be 1, and only compute the gradient of this potential on the edges connecting the 2 sides. In our case, we simply set one side of the $\mathring{x}$ to be the 0 vector and the other side of the $\mathring{x}$ to be a vector from $se(3)$. By only computing the strain of this potential on the edges connecting the 2 sides, we created a locally exact strain with no global potential exist on non-trivial topology. Doing this for all $2g$ generators and all $6$ basis elements of
$\mathfrak{se}(3)$ gives $12g$ fields, matching $h^1$.
These $12g$ fields span $H^1$, but they are not yet the harmonic representatives. To get
those, run Algorithm~\ref{alg:decomp} on each one and keep $\bH$; since the algorithm is
$\star_1$-orthogonal, what comes back is the $\star_1$-harmonic representative of the same
class.

\section{Discrete Elastic Complex on Surfaces with Boundaries}
\label{appendix_discrete-boundary}

The discrete Gauss-Codazzi equation is a condition purely on the interior points of the surface. Our variational Gauss-Codazzi equation is also a constraint on the interior points. 
We have previously covered in the continuous theory that there are harmonic strains with natural boundary conditions or clamped boundary conditions. We can also have harmonic strains with a separate pinned boundary condition.
Each boundary condition requires different treatment.

\begin{figure}[h]
\centering
\includegraphics[scale=0.4]{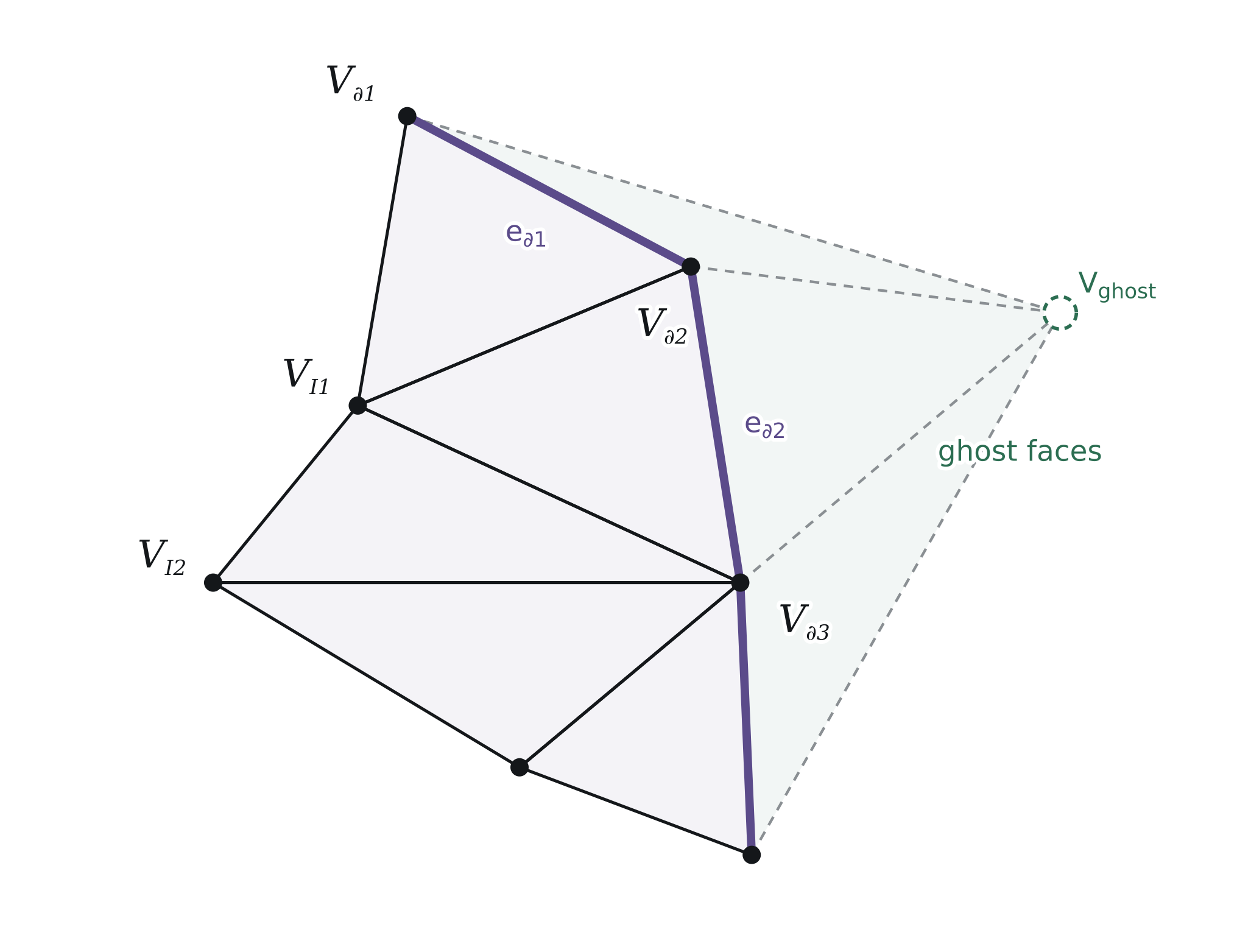}
\caption{A bounded patch. Boundary vertices and edges are $v_{\partial_i}$,
$e_{\partial_i}$; interior ones are $v_{I_j}$, $e_{I_j}$. Note that $v_{I_1}$ is adjacent
to a boundary triangle while $v_{I_2}$ is not.}
\label{fig:boundary}
\end{figure}

We will refer to the vertices and edges on the boundary as $v_{\partial i}$ and $e_{\partial i}$; and the interior vertices and edges as $v_{Ij}$ and $e_{Ij}$. 
Among all the interior vertices, we also need to differentiate between vertices $v_{I_1}$ and $v_{I_2}$, where $v_{I_1}$ is directly adjacent to a boundary triangle and $v_{I_2}$ is further in the interior. 
Essentially we divide the edges into sets $E_I \oplus E_\partial$ and $V_I \oplus V_\partial = (V_{II} \oplus V_{I\partial}) \oplus V_\partial$. 

We will refer to boundary strecthing as $\mathring{l}_\partial$, and the set of degress of freedom of those boundary strecthing strain as $E_{\partial_\text{stretch}}$,and boundary bending strain as $E_{\partial_\text{bend}}$. 
$\mathring{l}_\partial \in E_{\partial_\text{stretch}}$, and $\mathring{\beta}_\partial \in E_{\partial_\text{bend}}$.

To compute absolute cohomology, we had to pick an equivalence class of $\begin{bmatrix}
    \mathring{\bl}\\ \mathring{\bbeta}
\end{bmatrix}$ since we are quotieting out the boundary bending. We need no such treatment for the clamped case but we also need this for the pinned case.

The relative cohomology $H^k(M, \partial M)$ measures the gap between closed and exact among
the strains and stresses that vanish on the boundary. 
Algebraically, computing a relative cohomology means identifying a sub-complex. In our
discretization both clamped cohomology and pinned cohomology are taken relative to the \emph{same}
sub-complex, with boundary bending quotiented out for the pinned case. 

\subsection{ Discrete Absolute Cohomology}

Since the boundary bending is meaningless in the case of free-boundary condition, the space of strains that we want to decompose is in fact $\begin{bmatrix}\mathring{\bl}_I \\ \mathring{\bl}_\partial \\ \mathring{\bbeta}_I \\ \mathring{\bbeta}_\partial\end{bmatrix} /\begin{bmatrix} \mathring{\bbeta}_\partial\end{bmatrix}$. 
Rather than working with the quotient directly, we fix a gauge and work with a representative.
We take the representative in which the bending on every boundary edge vanishes. Our operator $D_\text{gauge }$ needs to reflect this choice.
Then, for the discrete $GC$ operator, we need to make sure that we construct a consistent mapping from this equivalent class of strain to the correct equivalent class of Airy potentials. We enforce this by defining $GC_\text{gauge}$, which prescribe a single equation
per boundary vertex stating that the adjacent boundary bending vanishes. Below is the matrix construction of these two operators, ordering the
columns as stretching, interior bending, boundary bending,

\begin{align}
    {\rm D}_{\rm gauge}\coloneqq
    \begin{bmatrix}
        {\rm D_L}\\
        {\rm D_B}\big|_{E_I}\\
        0
    \end{bmatrix}\in\RR^{(2\sfE)\times(3\sfV)} .
\end{align}

\begin{align}
    \GC_{\rm gauge} \coloneqq
    \begin{bmatrix}
        {\rm GC_L} & {\rm GC_B}\big|_{E_I} & 0\\[2pt]
        0 & 0 & {\rm GC}_{\rm gauge}
    \end{bmatrix}
    \in \RR^{(3|\mathsf{intV}| + |\mathsf{bdryV}|) \times (2\sfE)} ,
\end{align}
where ${\rm GC}_{\rm gauge}(i,j) = 1$ when boundary vertex $i$ is paired with boundary
edge $j$.
This means that the only strain/stress that will satisfy the closed condition are the ones that come from displacement and have exactly 0 bending on the boundary, effectively picking the right representative.

\begin{remark}
    \(GC_{\rm gauge}\) is not looking at integrability at boundary vertex; its purpose is to pick one equivalent class for us to work with. Indeed by construction, \(GC_{\rm essential}D_{\rm gauge} = 0\).  And conversely, if \(GC_{\rm essential}E = 0\) then the boundary bending part of \(E\) must be zero, and it is (locally) in the image of \(D_{\rm gauge}\).  
\end{remark}

When solving for the ellipitic PDEs constructed with these two opeartors, there is no extra treatment for the boundary condition required. Solving the discrete linear system, the exact part inherits the boundary values of the input and the coexact part satisfies
the essential condition.  In particular, the remainder is a harmonic field with natural boundary
conditions. 

$\rm GC_\text{gauge}  \rm D_\text{gauge}$ forms a chain complex, and we can get the following decomposition:

\begin{proposition}[H natural]
For the free boundary we
work on the quotient $\mathcal{E}/\mathcal{E}^{\rm b}_\partial$, with ${\rm D}_{\rm gauge}$
and $\GC_{\rm gauge}$
    \begin{align}
\begin{bmatrix}\mathring{\bl}_I \\ \mathring{\bl}_\partial \\ \mathring{\bbeta}_I \\ \textbf{0}\end{bmatrix} 
\;=\; {\rm D}_{\rm gauge}\, \mathring\bx
\;\oplus\; \bH_{\rm natural}
\;\oplus\; \star_1^{-1}\,\GC_{\rm gauge}^{\top}\star_2\, \vcat{\mathring \br_I}{\textbf{0}} ,
\label{eq:bdrydecomposition-absolute}
\end{align}
with $\bH_{\rm natural}$ characterised by $\GC_{\rm gauge}\bH = 0$ and
${\rm D}_{\rm gauge}^{\top}\star_1 \bH = 0$.
\end{proposition}
Here \(\bH_{\rm natural}\) corresponds to \(\cH_\tractionfree\fX^1(M)\) of the continuous theory.

\subsubsection{dimension and basis }
 By counting ranks of this pair of matrices, we get the dimension of $(H^0, H^1, H^2) = (6\;, 6(2g+b-1), \;0)$, exactly the dimension of (discrete) absolute sheaf cohomology of $\mathfrak{se}(3)$.

Absolute cohomologies are the cohomologies dual to the relative homologies, which are the cross-sections on the surface. To find a basis, one could run tree-cotree to find a relative homology generator loop \cite{Chern2018}, and prescribe the 6 mode of infinitesimal rigid-body motion across it.

\subsection{Discrete Clamped Cohomology}
\label{sec:clamped}
 
A clamped boundary prescribes all essential boundary conditions: $\mathring x$ and $\mathring x \cdot \vec{n}$. We model the clamp as a rigid \emph{ghost face} attached
across every boundary edge, whose \emph{ghost vertex} $v_\text{ghost}$ never moves
(Figure~\ref{fig:boundary}). The clamped complex is then the relative complex of $M$
with respect to a subcomplex $S$ which we will describe next.

\subsubsection{Choice of Subcomplex}

Our choice of the sub-complex to be quotient out consists of the ghost vertices outside of the mesh, as well as all the related degrees of freedom. First, consider the meaning of $GC$. For $\mathring{r}$ on each vertex, it aggregates bending mode on edges directly adjacent to it, and the stretching mode on the edges across the face it is connected to. This means that if we were to apply $GC$ on a ghost vertex outside the mesh, for boundary edges, only it's strecthing mode will touch the ghost vertex. This means that the strain/stress degree of freedom that we need to remove are exactly these boundary strecthing mode. As shown in the illustration, only the bending mode of the blue edges will touch $v_\text{ghost}$, while the strecthing modes don't. Correspondingly, we also need to remove all the 3 degrees of freedom of $\mathring{x}$ on boundary vertices, because they touches all the boundary strecthing mode, and they will create pinned boundary condition.
 
Removing $\mathring{r}$ at the ghost vertices therefore forces us to remove the
boundary stretching modes. Each boundary stretching mode in turn reads only the
displacements of its two endpoints, so we must also remove all three components of
$\mathring{x}$ at every boundary vertex; this is what pins the boundary.

To summarize, the DOFs of the sub-complex that we need to remove are: (boundary vertices' $\mathring{x}$, boundary edge strecthing mode, ghost vertices's $\mathring{r}$). We call this sub-complex $S$.



\subsubsection{Constructing ${\rm D}_{\rm clamped}$ and ${\rm GC}_{\rm clamped}$}
Ordering the rows as interior stretching, interior bending, boundary bending,
\begin{align}
    {\rm D}_{\rm clamped}\coloneqq
    \begin{bmatrix}
        {\rm D_L}\big|_{E_I \times V_I}\\[2pt]
        {\rm D_B}\big|_{E_I \times V_I}\\[2pt]
        {\rm D}_{\partial\beta}\big|_{E_\partial \times V_I}
    \end{bmatrix}
    \in\RR^{(|E_I| + |E|) \times 3|V_I|} .
    \label{eq:Dclamped}
\end{align}
The columns run over interior vertices only. Removing the boundary displacements is what
imposes the pinned half of the clamped condition, and it is already why $h^0$ will come out
zero: no rigid motion survives that fixes the boundary. The boundary stretching rows are
gone as well, where on a closed surface every edge contributes one; they come back as a
column deletion in $\GC$ below, and it is the two deletions together that keep the pair a
complex.

The first two blocks are restrictions of the closed-surface blocks. The entries of  ${D_{\partial\mathring{\beta}}}$ is essentially the same formula as how the interior edge bendings are calculated, but with the contribution of the ``ghost face'' removed. It now has contribution from one face only. For a
boundary edge $e = (i,j)$ with incident face $f$ and opposite vertex $k$,
\begin{align}
    \mathring{\beta}_{\partial e}
    = \frac{1}{\ell_e}\,\vec{n}_f \cdot
      \Bigl(
        \cot\alpha^{f}_{j}\,\mathring{x}_i
        + \cot\alpha^{f}_{i}\,\mathring{x}_j
        - \bigl(\cot\alpha^{f}_{i} + \cot\alpha^{f}_{j}\bigr)\mathring{x}_k
      \Bigr) ,
    \label{eq:bdrybend}
\end{align}.

Equation~\eqref{eq:bdrybend} vanishes if and only if the clamped boundary condition is
satisfied, which is what makes this discretization convenient. Singling out the
$\mathring{x}\cdot\vec{n}$ degree of freedom directly would require a local coordinate
system at every boundary vertex, and even then the frame is only approximate, since
neighbouring vertices have different tangent planes. Equation~\eqref{eq:bdrybend} reaches
the same condition without ever the need of defining a point normal.

Then for $\rm GC_\text{clamped}$, we simply construct it as if all vertices are interior vertices. This means that, for a boundary vertex,$v_{\partial_0}$ this $GC$ operator is measuring can we rotate 1 boundary edge around the faces $v_{\partial_0}$ is adjacent to to the other boundary edge.
\begin{align}
    \GC_{\rm clamped}\coloneqq
    \begin{bmatrix}
        {\rm GC_L}\big|_{V \times E_I} &
        {\rm GC_B}\big|_{V \times E}
    \end{bmatrix}
    \in\RR^{3|V| \times (|E_I| + |E|)} .
    \label{eq:GCclamped}
\end{align}
The boundary stretching columns are deleted, as we remove any DOF in subcomplex $S$.

$ \rm D_{\rm clamped}$ and $\GC_{\rm clamped}$ are our operator constrained to $M \setminus S$, and they form a chain complex.

\begin{remark}
    Note that our convention for the label ``clamped'' is different from that in the continuous theory.
\end{remark}

\begin{remark}
    Even though our clamped boundary condition for displacements only restrict those on boundary vertices to be zero, the linear system \(D_\clamped^T\star_1 D_\clamped\mathring\bx = {\rm external force}\) converges to a displacement with continuous clamped condition, due to the second order nature of the \(D\) operator.
\end{remark}

When we set the boundary displacement to be 0, it will converge to the clamped case as we refine the mesh.
Computationally, for finite resolution,
to compute the clamped relative cohomology, we simply modify the hodge star $\star_1$ where we multiply the boundary bending weight with a large number, making it stiff. We call this modified metric $\star_{1_\text{clamped}}$. 

To get a clamped exact strain/stress, we solve for 
\begin{equation}
\label{eq:clampedsolve}
    \rm D^\top \star_{1_\text{clamped}} {\rm D} \mathring{\bx} = \star_1 \begin{bmatrix}
    \mathring{\bl}\\\mathring{\bbeta}
\end{bmatrix},
\end{equation}
note that the Dirichelet boundary condition is implicitly imposed on the ghost vertices. The difference of hodge star here is important, as this treatment will give us the truely clamped $\mathring \bx$

But to get the harmonic strain, we need to subtract the exact strain solved by 
\begin{equation}
    \rm D^\top \star_{1_\text{clamped}} {\rm D} \mathring{\bx} = \star_{1_\text{clamped}} \begin{bmatrix}
    \mathring{\bl}\\\mathring{\bbeta}
\end{bmatrix}
\end{equation}

To obtain the clamped harmonic strain, we also need to remove a free-boundary condition coexact part, which can be solved through the following with $\mathfrak{se}(3)$ pinned:

$$\GC_\text{clamped}\,\star_{1_\text{clamped}}^{-1}\GC_\text{clamped}^{\top}\star_2\,\mathring{\br} \;=\; \GC_\text{clamped}\,\begin{bmatrix}
    \mathring{\bl}\\\mathring{\bbeta}
\end{bmatrix}.$$

We have the following decomposition: 
\begin{proposition}[H clamped]
For the clamped boundary the strain space is not a quotient, but the boundary stretching
splits off as its own summand, and the metric is $\star_{1_{\rm clamped}}$:
\begin{align}
\begin{bmatrix}\mathring{\bl}_I \\ \textbf{0} \\ \mathring{\bbeta}_I \\ \mathring{\bbeta}_\partial\end{bmatrix} 
\;=\;  {\rm D}_{\rm clamped}\, \mathring\bx
\;\oplus\; \bH_{\rm clamped}
\;\oplus\; \star_{1_{\rm clamped}}^{-1}\,\GC_{\rm clamped}^{\top}\star_2\, \mathring \br ,
\label{eq:hodge-clamped}
\end{align}
with $\GC_{\rm clamped}\bH = 0$ and
${\rm D}_\text{ clamped}^{\top}\star_{1_{\rm clamped}} \bH = 0$. Note that $\rm D_\text{clamped} \mathring \bx$ converges to clamped displacement as the mesh refines, and if one can use \ref{eq:clampedsolve} for a clamped displacement under finite resolution.
\end{proposition}

\subsubsection{dimension and basis}  By counting ranks
of this pair of matrices, we get the dimension of $(H^0, H^1, H^2) = (0,\; 6(2g + b -1), \;6)$, exactly the dimension of (discrete) relative cohomology of $\mathfrak{se}(3)$ with pinned boundary condition. A set of basis can be computed numerically via svds or by pumping $\mathfrak{se}(3)$ across a absolute homology generator loop, and compute the decomposition.


\subsection{Discrete Pinned Cohomology}

Recall a pinned boundary constraints the boundary displacement while allowing free bending.  The condition corresponds to removing the bending DOFs on the boundary. This requires same treatment as for the absolute (free boundary) case, where computationally will need to pick a gauge and make it consistent with $D$ and $GC$. Let's for now just call them $D_\text{pinned}$ and $GC_\text{pinned}$.

The sub-complex to remove is the same as the clamped case, were the boundary ring of vertices and boundary edge's strecthing mode will touch the "ghost" vertices' computation of the Gauss-Codazzi constraints. 

Then, the relative cohomology of the pinned case can essentially be understood as the abosolute cohomology of the interior part plus the boundary degrees of freedom. We removed the boundary vertices in $D$ and we also quotient out the boundary Airy potentials on the vertices for $GC$, so:

$$H^0, H^1, H^2 = (0, 6(2g+b -1) + \text{extra boundary DOFs}, 0)$$. 

There are $2|V_\partial|$ of them (3 DOFs per vertex, minus one stretch constraint per boundary edge). 6 of them correspond to the movement generated by $\mathfrak{se}(3)$, which are consistent with the interior motion that generates nothing. That leaves $2|V_\partial| - 6$ boundary classes. On top of these are the $6(2g + b - 1)$ topological classes of the absolute cohomology, so 
$\dim \bH_{\text{pinned}} = 6\,(2g + b - 1) + 2 |V_\partial| - 6$

This gives us the following decomposition:

\begin{proposition}[H pinned]
\begin{align}
\begin{bmatrix}\mathring{\bl}_I \\ \textbf{0} \\ \mathring{\bbeta}_I \\ \textbf{0}\end{bmatrix} 
\;=\;  {\rm D}_{\rm pinned}\, \mathring\bx
\;\oplus\; \bH_{\rm pinned}
\;\oplus\; \star_{1}^{-1}\,\GC_{\rm pinned}^{\top}\star_2\, \vcat{\mathring \br_I}{\textbf{0}} ,
\label{eq:hodge-pinned}
\end{align}
with $\bH_{\rm pinned}$ being $\bH_{\rm absolute}$ on the interior and the extra boundary DOFs.
\end{proposition}

\end{document}